\documentclass[amssymb,twocolumn,aps,prd,twocolumn,superscriptaddress,floatfix,nopacs,preprintnumbers,nofootinbib,10pt]{revtex4-1}
\usepackage[utf8]{inputenc}

\usepackage{graphicx}
\usepackage[figuresright]{rotating}
\usepackage{bm,amsmath,amssymb}
\usepackage[mathscr]{eucal}
\usepackage{multirow}
\usepackage{xcolor}
\usepackage{mathrsfs}
\usepackage{mathtools}
\usepackage{slashed}
\usepackage{graphics}
\usepackage{graphicx}

\usepackage[normalem]{ulem}
\usepackage{xcolor}
\definecolor{lcolor}{rgb}{0.5,0,0}
\definecolor{citcolor}{rgb}{0,0.3,0.0}
\usepackage{nicefrac}
\usepackage[breaklinks,colorlinks,urlcolor=blue,citecolor=citcolor,linkcolor=lcolor]{hyperref}
\usepackage{mciteplus}

\usepackage{subfigure}
\usepackage{dsfont}
\usepackage{longtable}
\usepackage{bbm} 
\usepackage{float}
\usepackage{tcolorbox}
\usepackage{lipsum}
\usepackage{cancel}
\usepackage{enumerate}

\newcommand{\Hcal}{\mathcal{H}}
\newcommand{\Ocal}{\mathcal{O}}
\newcommand{\Mcal}{\mathcal{M}}
\newcommand{\Fcal}{\mathcal{F}}
\newcommand{\Acal}{\mathcal{A}}
\newcommand{\Ncal}{\mathcal{N}}
\newcommand{\Tcal}{\mathcal{T}}
\newcommand{\Scal}{\mathcal{S}}
\newcommand{\Ccal}{\mathcal{C}}
\newcommand{\Pcal}{\mathcal{P}}

\newcommand{\etL}[1]{\boldsymbol{\epsilon}_{\perp #1}}
\newcommand{\etU}[1]{\boldsymbol{\epsilon}_{\perp}^{#1}}

\newcommand{\ztL}[1]{\boldsymbol{z}_{\perp #1}}

\newcommand{\rtL}[1]{\boldsymbol{r}_{\perp #1}}
\newcommand{\rtU}[1]{\boldsymbol{r}_{\perp}^{#1}}

\newcommand{\btL}[1]{\boldsymbol{b}_{\perp #1}}

\newcommand{\btCL}[1]{\boldsymbol{b}'_{\perp #1}}

\newcommand{\rtCL}[1]{\boldsymbol{r}'_{\perp #1}}

\newcommand{\rtC}{\boldsymbol{r}'_{\perp}}

\newcommand{\ptU}[1]{\boldsymbol{p}_{\perp}^{#1}}

\newcommand{\gammatU}[1]{\boldsymbol{\gamma}_{\perp}^{#1}}

\newcommand{\vect}[1]{\boldsymbol{#1}_{\perp}}
\newcommand{\kt}{\vect{k}}
\newcommand{\pt}{\vect{p}}

\newcommand{\lt}{\vect{l}}

\newcommand{\bt}{\vect{b}}
\newcommand{\Bt}{\vect{B}}

\newcommand{\xt}{\vect{x}}
\newcommand{\yt}{\vect{y}}
\newcommand{\zt}{\vect{z}}

\newcommand{\rt}{\vect{r}}

\newcommand{\der}{\mathrm{d}}

\newcommand{\Tr}{\mathrm{Tr}}

\begin{document}

\author{Vincent Cheung}
\email{cheung27@llnl.gov}
\affiliation{
Nuclear and Chemical Sciences Division, Lawrence Livermore National Laboratory, Livermore, California 94551, USA}
\author{Zhong-Bo Kang}
\email{zkang@physics.ucla.edu}
\affiliation{Department of Physics and Astronomy, University of California, Los Angeles, CA 90095, USA}
\affiliation{Mani L. Bhaumik Institute for Theoretical Physics, University of California, Los Angeles, CA 90095, USA}
\author{Farid Salazar}
\email{faridsal@uw.edu}
\affiliation{Institute for Nuclear Theory, University of Washington, Seattle WA 98195-1550, USA}
\affiliation{Nuclear Science Division, Lawrence Berkeley National Laboratory, Berkeley, CA 94720, USA}
\affiliation{Physics Department, University of California, Berkeley, CA 94720, USA}
\affiliation{Department of Physics and Astronomy, University of California, Los Angeles, CA 90095, USA}
\affiliation{Mani L. Bhaumik Institute for Theoretical Physics, University of California, Los Angeles, CA 90095, USA}
\author{Ramona Vogt}
\email{rlvogt@lbl.gov}
\affiliation{
Nuclear and Chemical Sciences Division, Lawrence Livermore National Laboratory, Livermore, California 94551, USA}
\affiliation{
Department of Physics and Astronomy, University of California, Davis, Davis, CA 95616, USA}

\title{Direct quarkonium production in DIS from a joint CGC and NRQCD framework}

\begin{abstract}
We compute the differential cross section for direct quarkonium production in high-energy electron-nucleus collisions at small $x$. Our computation is performed within the nonrelativistic QCD factorization formalism that separates the calculation into short distance coefficients and long distance matrix elements that depend on the color and spin of the state.  We obtain the short distance coefficients of the production of the heavy quark pair within the framework of the Color Glass Condensate effective field theory, which resums coherent multiple interactions of the heavy quark pair with the nucleus to all orders. Our results are expressed as the convolution of perturbatively calculable perturbative functions with multi-point light-like
Wilson line correlators.  In the correlation limit, we establish the correspondence between our CGC formulation with calculations employing the transverse
momentum dependent (TMD) framework.  We extend this correspondence by resumming kinematic power corrections within the improved TMD framework, which interpolates between the TMD formalism and $k_\perp$-factorization formalism. We present a detailed numerical analysis, focusing on $J/\psi$ production in the kinematics accessible at the future Electron-Ion Collider, highlighting the importance of genuine higher-order saturation contributions when the electron collides with a large nucleus.

\end{abstract}

\maketitle


\section{Introduction}

Over the last few decades, high-energy nuclear and particle physics collider experiments have extensively studied the landscape of quantum chromodynamics (QCD). Major efforts have been devoted to elucidating the structure of protons and nuclei in terms of their fundamental constituents: quarks and gluons, collectively known as partons. It is well known that the probability of finding gluons that carry momentum fraction $x$ of the hadron rapidly grows at smaller values of $x$, and it is conjectured that recombination effects in QCD result in gluon saturation suppressing this growth \cite{Gribov:1984tu,Mueller:1985wy}. The large gluon occupation number at small $x$ suggests that a more natural characterization of the degrees of freedom of this regime is in terms of dense fields instead of the usual partonic picture. The Color Glass Condensate (CGC) is an effective field theory of QCD which provides a systematic way to study the dynamics of these fields and their consequences on particle production across different colliding systems \cite{Iancu:2003xm,Gelis:2010nm,Kovchegov:2012mbw,Albacete:2014fwa,Blaizot:2016qgz,Morreale:2021pnn}. An important feature of the CGC is the emergence of a semi-hard, energy and nuclear-size dependent, momentum scale $Q_s$ known as the saturation scale. Momentum modes of the gluon fields with transverse momenta less than or comparable to the saturation scale are suppressed, which in turn has an imprint on particle production. A promising tool to probe the signatures of gluon saturation is the direct production of quarkonia in high-energy collisions.  The mass of the quarkonium state, such as $J/\psi$, is similar in magnitude to the expected saturation scale reached at small $x$ in collider experiments while still sufficiently hard to allow for a controllable perturbative expansion.   

Various research efforts have been carried out to study quarkonium produced with large transverse momentum within the collinear perturbative QCD formalism jointly with different mechanisms for the formation of the quarkonium, such as nonrelativistic QCD (NRQCD) \cite{Bodwin:1994jh,Butenschoen:2009zy,Nayak:2006fm,Ma:2013yla,Ma:2014eja}, the color evaporation model \cite{Gavai:1994in,Schuler:1996ku,Nelson:2012bc,Lansberg:2006dh}, and the improved color evaporation model \cite{Ma:2016exq,Cheung:2017loo,Cheung:2017osx,Cheung:2018tvq,Cheung:2018upe,Cheung:2021epq,Cheung:2022nnq,Cheung:2024bnt}. The collinear pQCD approach has been extended to incorporate next-to-leading power corrections \cite{Kang:2011mg,Kang:2014tta,Kang:2014pya,Ma:2014svb,Lee:2022anw,Fleming:2012wy} as well as next-to-leading order corrections in $\alpha_s$ \cite{Butenschoen:2010rq,Ma:2010yw,Gong:2008sn,Flore:2020jau,ColpaniSerri:2021bla}. On the other hand, in the low transverse momentum regime, quarkonium production has been studied using the transverse momentum dependent (TMD) factorization formalism \cite{Mukherjee:2015smo,Mukherjee:2016cjw,Bacchetta:2018ivt,Boer:2021ehu,Kishore:2018ugo,Boer:2020bbd,Kishore:2021vsm,Boer:2023zit} (see \cite{DAlesio:2019qpk,Kishore:2022ddb,Maxia:2024cjh,Chakrabarti:2022rjr} for quarkonium accompanied by a jet or a photon), and soft-collinear effective theory \cite{Fleming:2006cd,Fleming:2019pzj,Echevarria:2024idp}. 

In the forward/small-$x$ regime, quarkonium production has been investigated within the $k_\perp$ factorization formalism \cite{Catani:1990eg,Collins:1991ty,Levin:1991ry,Baranov:2011ib,Babiarz:2019mag,Babiarz:2020jkh} in terms of unintegrated gluon distributions and off-shell gluon partonic matrix elements. The unintegrated gluon distribution obeys the Balitsky-Fadin-Kuraev-Lipatov (BFKL) equation \cite{Lipatov:1976zz,Kuraev:1977fs,Balitsky:1978ic} resumming large energy logarithms (for other approaches to high-energy factorization following the collinear framework see \cite{Lansberg:2021vie,Lansberg:2023kzf,Celiberto:2023fzz}). On the other hand, at sufficiently low values of $x$ or in nuclear environments, we expect that a more appropriate description of particle production is provided by the CGC/saturation formalism, which captures the physics of multiple scattering as well as non-linear QCD evolution equations the Jalilian-Marian-Iancu-McLerran-Weigert-Leonidov-Kovner (JIMWLK) equations \cite{JalilianMarian:1996xn,JalilianMarian:1997dw,Kovner:2000pt,Iancu:2000hn,Iancu:2001ad,Ferreiro:2001qy} and their mean field approximation the Balitsky-Kovchegov (BK) equation \cite{Balitsky:1995ub,Kovchegov:1999yj}. In the CGC effective theory, direct quarkonium production studies have focused on high-energy proton-proton and proton-nucleus\footnote{Nuclear modification to quarkonium production in a cold QCD environment have also been investigated in \cite{Ferreiro:2008wc,Ferreiro:2013pua} following a collinear or a $k_\perp$ factorized approach combined with the nuclear modification of the gluon densities in nuclei (see also \cite{Lansberg:2016deg} for a more general approach).} collisions at RHIC and the LHC \cite{Kharzeev:2005zr,Kharzeev:2008nw,Dominguez:2011cy,Kharzeev:2012py,Fujii:2013gxa,Ducloue:2015gfa,Boussarie:2017oae,Levin:2019fvb,Qiu:2013qka,Watanabe:2015yca,Ma:2018qvc,Stebel:2021bbn} (see also \cite{Boussarie:2018zwg} for diffractive studies). In particular, the combined framework of CGC + NRQCD developed in \cite{Kang:2013hta}, and its subsequent phenomenological studies have successfully described the particle spectra in the semi-hard regime, $p_\perp \lesssim 5\ \mathrm{GeV}$, as well as the rapidity distribution \cite{Ma:2014mri,Ma:2015sia,Ma:2017rsu,Ma:2018bax,Salazar:2021mpv,Gimeno-Estivill:2024gbu}. Meanwhile, most studies of quarkonium production in deep inelastic scattering (DIS) and photo-production within the saturation framework have been devoted to diffractive production \cite{Munier:2001nr,Kowalski:2003hm,Kowalski:2006hc,Lappi:2010dd,Goncalves:2011vf,Mantysaari:2016ykx,Goncalves:2017wgg,Cepila:2019skb,Boussarie:2019vmk,Mantysaari:2020lhf,Xing:2020hwh,Bendova:2020hbb,Hentschinski:2020yfm,Mantysaari:2022sux,Mantysaari:2022kdm,Boer:2023mip,Mantysaari:2023prg,Mantysaari:2023xcu,Benic:2023ybl,Brandenburg:2024ksp}.

\begin{center}
\begin{figure}
    \centering
    \includegraphics[width=0.45\textwidth]{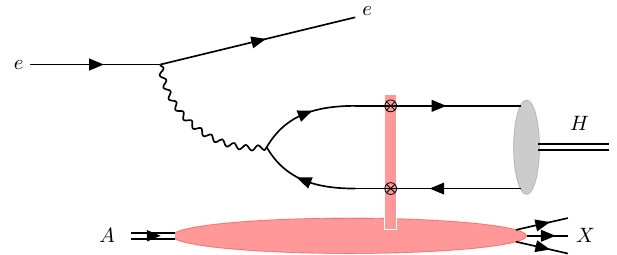}
    \caption{Schematic diagram of quarkonium production in high-energy electron-nucleus collisions. The elongated red oval represents the CGC effective interaction of the heavy quark-antiquark pair with the nuclear target. The gray oval represents the formation of the quarkonium in NRQCD.} 
    \label{fig:schematic-quarkonium-DIS-CGC-NRQCD}
\end{figure}
\end{center}

In this paper, we compute, for the first time, the direct quarkonium production in electron-nucleus collisions at small $x$ within the joint CGC + NRQCD framework (see Fig.\,\ref{fig:schematic-quarkonium-DIS-CGC-NRQCD}). Our computation of the short distance coefficients in the CGC allows us to resum the coherent multiple interactions of the heavy quark pair to all orders in the small-$x$ gluon background field of the nucleus. This paper is organized as follows. In Sec.\,\ref{sec:theoretical_framework} we briefly introduce our conventions and the kinematic variables for the process under consideration. We present a convenient decomposition of our differential cross section expressed in terms of the density matrix for quarkonium production in virtual photon-nucleus collision. We then review the basic theoretical tools for our computation: the CGC effective field theory and the NRQCD formalism. 

We review the computation of the leading order amplitude for the production of the heavy quark pair within the CGC in Sec.\,\ref{sec:QQbar_CGC}. The computations of the NRQCD short distance coefficients for the differential cross section of quarkonium production are performed in Sec.\,\ref{sec:projectionQQbar}, where we include both color octet and singlet contributions, as well as $S$ and P wave states. Our results are expressed as the convolution of a color-dependent CGC distribution which encodes the properties of the scattering of the heavy quark pair with the small-$x$ background field of the target, and perturbative functions which depend on the polarization of the virtual photon and the spin state of the heavy quark pair.  We comment on the origin of $k_\perp$-factorization breaking when the saturation scale $Q_s^2$ is comparable to the other semi-hard scales in the process.

In Sec.\,\ref{sec:TMD_factorization} we show that in the small $p_\perp$ limit our results are consistent with those obtained within TMD factorization at small $x$ provided the saturation scale $Q_s^2$ is also sufficiently smaller than the hard scales $Q^2$ and $M_{\mathcal{Q}}^2$. In this limit, the differential cross section is expressed as the product of hard function and the small-$x$ Weizsäcker-Williams (WW) transverse-momentum-dependent gluon distribution,  consistent with the results obtained in \cite{Bacchetta:2018ivt}. Furthermore, following the strategy in \cite{Boussarie:2020vzf,Boussarie:2021ybe} we resum the kinematic twists ($p_\perp^2/Q^2$ and $p_\perp^2/M_{\mathcal{Q}}^2$) to all orders, obtaining analytic expressions for the improved TMD (ITMD) hard functions. 
We present a numerical analysis of our results in Sec.\,\ref{sec:numerical_analysis} where we study the short distance coefficients relevant for $J/\psi$ production, as well as the differential cross section. We study their dependencies on the transverse momentum $p_\perp$ and the virtuality $Q$. We compare the results of the full CGC, TMD, and ITMD calculations. We present our conclusions and potential avenues for future work in Sec.\,\ref{sec:summary_outlook}.

Lastly, our manuscript is supplemented with several appendices. In Appendix\,\ref{app:projections} we briefly outline the computation of the spin projections necessary to obtain the perturbative factors. Then, in Appendix\,\ref{app:hard_factors} we collect the final results for the perturbative functions in the full CGC calculation, as well as the hard functions in the TMD and ITMD limits.  In Appendix\,\ref{app:Gaussian} we briefly review the computation of light-like Wilson line correlators with the Gaussian approximation.  Finally, in Appendix \ref{app:additional_numerical_results} we provide supplementary numerical results for the short distance coefficients.

\section{Theoretical Framework}

\label{sec:theoretical_framework}
We begin this section by defining the kinematics and notations employed throughout this paper and remind the reader of the decomposition for particle production in DIS expressed in terms of the sub-hadronic virtual photon-nucleus scattering. We then briefly review the basic elements of the CGC effective theory and the NRQCD formalism that we will use to compute direct quarkonium production.

\subsection{Kinematics and notations}

The 4-momenta of the nucleus and the exchanged space-like photon are denoted by $P_A$ and $q$ respectively. We work in a frame where they move along the $z$-axis, the photon has a large $q^+$ component, and the nucleus has a large $P_A^-$ component,
\begin{align}
    P_A^\mu &= \left(0, P_A^-, \vect{0} \right) \,, \\
    q^\mu &= \left( q^+, -\frac{Q^2}{2q^+}, \vect{0} \right) \,.
    \label{eq:definition_PAq}
\end{align}
We ignore the mass of the nucleus, and $Q^2=-q^2$ is the virtuality of the photon.

Let $p_1$ and $p_2$ be the 4-momenta of heavy quark and antiquark, respectively.  We define $p$ as the total momentum and $k$ as half the relative momentum,
\begin{align}
    p &= p_1 + p_2 \,, \nonumber \\
    k &= \frac{1}{2}(p_1 - p_2) \,.
    \label{eq:definition_pq}
\end{align}
The on-shell conditions,
\begin{align}
    p_1^2 = \left( \frac{p}{2} + k\right)^2 = m_Q^2 \,, \nonumber \\
    p_2^2 = \left( \frac{p}{2} - k\right)^2 = m_Q^2 \,,
\end{align}
where $m_Q$ is the mass of the heavy quark, demand
\begin{align}
    p^\mu k_{\mu} &= 0 \,,\nonumber \\
    p^2 &= 4 (m_Q^2 - k^2) \,.
\end{align}
The total momentum $p$ will correspond to the momentum of the produced quarkonium, and the relative momentum $k$ provides an expansion parameter in NRQCD to calculate the different states of orbital angular momentum of the heavy quark-antiquark pair. The quarkonium mass $M_{\mathcal{Q}} $ equals twice the mass of the heavy quark $m_Q$. Thus after expansion around $k=0$ we have $p^2 = M_{\mathcal{Q}}^2$.

\subsection{Lepton-hadron tensor decomposition in the photon polarization basis}
We compute direct quarkonium production $H$ in deep inelastic electron-nucleus scattering at small $x$
\begin{align}
    e(k_e) + A(P_A) \to e(k_e') + H(p) + X \,. \label{eq:DIS}
\end{align}
As is conventional in small-$x$ calculations, we compute the sub-hadronic process:
\begin{align}
    \gamma^{*}(q,\lambda) + A(P_A) \to  H(p) + X \,, \label{eq:gammaA}
\end{align}
where $\lambda$ denotes the polarization of the virtual photon. 

Following the decomposition in Sec.~2 in  \cite{Mantysaari:2020lhf}, the DIS process in Eq.\,\eqref{eq:DIS} and the $\gamma^{*}+A$ process in Eq.\,\eqref{eq:gammaA} are related by
\begin{widetext}
\begin{align}
    \frac{\der \sigma^{eA\to e H +X}}{\der Q^2 \der y \der p_\perp^2  \der \phi_{e H}}  = \frac{\alpha_{\mathrm{em}}}{ 2\pi^2 Q^2 y  }  
    &\left\{ (1-y) \frac{\der \sigma^{\gamma^{*} A\to H+X}_{\mathrm{L}}}{\der p_\perp^2 } + \frac{1}{2} \left[1 +(1-y)^2 \right] \frac{\der \sigma^{\gamma^{*} A\to H+X}_{\mathrm{T}}}{\der p_\perp^2 }  \right. \nonumber \\ 
    & \left. + \sqrt{2(1-y)} (2-y)  \frac{\der \sigma^{\gamma^{*} A\to H+X}_{\mathrm{TL}}}{\der p_\perp^2} \cos \phi_{e H}   +  (1-y) \frac{\der \sigma^{\gamma^{*} A\to H+X}_{\mathrm{\mathrm{Tflip}}}}{\der p_\perp^2} \cos 2 \phi_{e H}   \right\} \,,
    \label{eq:decomposition_eA_gammaA}
\end{align}
\end{widetext}
where $\phi_{eH}=\phi_e-\phi_H$ is the relative azimuthal angle between the electron and the produced quarkonium, $y$ is the inelasticity $y = (q\cdot P_A)/(k_e\cdot P_A)$ and $x_{\rm Bj}=Q^2/(y s)$. In the decomposition in Eq.\,\eqref{eq:decomposition_eA_gammaA},  we introduce the following shorthand notation:
\begin{align}
    \der \sigma^{\gamma^{*} A\to H+X}_{\mathrm{L}} & = \der \sigma^{\gamma^{*} A\to H+X}_{0,0} \,, \nonumber \\
   \der \sigma^{\gamma^{*} A\to H+X}_{\mathrm{T}} & = \overline{\sum} \der \sigma^{\gamma^{*} A\to H+X}_{\lambda, \lambda} \nonumber \\ 
   \der \sigma^{\gamma^{*} A\to H+X}_{\mathrm{TL}} & =  \overline{\sum} \der \sigma^{\gamma^{*} A\to H+X}_{0, \lambda} \,, \nonumber \\
   \der \sigma^{\gamma^{*} A\to H+X}_{\mathrm{Tflip}} & =  \overline{\sum} \der \sigma^{\gamma^{*} A\to H+X}_{-\lambda, \lambda} \,,
    \label{eq:density_matrix_gammaA}
\end{align}
where $\der \sigma^{\gamma^{*} A\to H+X}_{\lambda\lambda'}$ is the ``density matrix" for direct quarkonium production in $\gamma^{*} A$ scattering. $\lambda$ and $\lambda'$ refer to the polarization of the virtual photon in the amplitude and complex conjugate amplitude respectively. We also introduce the shorthand notation $\overline{\sum} = \frac{1}{2}\sum_{\lambda = \pm 1}$. The diagonal elements $\lambda=\lambda'$ correspond to the differential cross section for quarkonium production in the scattering of a nucleus with a virtual photon with definite polarization $\lambda$ and the off-diagonal terms correspond to quantum interference terms which are accessible in the DIS azimuthal correlations as seen in Eq.\,\eqref{eq:decomposition_eA_gammaA}. The evaluation of the expressions in Eq.\,\eqref{eq:density_matrix_gammaA} within the joint CGC + NRQCD framework is one of the principal results of this manuscript.

In this manuscript, we work in light-cone gauge, $A^+=0$, of the photon field where the polarization vectors are
\begin{align}
    \epsilon^{\mu}(q,\lambda=0) = \left(0, \frac{Q}{q^+}, \vect{0} \right) \,,
\end{align}
for longitudinally polarized photons, and
\begin{align}
    \epsilon^{\mu}(q,\lambda=\pm 1) = \left( 0,0, \vect{\epsilon}^{\lambda} \right) \,,
\end{align}
for transversely polarized photons, with two-dimensional transverse vector $\vect{\epsilon}^{\lambda}= (\cos(\phi), i\lambda \sin(\phi))$, where $\phi$ is an arbitrary angle, whose dependence will disappear at the level of the cross-section. To obtain the specific form in Eq.\,\eqref{eq:decomposition_eA_gammaA} we have chosen $\phi=\phi_H$ the azimuthal angle of the produced quarkonium \footnote{In \cite{Mantysaari:2020lhf} we chose $\phi = 0$ in such case one has to extract the angular dependence from the sub-hadronic matrix elements.}. With this choice the expressions in Eq.\,\eqref{eq:density_matrix_gammaA} are independent of the angle $\phi_H$, and all the angular dependence is explicit in the azimuthal modulations $\cos(n\phi_{eH})$ in Eq.\,\eqref{eq:decomposition_eA_gammaA}.

\subsection{Color Glass Condensate}
\label{sec:Feynmanrules}

In the Color Glass Condensate, the large $x$ partons (with rapidities $Y<Y_0$, where $Y = \ln(1/x)$) of the nucleus are integrated out and effectively treated as stochastic classical color charge density sources $\rho_A$. For a fast-moving nucleus along the minus component of the light-cone direction, the color sources generate a current density of the form
\begin{align}
    J^\mu(x^+,\xt) = \delta^{\mu-} \rho_A(x^+,\xt)\,,
    \label{eq:current_CGC}
\end{align}
where the sub-eikonal components of the current are neglected. In turn, this current generates the gauge field $A^{\mu}$ (referred as to the background field) which represents the small-$x$ content (partons with rapidities $Y>Y_0$) of the nucleus \cite{McLerran:1993ni,McLerran:1993ka,McLerran:1994vd,Ayala:1995kg,Ayala:1995hx}. In the CGC, the expectation value of any observable is computed from the path integral
\begin{align}
    \left \langle \Ocal \right \rangle = \int D \rho_{A} W_{Y_0}[\rho_A] \frac{\int^{Y_0} [\mathcal{D} A] \Ocal e^{i \Scal[A,\rho_A]}}{\int^{Y_0} [\mathcal{D} A] e^{i \Scal[A,\rho_A]}} \,,
\end{align}
where $W_{Y_0}[\rho_A]$ is a gauge-invariant weight functional for the distribution of the color charges $\rho_A$. The invariance of the physical observables on the arbitrary rapidity cutoff $Y_0$ results the JIMWLK  renormalization group evolution (RGE) equations \cite{JalilianMarian:1996xn,JalilianMarian:1997dw,Kovner:2000pt,Iancu:2000hn,Iancu:2001ad,Ferreiro:2001qy}.

In the semi-classical approximation, the small-$x$ color field is obtained in the saddle point approximation of the path integral by solving the classical Yang-Mills equations $[D_{\mu}, F^{\mu\nu}] = J^{\nu}$, where the current is given by Eq.\,\eqref{eq:current_CGC}. In the light-cone gauge, $A^+=0$, these equations have the solution
\begin{align}
    A^\mu(x^+,\xt) = \delta^{\mu-} \alpha_A(x^+,\xt)\,,
\end{align}
where $\alpha_A(x^+,\xt)$ satisfies the Poisson equation
\begin{align}
    \nabla_\perp^2 \alpha_A(x^+,\xt) = -\rho_A(x^+,\xt) \,.
\end{align}
Corrections beyond the semi-classical approximation can be performed systematically in perturbation theory. An important subset of these contributions yields large rapidity logarithms of the form $\alpha_s^n (Y-Y_0)^n$, where $Y$ is the physical rapidity of the observable. These potentially large corrections can be absorbed by the JIMWLK renormalization group evolution.

\begin{center}
\begin{figure}[H]
    \centering
    \includegraphics[width=0.4\textwidth]{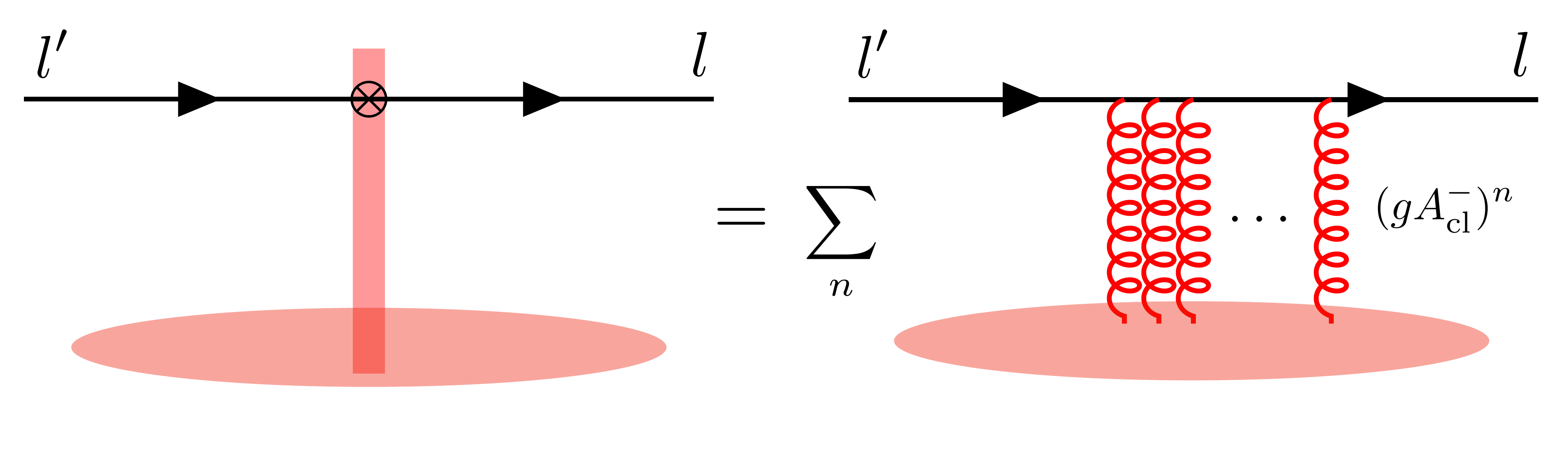}
    \caption{The effective vertex for the quark interaction with the CGC background field, accounting for multiple (eikonal) scattering off the classical gauge fields 
    $A_{\rm cl}^-$.} 
    \label{fig:CGC_effective_vertex}
\end{figure}
\end{center}
A key ingredient for computing scattering amplitudes in the CGC is the effective vertex for the eikonal interaction of fast-moving colored charged partons with the background (see Fig.\,\ref{fig:CGC_effective_vertex}). For a quark one has
\begin{align}
    \mathcal{T}^q_{\sigma\sigma',ij}(l,l') &=  (2\pi) \delta(l^+ -l'^+) \gamma^+_{\sigma\sigma'} \nonumber \\
    & \times \int \der^2\vect{z} e^{-i(\lt-\lt')\cdot \vect{z}} V_{ij}(\vect{z}) \,,
    \label{eq:effective_vertex}
\end{align}
while for an antiquark the interaction is given by
\begin{align}
    \mathcal{T}^{\overline{q}}_{\sigma\sigma',ij}(l,l') &=  -(2\pi) \delta(l^+ -l'^+) \gamma^+_{\sigma\sigma'}  \nonumber \\
    & \times \int \der^2\vect{z} e^{-i(\lt-\lt')\cdot \vect{z}} V_{ij}^{\dagger}(\vect{z}) \,,
    \label{eq:effective_vertex-anti}
\end{align}
where $l$ and $l'$ are the outgoing and incoming momenta of the quark (antiquark). The indices $i,j$ represent the color state of the outgoing and incoming quark (antiquark) respectively, and $\sigma,\sigma'$ are their Dirac indices. The light-like Wilson line in the fundamental representation appearing in the effective CGC vertex is given by the SU(3) matrices
\begin{align}
    V_{ij}(\vect{z}) &= \Pcal \exp{ \left( ig \int_{-\infty}^\infty dz^+ A^{-,c}  (z^+,\vect{z}) t^c_{ij}  \right)}\,,
\end{align}
where $A^-_c$ is the background field, $t^c_{ij}$ are the generators of SU(3) in the fundamental representations, and $\Pcal$ stands for path ordering. A similar effective vertex exists for the interaction of a gluon with the background field, this time in terms of the light-like Wilson line in the adjoint representation. However, we will not include it here as it will not appear in the calculations in this manuscript. 

\subsection{Nonrelativistic QCD}

\label{sec:NRQCD}

In NRQCD the production of quarkonium $H$ is computed by first evaluating the short distance coefficients, $\der \hat{\sigma}^{\kappa}$, for the production of a heavy quark pair in a given quantum state $\kappa = {}^{2S+1} L^{[c]}_{J} $. These states have definite spin S, orbital angular momentum $L$, total angular momentum $J$, and color state $[c]$. The short distance coefficients are then weighted by nonperturbative long-distance matrix elements (LDMEs), $\langle \mathcal{O}^{H}_{\kappa}\rangle$, and summed,
\begin{align}
    \der \sigma_H = \sum_{\kappa} \der \hat{\sigma}^{\kappa} \langle \mathcal{O}^{H}_{\kappa} \rangle \,. 
    \label{eq:NRQCD_decomp}
\end{align}
For example, for $J/\psi$ production one has
\begin{align}
\label{eq:cs}
    \der \sigma_{J/\psi} = \sum_{\kappa} \der \hat{\sigma}^{\kappa}  \langle \mathcal{O}^{J/\psi}_{\kappa} \rangle \,,
\end{align}
where only three color octet states and one color singlet state contribute to $J/\psi$ production ($\kappa = \{  {}^{1} \mathrm{S}^{[8]}_0\,,  {}^{3} \mathrm{S}^{[8]}_1\,, {}^{3} \mathrm{P}^{[8]}_J\,, {}^{3} \mathrm{S}^{[1]}_1 \}$). 

To compute the short distance coefficients $\der \hat{\sigma}^{\kappa}$ we need the projection of the $Q\overline{Q}$ amplitude to the specific quantum state $\kappa$. Following \cite{Kang:2013hta}, the amplitude of the short distance coefficient is
\begin{widetext}
    \begin{align}
    \Mcal^{\lambda, \kappa,J_z}(p)  
    & = \frac{1}{\sqrt{m_Q}} \sum_{\substack{L_z,S_z\\
    s,\overline{s},i,\overline{i}}} \langle L L_z; S S_z | J J_z \rangle \left \langle \frac{1}{2} s; \frac{1}{2} \overline{s} \Big | S S_z  \right \rangle \langle 3i;\overline{3}\overline{i} | (1,8c) \rangle \times \begin{cases}
        \Mcal^{\lambda}_{s \overline{s}, i\overline{i}}(p,0) & \mathrm{if}\ \kappa\ \mathrm{is} \ \mathrm{S\ wave}\,, \\
        \epsilon^{*}_{\beta}(L_z) \frac{\partial \Mcal^{\lambda}_{s \overline{s}, i\overline{i}}(p,k)}{\partial k_\beta}  \Big |_{k=0} & \mathrm{if}\ \kappa\ \mathrm{is} \ \mathrm{P \ wave} \,.
    \end{cases}
    \label{eq:projection_QQbar}
\end{align}
\end{widetext}
where $\Mcal^{\lambda}_{s \overline{s}, i\overline{i}}(p,k)$ is the amplitude for direct $Q\overline{Q}$ production by virtual photon-nucleus scattering in the CGC.  The polarization of the virtual photon is denoted by $\lambda$, while $s$ ($\overline{s}$) and $i$ ($\overline{i}$) denote the spin and color indices of the quark (antiquark) respectively.

As in \cite{Kang:2013hta}, we use the following normalization convention for the color states: $\langle 3i;\overline{3}\overline{i} | 1 \rangle = \delta^{i\overline{i}} /\sqrt{N_c} $ and $\langle 3i;\overline{3}\overline{i} | 8c \rangle = \sqrt{2} t_c^{i\overline{i}}$.

\section{Heavy-quark pair production in the CGC}

\label{sec:QQbar_CGC}

\begin{figure}
    \centering
    \includegraphics[width=0.45\textwidth]{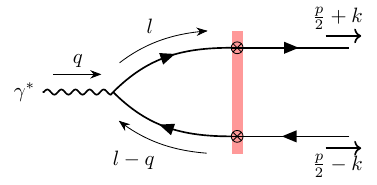}
    \caption{Leading order $Q\overline{Q}$ production by virtual photon scattering with the background field of the nucleus in the CGC. The red rectangle represents the effective interaction of the quark and antiquark with the CGC background field.}
    \label{fig:QQbar_prod_CGC}
\end{figure}
At leading order, the quarkonium is produced by the splitting of the virtual photon into a quark-antiquark pair $Q\overline{Q}$ which scatters off the nucleus and then emerges as a quarkonium state. The leading order diagram for $Q \overline Q$ production in virtual photon-nucleus collisions within the CGC is shown in Fig.~\ref{fig:QQbar_prod_CGC}. Employing standard QCD+QED Feynman rules with the effective vertex in \ref{sec:Feynmanrules}, the leading order scattering amplitude is
\begin{widetext}
    \begin{align}
    \Scal^{\lambda}_{s \overline{s}, i\overline{i}}(p,k)  = \int \frac{\der^4 l}{(2\pi)^4} \overline{u}_s\left( \frac{p}{2} + k \right) \Tcal^q_{ik}\left(\frac{p}{2}+k,l\right) S_0(l) (-ie e_Q \slashed{\epsilon}(q,\lambda))  S_0(l-q) \Tcal^{\overline{q}}_{k\overline{i}}\left(l-q,-\frac{p}{2}+k\right) v_{\overline{s}}\left(\frac{p}{2} - k\right) \,,
    \label{eq:Smatrix}
    \end{align}
\end{widetext}
where $e_Q$ is the fractional charge of the quark. The vacuum fermion propagator is $S_0(l)=i(\slashed{l}+m_Q)(l^2 - m_Q^2 + i\epsilon)^{-1}$ and $\mathcal{T}^{q(\overline q)}$ are the effective vertices in Eqs.\,\eqref{eq:effective_vertex} and \eqref{eq:effective_vertex-anti}. For simplicity, we do not explicitly write the Dirac indices.

We introduce the (reduced) amplitude $\Mcal^{\lambda}_{s \overline{s}, i\overline{i}}(p,k)$, 
\begin{align}
    &(2\pi) \delta(q^+ - p^+) \Mcal^{\lambda}_{s \overline{s}, i\overline{i}}(p,k)   
 \nonumber \\
 & = \Scal^{\lambda}_{s \overline{s}, i\overline{i}}(p,k) - \Scal^{\lambda}_{s \overline{s}, i\overline{i}}(p,k) \Big |_{\rm{nonint}} \,,
    \label{eq:reducedM}
\end{align}
where we subtract the noninteracting contribution, corresponding to equating the Wilson lines with the identity matrix $V(\xt) = \mathbbm{1}$ and $V^\dagger(\yt) = \mathbbm{1}$.  We also factor out an overall light-cone momentum conserving delta function $(2\pi) \delta(q^+ - p^+)$. Putting these ingredients together, the reduced amplitude for $\gamma_{\lambda}^{*} +A \to Q\overline{Q} + X$ production is
\begin{align}
    &\Mcal^{\lambda}_{s \overline{s}, i\overline{i}}(p,k)  = \frac{e e_Q q^+}{\pi} \int \der^2 \bt \int \der^2 \rt e^{-i \kt \cdot \rt}  e^{-i \pt \cdot \bt} \nonumber \\
    & \times \overline{u}_s\left( \frac{p}{2} + k \right) \Ncal^{\lambda}(p,k  ;\rt) v_{\overline{s}}\left(\frac{p}{2} - k\right)  \nonumber \\
    & \times \left[V\left(\bt + \frac{\rt}{2} \right) V^\dagger \left(\bt - \frac{\rt}{2} \right) -\mathbbm{1} \right]_{i\overline{i}} \,,
    \label{eq:amplitude_QQbar}
\end{align}
where we introduce the perturbative function $\Ncal^{\lambda}$
\begin{align}
    &\Ncal^{\lambda}(p,k;\rt) \nonumber \\
    & =  \int \frac{\der^4 l}{(2\pi)^2} \frac{
    -i (2q^+) T^{\lambda}(l) \delta\left(l^+ - \frac{p^+}{2} - k^+ \right) e^{i \lt \cdot \rt } }{\left[l^2 - m_Q^2 + i \epsilon \right] \left[(l-q)^2 - m_Q^2 + i \epsilon \right]} \,,
    \label{eq:hard_factor}
\end{align}
with Dirac-Lorentz structure
\begin{align}
    T^{\lambda}(l) = \frac{1}{(2q^+)^2}  \left[ \gamma^+ (\slashed{l} + m_Q) \slashed{\epsilon}(q,\lambda) (\slashed{l} -\slashed{q} + m_Q) \gamma^+ \right] \,.
    \label{eq:Dirac-Lorentz}
\end{align}
The integral over the internal momenta $l$ can be easily worked out, and the results for longitudinally and transversely polarized photons are
\begin{align}
    &\Ncal^{\lambda=0}(p,k; \rt) \nonumber \\
    &= -  \left(\frac{1}{2} + \xi \right)\left(\frac{1}{2} - \xi \right)  Q K_0(\overline{Q}_{\xi} |\rt|)  \frac{\gamma^+}{q^+} \,,
    \label{eq:NcalL}
    \\
    &\Ncal^{\lambda=\pm 1}(p,k ; \rt) \nonumber \\
    &=
    -\frac{i \rtL{\alpha}}{|\rt|} \overline{Q}_{\xi} K_1(\overline{Q}_{\xi} |\rt|) \etL{\beta}^{\lambda}   \left[ \frac{1}{4} [\gammatU{\alpha},\gammatU{\beta}]  + \xi \delta_\perp^{\alpha\beta} \right]  \frac{\gamma^+}{q^+}   \nonumber \\
    &- \frac{1}{2} m_Q  K_0(\overline{Q}_{\xi} |\rt| ) \etL{\alpha}^{\lambda} \gammatU{\alpha} \frac{\gamma^+}{q^+}  \,,
    \label{eq:NcalT}
\end{align}
respectively, where $\xi = k^+/q^+$ and we introduce the effective virtuality
\begin{align}
    \overline{Q}_{\xi}^2 = \left(\frac{1}{2} + \xi \right)\left(\frac{1}{2} - \xi \right)Q^2 + m_Q^2 \,.
\end{align}

\section{Quarkonium production in CGC + NRQCD}
\label{sec:projectionQQbar}

The amplitude for the short distance coefficients in the CGC is obtained by combining the results in Eqs.\,\eqref{eq:amplitude_QQbar} and \eqref{eq:projection_QQbar}, giving
\begin{align}
    & \Mcal^{\lambda, \kappa,J_z}(p,k) \nonumber \\
    &= \frac{e e_Q q^+}{\pi} \int \der^2 \rt    \Fcal^{\lambda, \kappa,J_z}(p,\rt) \int \der^2 \bt e^{-i \pt \cdot \bt} \nonumber \\
    & \times \Tr\left[ \left( V\left(\bt + \frac{\rt}{2} \right) V^\dagger \left(\bt - \frac{\rt}{2} \right) -\mathbbm{1} \right) \Ccal^{\kappa} \right] \,,
    \label{eq:amplitude_projected}
\end{align}
where we define the color projector
\begin{align}
    \Ccal^k = \begin{cases}
        \mathbbm{1}/\sqrt{N_c} & \mathrm{if}\ \kappa\ \mathrm{is} \ \mathrm{singlet} \,, \\
        \sqrt{2} t_c & \mathrm{if}\ \kappa\ \mathrm{is} \ \mathrm{octet} \,,
    \end{cases}
    \label{eq:ColorProjector}
\end{align}
and the perturbative functions
\begin{widetext}
    \begin{align}
    \Fcal^{\lambda, \kappa,J_z}(p,\rt) =& \sum_{L_z,S_z}  \langle L L_z; S S_z | J J_z \rangle  \times \begin{cases}
        \Tr\left[ \Pi^{SS_z}(p,k=0)  \Ncal^{\lambda}(p, k=0; \rt)  \right] & \mathrm{if}\ \kappa\ \mathrm{is} \ \mathrm{S\ wave} \,, \\
        \epsilon^{*}_{\mu}(L_z) \frac{\partial}{\partial k_\mu} \left\{ e^{-i \kt \cdot \rt} \Tr\left[ \Pi^{SS_z}(p,k)  \Ncal^{\lambda}(p, k; \rt)  \right]\right \} \Big |_{k=0} & \mathrm{if}\ \kappa\ \mathrm{is} \ \mathrm{P\ wave} \,,
    \end{cases} 
    \label{eq:F-pert-factor}
\end{align}
\end{widetext}
with the covariant spin projectors
\begin{align}
    & \Pi^{SS_z}(p,k) \nonumber \\
    & = \frac{1}{\sqrt{m_Q}}  \sum_{s,\overline{s}} \left \langle \frac{1}{2} s; \frac{1}{2} \overline{s} \Big | S S_z  \right \rangle v_{\overline{s}}\left(\frac{p}{2} - k \right) \overline{u}_{s} \left(\frac{p}{2} + k\right) \,.
    \label{eq:covariant-spin-projector}
\end{align}
The differential cross section\footnote{More precisely $\der \hat{\sigma}^{\kappa}_{\lambda\lambda'}$ is a density matrix in the polarization state of the virtual photon (see the discussion below Eq.\,\eqref{eq:density_matrix_gammaA}).} of the production of a heavy quark pair
with spin state and color state $\kappa$ in a virtual photon-nucleus collision is
\begin{align}
    \frac{\der \hat{\sigma}^{\kappa}_{\lambda\lambda'}}{\der p_\perp^2} =  \frac{1}{4\pi (2p^+)^2}  \frac{1}{N^{\kappa}} \sum_{J_z} \left \langle \Mcal^{\lambda, \kappa,J_z}(p) \overline{\Mcal}^{\lambda', \kappa,J_z}(p) \right \rangle_Y \,,
    \label{eq:diff-xsec}
\end{align}
where the factor $N^{\kappa} = (2J+1) N_{\rm{color}}$ (with $N_{\rm{color}} = 1$ for the singlet and 8 for the octet) is included to average over the number of states for a given $\kappa$. Combining Eqs.\,\eqref{eq:amplitude_projected} and \,\eqref{eq:diff-xsec}, we can cast the differential cross section as
\begin{widetext}
\begin{align}
    \frac{\der \hat{\sigma}^{\kappa}_{\lambda\lambda'}}{ \der p_\perp^2} &=   \alpha_{\rm{em}} e_Q^2 \int \frac{\der^2 \rt}{2\pi} \int \frac{\der^2 \rt'}{2\pi} \overline{\sum_{J_z}}  \Fcal^{\lambda, \kappa,J_z}(p,\rt) \Fcal^{\dagger\lambda', \kappa,J_z}(p,\rt')  \nonumber \\
    & \times \int \der^2 \bt \int \der^2 \bt'  e^{-i \pt \cdot (\bt - \bt')} \overline{\Xi}_{Y}^{\kappa}\left(\bt+ \frac{\rt}{2},\bt- \frac{\rt}{2};\bt'- \frac{\rt'}{2},\bt'+ \frac{\rt'}{2} \right) \,,
    \label{eq:xsec_photonA_projected_0}
\end{align}
\end{widetext}
where the fine structure constant is $\alpha_{\rm{em}} = e^2/(4\pi)$ and we define
\begin{align}
    \overline{\sum_{J_z}} = \frac{1}{(2J+1)} \,, \quad \quad \sum_{J_z}\overline{\Xi}_{Y}^{\kappa} = \frac{1}{N^{\mathrm{color}}} \Xi_{Y}^{\kappa} \,.
\end{align}
The color-projected correlator of light-like Wilson lines is
\begin{align}
    &\Xi_{Y}^{\kappa}(\xt,\yt,\yt'\xt') 
     = \left \langle \Tr\left[ \left( V\left(\xt \right) V^\dagger \left(\yt \right) -\mathbbm{1} \right) \Ccal^{\kappa} \right] \right. \nonumber \\
    & \times \left. \Tr\left[ \left( V\left(\yt' \right) V^\dagger \left(\xt' \right) -\mathbbm{1} \right) \Ccal^{\kappa} \right] \right \rangle_Y \,.
    \label{eq:ColorCorrelator}
\end{align}
We can cast our final result in the compact form:
\begin{align}
    \frac{\der \hat{\sigma}^{\kappa}_{\lambda\lambda'}}{ \der p_\perp^2}  & =    \int \frac{\der^2 \rt}{2\pi} \int \frac{\der^2 \rt'}{2\pi} \Gamma^{\kappa}_{\lambda\lambda'}(\pt,Q;\rt,\rt') \nonumber \\
    & \times \mathcal{G}^{\kappa}_{Y}(\pt;\rt,\rt') \,,
    \label{eq:Jpsi-gammaA-DIS}
\end{align}
where we define CGC distribution
\begin{align}
    &\mathcal{G}^{\kappa}_{Y}(\pt;\rt,\rt') = \int \der^2 \bt \int \der^2 \bt' e^{-i \pt \cdot (\bt-\bt')}  \nonumber \\
    & \times \overline{\Xi}_{Y}^{\kappa}\left(\bt+ \frac{\rt}{2},\bt- \frac{\rt}{2};\bt'- \frac{\rt'}{2},\bt'+ \frac{\rt'}{2} \right) \,.
    \label{eq:color-correlator}
\end{align}
The perturbative function $\Gamma$ in Eq.~(\ref{eq:Jpsi-gammaA-DIS}) is defined as
\begin{align}
    &\Gamma_{\lambda\lambda'}^{\kappa}(\pt,Q;\rt,\rt') \nonumber \\
    & = \alpha_{\rm{em}} e_Q^2 \overline{\sum_{J_z}} \Fcal^{\lambda, \kappa,J_z}(p,\rt) \Fcal^{\dagger\lambda', \kappa,J_z}(p,\rt') \,.
    \label{eq:GammahardFactor}
\end{align}
This expression provides the short distance coefficients of the production of $Q \overline Q$ pairs in virtual-photon nucleus collisions within the CGC framework. The heavy quark pair spin and virtual photon polarization dependence are fully encoded in the perturbative functions $\Gamma$. Explicit expressions for these functions for all polarization and spin combinations are given in Appendix \ref{app:hard_CGC}.  The multiple eikonal scattering of the $Q\overline{Q}$ is encoded in the nonperturbative function $\mathcal{G}^{\kappa}_{Y}(\pt;\rt,\rt')$ which depends only on the color state of the heavy quark pair.  The color singlet is 
\begin{align}
    &\mathcal{G}^{[1]}_{Y}(\pt;\rt,\rt') = \int \der^2 \bt \int \der^2 \bt' e^{-i \pt \cdot (\bt-\bt')}  \nonumber \\
    & \times N_c \left[ S_{Y;\xt,\yt,\yt',\xt'}^{(2,2)}- S_{Y;\xt,\yt}^{(2)}  - S_{Y;\yt',\xt'}^{(2)} + 1 \right] \,, \label{eq:singlet_correlator} 
\end{align}
while for the color octet we have
\begin{align}
    &\mathcal{G}^{[8]}_{Y}(\pt;\rt,\rt') = \int \der^2 \bt \int \der^2 \bt' e^{-i \pt \cdot (\bt-\bt')}  \nonumber \\
    & \times  \frac{N_c}{N_c^2-1}  \left[S_{Y;\xt,\yt\yt',\xt'}^{(4)} - S_{Y;\xt,\yt,\yt',\xt'}^{(2,2)} \right] \label{eq:octet_correlator} \,,
\end{align}
where we introduce the dipole, double-dipole, and quadrupole correlators of light-like Wilson lines,
\begin{align}
    &S_{Y;\xt,\yt}^{(2)} =\frac{1}{N_c} \left \langle \Tr\left[ V(\xt) V^\dagger(\yt) \right] \right \rangle_Y \,, \nonumber \\
    &S_{Y;\xt,\yt;\yt',\xt'}^{(2,2)} \nonumber \\
    &=\frac{1}{N_c^2} \left \langle \Tr\left[ V(\xt) V^\dagger(\yt) \right] \Tr\left[ V(\yt') V^\dagger(\xt') \right]  \right \rangle_Y \,, \nonumber \\
    & S_{Y;\xt,\yt;\yt',\xt'}^{(4)} \nonumber \\
    &=\frac{1}{N_c} \left \langle \Tr\left[ V(\xt) V^\dagger(\yt)  V(\yt') V^\dagger(\xt') \right]  \right \rangle_Y \,,
\end{align}
respectively. The calculation of quarkonium production in CGC + NRQCD has been previously carried out for the singlet channel (diffractive) ${}^{3} \mathrm{S}^{[1]}_{1}$.  We have verified that our results in this channel are consistent with those presented in \cite{Lappi:2020ufv}.

It is evident from the convolution in Eq.\,\eqref{eq:Jpsi-gammaA-DIS} that quarkonium production in DIS within the CGC does not satisfy $k_\perp$ factorization \cite{Catani:1990eg,Collins:1991ty,Levin:1991ry} . To see this more clearly, we express Eq.\,\eqref{eq:Jpsi-gammaA-DIS} in momentum space,
\begin{align}
    \frac{\der \hat{\sigma}^{\kappa}_{\lambda \lambda'}}{ \der p_\perp^2}  &=    \int \frac{\der^2 \lt}{2\pi} \int \frac{\der^2 \lt'}{2\pi}  \widetilde{\Gamma}_{\lambda\lambda'}^{\kappa}(\pt,Q;\lt,\lt') \nonumber \\
    & \times \widetilde{\mathcal{G}}^{\kappa}_{Y}(\pt;\lt,\lt') \,.
\end{align}
The transverse momentum transferred from the background field to the quark and antiquark are respectively $\pt/2 + \lt$  and $\pt/2 - \lt$ in the amplitude, with similar expressions in terms of $\lt'$ for the momenta flow in the conjugate amplitude. Our conclusions for quarkonium production in DIS are analogous to the observation that multiple scattering breaks $k_\perp$-factorization in the production of open heavy-flavor and quarkonium in proton-nucleus collisions in the CGC \cite{Gelis:2003vh,Blaizot:2004wv,Fujii:2006ab,Fujii:2013yja,Fujii:2013gxa}. 

We end this section by providing the complete expression for direct quarkonium production in DIS within the joint CGC+NRQCD framework,
\begin{widetext}
\begin{align}
    &\frac{\der \sigma_{\mathrm{CGC}}^{eA\to e H +X}}{\der Q^2 \der y \der p_\perp^2  \der \phi_{e H}}  = \frac{\alpha_{\mathrm{em}}}{ 2\pi^2 Q^2 y }   \sum_{\kappa} \langle \mathcal{O}^{H}_{\kappa} \rangle  \int \frac{\der^2 \rt}{2\pi} \int \frac{\der^2 \rt'}{2\pi} \mathcal{G}^{\kappa}_{Y}(\pt;\rt,\rt') \nonumber \\ 
    & \times \Bigg\{ (1-y) \Gamma^{\kappa}_{\mathrm{L}}(\pt,Q;\rt,\rt')  + \frac{1}{2} \left[1 +(1-y)^2 \right] \Gamma^{\kappa}_{\mathrm{T}}(\pt,Q;\rt,\rt') \nonumber \\
    & + \sqrt{2(1-y)} (2-y) \Gamma^{\kappa}_{\mathrm{TL}}(\pt,Q;\rt,\rt') \cos \phi_{e H}   +  (1-y) \Gamma^{\kappa}_{\mathrm{Tflip}}(\pt,Q;\rt,\rt')\cos 2 \phi_{e H}   \Bigg\}  \,,
    \label{eq:CGC+NRQCD-DIS_quarkonium}
\end{align}
\end{widetext}
where the small-$x$ CGC distributions are given by Eqs.\,\eqref{eq:singlet_correlator} and \eqref{eq:octet_correlator}. The list of all perturbative functions can be found in Appendix \ref{app:hard_CGC}. Equation~\eqref{eq:CGC+NRQCD-DIS_quarkonium} is the main result of this manuscript. We present a numerical study in Sec.\,\ref{sec:numerical_analysis}. Before doing so, we study its correlation limit and show that the resulting expression is consistent with TMD factorization. We will also study the improved TMD framework which will extend the regime of applicability of the correlation limit to large values of $p_\perp$.

\section{Correlation limit and beyond}

\label{sec:TMD_factorization}

In this section, we first examine the correlation limit, which amounts to performing a derivative expansion of the light-like Wilson line correlators in the CGC distribution \cite{Dominguez:2011wm}. Kinematically, this limit corresponds to the phase space where the quarkonium transverse momentum $p_\perp^2$ and the intrinsic saturation scale $Q_s^2$ are much smaller than $Q^2 + M_{\mathcal{Q}}^2$, where $M_{\mathcal{Q}} = 2 m_Q$ is the mass of the quarkonium. In this limit, the short distance coefficients factorize into a hard function times the gluon Weizsäcker-Williams TMD. These results have been obtained directly within the TMD formalism in \cite{Bacchetta:2018ivt}, so studying this limit of our calculation provides a strong cross check on our results. We then study the Improved TMD (ITMD) expansion which does not impose a constraint on $p_\perp^2$, but requires the hard scales $Q^2+ M_{\mathcal{Q}}^2$ to be larger than the saturation scale $Q_s^2$. 

The $k_\perp$-factorization approach involving unintegrated gluon distributions and off-shell matrix elements of the partonic sub-process can be recovered from the ITMD approach (and more generally the full CGC result) in the limit in which the saturation scale is much smaller than the (semi-)hard scale of the process. 

\subsection{Correlation limit}
In the limit $p_\perp^2, Q_s^2 \ll Q^2 + M_{\mathcal{Q}}^2$, the convolution in Eq.\,\eqref{eq:Jpsi-gammaA-DIS} are dominated by small dipole sizes $\rt$ and $\rt'$ due to the suppression from the perturbative functions $\Gamma$. The leading contribution in the dipole size expansion is \cite{Dominguez:2011wm}
\begin{align}
    &V\left(\bt + \frac{\rt}{2} \right) V^\dagger\left(\bt - \frac{\rt}{2} \right) \nonumber \\
    &\approx \mathbbm{1} +  \rtL{\alpha} V(\bt)  \left( \partial_\perp^{\alpha} V^\dagger(\bt) \right) \,.
    \label{eq:correlation-expansion}
\end{align}
At leading power, the singlet contribution vanishes since $\Tr[V(\bt)  \left( \partial_\perp^{\alpha} V^\dagger(\bt) \right) ] = 0$ \footnote{Contributions to the color singlet have been studied beyond the leading power, see e.g. \cite{Rodriguez-Aguilar:2023ihz,Rodriguez-Aguilar:2024efj,Kar:2023jkn} in the context of the diffractive production of jets.} while the octet contribution is
\begin{align}
    & \overline{\Xi}_{Y}^{[8]}\left(\bt+ \frac{\rt}{2},\bt- \frac{\rt}{2};\bt'- \frac{\rt'}{2},\bt'+ \frac{\rt'}{2} \right) \nonumber \\
    & \approx  \frac{1}{N_c^2-1}\rtL{\alpha} \rtCL{\alpha'} g^2 \left \langle  \Tr\left[ A_\perp^{\alpha}(\bt)  A_\perp^{\alpha'}(\bt') \right] \right \rangle_Y \,, \label{eq:Octet_correlator_expansion}
\end{align}
where we define
\begin{align}
    A_\perp^{\alpha}(\bt)  = \frac{i}{g} V(\bt)  \left( \partial_\perp^{\alpha} V^\dagger(\bt) \right)\,,
    \label{eq:transverse_gauge_field}
\end{align}
which corresponds to the small-$x$ transverse gauge field in light-cone gauge $A^-=0$. Thus we have
\begin{align}
    \mathcal{G}^{[8]}_{Y}(\pt;\rt,\rt') & \approx \frac{\alpha_s (2\pi)^4}{2(N_c^2-1)}\rtL{\alpha} \rtCL{\alpha'} G^{\alpha \alpha'}_Y(\pt) \,,
    \label{eq:color_correlator_expansion}
\end{align}
where we introduce the (non-abelian) Weizsäcker-Williams gluon TMD\footnote{This expression for the WW gluon TMD is consistent with its operator definition at small-$x$ \cite{Dominguez:2011wm,Boussarie:2021ybe}.},
\begin{align}
    G^{\alpha \alpha'}_Y(\pt) & = \int \frac{\der^2 \bt  \der^2 \bt'}{2\pi^3} e^{-i \pt \cdot (\bt-\bt')}    \nonumber \\
    & \times \left \langle \Tr\left[ A_\perp^{\alpha}(\bt)  A_\perp^{\alpha'}(\bt') \right] \right \rangle_{Y} \,.
    \label{eq:WW_lowx}
\end{align}
The differential cross section in the correlation limit is
\begin{align}
    \frac{\der \hat{\sigma}_{\lambda\lambda'}^{\kappa}}{ \der p_\perp^2} \Big|_{\mathrm{TMD}} &=   
    H^{\kappa}_{\lambda \lambda',\alpha \alpha'}(Q) G^{\alpha \alpha'}_Y(\pt) \,,
    \label{eq:Jpsi-gammaA-TMD}
\end{align}
where the hard function is defined as
\begin{align}
    H^{\kappa}_{\lambda \lambda',\alpha \alpha'}(Q) &= \frac{\alpha_s (2\pi)^4}{2(N_c^2-1)} \lim_{p_\perp  \to 0} \int \frac{\der^2 \rt}{2\pi} \int \frac{\der^2 \rt'}{2\pi} \nonumber \\
    & \times \rtL{\alpha}  \rtCL{\alpha'} \Gamma_{\lambda\lambda'}^{\kappa}(\pt,Q;\rt,\rt') \,. \label{eq:TMD-hard-factor}
\end{align}
The explicit results for the hard functions $H$ are shown in appendix \ref{sec:hard_factor_TMD}.
The correlation limit corresponds to the product of the hard matrix element $\gamma g \rightarrow Q\overline Q$  with the Weizsäcker Williams (WW) gluon TMD. It is now clear that, due to this effective one-gluon exchange, only the octet contribution survives. It is customary to decompose the WW distribution into its trace and traceless components:
\begin{align}
    G^{\alpha \alpha'}_{Y}(\pt) = \frac{\delta_\perp^{\alpha \alpha'}}{2} G^{(0)}_{Y}(p_\perp )  + \frac{\Pi^{\alpha\alpha'}_\perp(\pt)}{2} h^{(0)}_{Y}(p_\perp ) \,,
    \label{eq:WW_decomposition}
\end{align}
which are respectively known as the unpolarized and linearly polarized distributions~\cite{Mulders:2000sh,Meissner:2007rx,Boer:2010zf}. The projector is defined as
\begin{align}
    \Pi^{\alpha\alpha'}_\perp(\pt) = \left( \frac{2\ptU{\alpha}\ptU{\beta}}{p_\perp^2}-\delta_\perp^{\alpha \alpha'} \right) \,.
    \label{eq:Pi-projector}
\end{align}
Combining the results in Eqs.\,\eqref{eq:Jpsi-gammaA-TMD},\,\eqref{eq:WW_decomposition},\,\eqref{eq:WW_decomposition}, and \eqref{eq:decomposition_eA_gammaA}, the differential cross section for quarkonium production in DIS in the correlation limit is
\begin{widetext}
    \begin{align}
    \frac{\der \sigma_{\rm TMD}^{eA\to e H +X}}{\der Q^2 \der y \der p_\perp^2  \der \phi_{e H}}  = \frac{\alpha_{\mathrm{em}}}{ 2\pi^2 Q^2 y }  
    &\sum_{\kappa} \langle \mathcal{O}^{H}_{\kappa} \rangle  \Bigg\{ \left[ (1-y) H^{\kappa}_{\rm L}(Q)  + \frac{1}{2} \left[1 +(1-y)^2 \right] H^{\kappa}_{\rm T}(Q) \right] G^{(0)}_Y(p_\perp ) \nonumber \\ 
    &  +  (1-y) H^{\kappa}_{\rm Tflip}(Q) h^{(0)}_Y(p_\perp ) \cos 2 \phi_{e H}   \Bigg\} \,,
    \label{eq:diff-cross section-TMD}
\end{align}
\end{widetext}
where the sum $\kappa$ runs over the octet contributions only. To obtain this expression we exploit the orthogonality of the projectors $\delta_\perp^{\alpha\alpha'}, \Pi^{\alpha\alpha'}_\perp(\pt)$ when contracting the WW gluon TMD with the hard functions given in Appendix \ref{sec:hard_factor_TMD}, and define
\begin{align}
    H^{\kappa}_{\rm L}(Q) &= \frac{1}{2} \delta_\perp^{\alpha\alpha'}H^{\kappa}_{\rm L,\alpha\alpha'}(Q) \,, \nonumber \\
    H^{\kappa}_{\rm T}(Q) &= \frac{1}{2} \delta_\perp^{\alpha\alpha'}H^{\kappa}_{\rm T,\alpha\alpha'}(Q) \,, \nonumber \\
    H^{\kappa}_{\rm Tflip}(Q) & = \frac{1}{2} \Pi^{\alpha\alpha'}_\perp(\pt) H^{\kappa}_{\rm Tflip,\alpha\alpha'}(Q) \,.
\end{align}
While the azimuthal-angle-integrated differential cross section is proportional to the unpolarized WW gluon TMD, to access the linearly polarized distribution, one has to determine the azimuthal correlation between the produced quarkonium and the scattered electron in DIS. We have verified that Eq.\,\eqref{eq:diff-cross section-TMD} is in agreement with the results obtained in \cite{Bacchetta:2018ivt} directly obtained from the TMD formalism, providing a non-trivial consistency of our results.

The saturation scale $Q_s$ is implicit in the WW gluon TMD, which can be computed explicitly in the McLerran-Venugopalan (MV) model \cite{Dominguez:2011br,Dominguez:2011wm}. The small-$x$ evolution of the WW gluon TMD is given by the JIMWLK equation \cite{Dominguez:2011gc} and in the dilute limit by the BFKL equation. The numerical solutions of the evolution of the WW gluon TMD exhibit geometric scaling \cite{Dumitru:2015gaa}. Therefore, in principle one can probe the physics of saturation by studying the $p_\perp$ dependence of the produced quarkonium, especially in the low $p_\perp \lesssim Q_s$ regime most sensitive to the physics of saturation. However, beyond the leading order, large Sudakov-type logarithms \cite{Mueller:2013wwa} arising from soft gluon emissions modify the production at low $p_\perp$ and must be considered in a realistic calculation.

\subsection{Improved TMD limit}
An improved TMD (ITMD) framework to extend the validity of the correlation/TMD limit at small $x$ was proposed in \cite{Kotko:2015ura} which interpolates between the TMD formalism at low values of $p_\perp$ and high-energy $k_\perp$-factorization formalism at large $p_\perp$. The ITMD framework accounts for the off-shellness of the small-$x$ gluon which enters the calculations of the hard functions. For dijet production in hadronic collisions and in DIS, the ITMD framework provides an excellent to the full CGC calculation when the saturation scale $Q_s$ is small \cite{Mantysaari:2019hkq,Boussarie:2021ybe,Fujii:2020bkl}. Instead of following the original approach proposed in \cite{Kotko:2015ura}, we directly obtain the results in the ITMD framework by a careful expansion of the light-like Wilson line correlators as outlined in \cite{Altinoluk:2019fui,Altinoluk:2019wyu} where kinematic and genuine saturation corrections are isolated. We start with the exact identity that expresses the pair of light-like Wilson lines as a parallel transport of the transverse gauge field \cite{Boussarie:2020vzf} (for a brief derivation of this result see Appendix A in \cite{Boussarie:2021ybe})
\begin{align}
    V(\xt)V^\dagger(\yt) = \exp\left\{ig \int_{\xt}^{\yt} \der \ztL{\alpha} A_\perp^{\alpha}(\zt)\right\} \,,
\end{align}
where the transverse gauge field was defined in Eq.\,\eqref{eq:transverse_gauge_field}. The ITMD is obtained from the expansion
\begin{align}
    V(\xt)V^\dagger(\yt) = \mathbbm{1} + ig \int_{\xt}^{\yt} \der \ztL{\alpha} A_\perp^{\alpha}(\zt) + \dots \,. \label{eq:dipole-ITMD-expansion}
\end{align}
This approximation is more general than Eq.\,\eqref{eq:correlation-expansion} which can be obtained by taking the small $\rt$ limit in Eq.\,\eqref{eq:dipole-ITMD-expansion}. 

The parallel transport in Eq.\,\eqref{eq:dipole-ITMD-expansion} is independent of the path connecting $\xt$ and $\yt$. The simplest choice is the straight path, defined by $\zt(\xi) = \bt -\xi \rt$ where $\xi \in [-1/2,1/2]$, so that
\begin{align}
    &V\left(\bt + \frac{\rt}{2} \right) V^\dagger\left(\bt - \frac{\rt}{2} \right) \nonumber \\
    & = \mathbbm{1} - \rtL{\alpha} \int_{-1/2}^{1/2} \der \xi A_\perp^{\alpha}(\bt - \xi \rt ) + \dots \,,
\end{align}
As in the correlation limit, the singlet contribution vanishes since $\Tr\left[A_\perp^{\alpha} \right] =0$, while the correlator for the octet is now
\begin{align}
     & \mathcal{G}^{[8]}_{Y}(\pt;\rt,\rt')  \approx   \frac{(2\pi)^4 \alpha_s}{2(N_c^2-1)} G^{\alpha\alpha'}_Y(\pt)  \nonumber \\
     & \times \int_{-1/2}^{1/2} \der \xi' \int_{-1/2}^{1/2} \der \xi e^{-i \pt \cdot (\xi\rt-\xi'\rt')} \rtL{\alpha} \rtCL{\alpha'}  \,.
     \label{eq:color_correlator_expansion-ITMD}
\end{align}
The first line corresponds to the standard contribution in the correlation limit, while the second term captures the so-called kinematic twists. The differential cross section in the ITMD expansion is
\begin{align}
    \frac{\der \hat{\sigma}^{\kappa}_{\lambda\lambda'}}{ \der p_\perp^2} \Big|_{\mathrm{ITMD}} &=    \Hcal^{\kappa}_{\lambda\lambda',\alpha \alpha'}(Q,\pt) G^{\alpha \alpha'}_Y(\pt) \,,
    \label{eq:Jpsi-gammaA-ITMD}
\end{align}
where the improved TMD hard functions are defined as
\begin{align}
  & \Hcal^{\kappa}_{\lambda\lambda',\alpha\alpha'}(Q,\pt) = \frac{\alpha_s (2\pi)^4}{2(N_c^2-1)} \nonumber \\
  & \times \int_{-1/2}^{1/2} \der \xi \int_{-1/2}^{1/2} \der \xi'   \int \frac{\der^2 \rt}{2\pi} \int \frac{\der^2 \rt'}{2\pi} \rtL{\alpha}  \rtCL{\alpha'}   \nonumber \\
  & \times  \Gamma_{\lambda\lambda'}^{\kappa}(\pt,Q;\rt,\rt') e^{-i \xi (\pt \cdot \rt) } e^{i \xi' (\pt \cdot \rt') }\,.
  \label{eq:ITMD-hard-factor}
\end{align}
The explicit results for these hard functions are collected in Appendix \ref{app:hard_factors_ITMD}. Compared to the correlation/TMD result in Eq.\,\eqref{eq:Jpsi-gammaA-ITMD} the hard functions in the improved TMD in Eq.~\eqref{eq:Jpsi-gammaA-TMD} are $\pt$-dependent (off-shell). They satisfy the expected property:
\begin{align}          
    \lim_{\pt \to 0}\Hcal^{\kappa}_{\lambda\lambda',\alpha\alpha'}(Q,\pt) = H^{\kappa}_{\lambda\lambda',\alpha\alpha'}(Q) \,.
\end{align}
The term on right-hand side is the on-shell hard function defined in Eq.~\eqref{eq:TMD-hard-factor}.

Due to the more complicated tensor structure of the hard functions in Eq.\,\eqref{eq:ITMD-hard-factor}, both unpolarized and linearly polarized WW gluon distributions contribute to all the elements of the density matrix $\der \hat{\sigma}^{\kappa}_{\lambda\lambda'}$, which in turn implies that both components of the WW gluon distribution contribute to the azimuthal-angle-integrated cross section as well as to the angular anisotropies in $e(k_e) + A(P_A) \to e(k_e') + H(p) + X$. In particular, in the ITMD expansion, there is a non-zero contribution to the $\cos \phi_{eH}$ modulation. For completeness, we write the full expression for the differential cross section
\begin{widetext}
\begin{align}
    &\frac{\der \sigma_{\mathrm{ITMD}}^{eA\to e H +X}}{\der Q^2 \der y \der p_\perp^2  \der \phi_{e H}}  = \frac{\alpha_{\mathrm{em}}}{ 2\pi^2 Q^2 y }   \sum_{\kappa} \langle \mathcal{O}^{H}_{\kappa} \rangle  G^{\alpha \alpha'}_Y(\pt) \Bigg\{ (1-y) \Hcal^{\kappa}_{\mathrm{L},\alpha \alpha'}(Q,\pt)  + \frac{1}{2} \left[1 +(1-y)^2 \right] \Hcal^{\kappa}_{\mathrm{T},\alpha \alpha'}(Q,\pt) \nonumber \\
    & + \sqrt{2(1-y)} (2-y) \Hcal^{\kappa}_{\mathrm{LT},\alpha \alpha'}(Q,\pt) \cos \phi_{e H}   +  (1-y) \Hcal^{\kappa}_{\mathrm{Tflip},\alpha \alpha'}(Q,\pt) \cos 2 \phi_{e H}   \Bigg\}  \,,
    \label{eq:diff-cross section-ITMD}
\end{align}
\end{widetext}
where the sum $\kappa$ runs over the octet contributions only. In the limit $p_\perp \ll Q, M_{J/\psi}$, Eq.\,\eqref{eq:diff-cross section-ITMD} reduces to the TMD result in Eq.\,\eqref{eq:diff-cross section-TMD}.

The virtue of the ITMD expansion is that, as in the TMD limit, all the physics of saturation is contained small-$x$ Weizsäcker-Williams gluon distribution. Yet, the ITMD result provides a good approximation to the full CGC result in Eq.\,\eqref{eq:CGC+NRQCD-DIS_quarkonium} in the limit where $Q_s^2 \ll Q^2$ or $Q_s^2 \ll M_{J/\psi}^2$ without imposing any constraint on $p_\perp^2$.

\section{Numerical analysis}
\label{sec:numerical_analysis}

Here we numerically compute the differential cross section for direct quarkonium production in the joint CGC + NRQCD formalism.
Our numerical studies focus on $J/\psi$ production, thus we include the channels $\kappa = \{  {}^{1} \mathrm{S}^{[8]}_0 ,  {}^{3} \mathrm{S}^{[8]}_1, {}^{3} \mathrm{P}^{[8]}_J, {}^{3} \mathrm{S}^{[1]}_1 \}$ . Furthermore, we employ heavy quark symmetry of the long distance matrix elements,
\begin{align}
    \langle\mathcal{O}^{J/\psi}_{ {}^{3} \mathrm{P}^{[8]}_J  }  \rangle = (2J+1) \langle\mathcal{O}^{J/\psi}_{ {}^{3} \mathrm{P}^{[8]}_0  }  \rangle \,,
    \label{eq:heavy-quark-symmetry}
\end{align}
which motivates to define the ``averaged" short distance coefficient \footnote{Following the definition of this SDC we define its corresponding perturbative function $\Gamma_{\lambda\lambda'}^{{}^{3} P_{\langle J\rangle}}$ in Eq.\,\eqref{eq:pert-factor-3PJ-average} with analogous constructions for the TMD and ITMD hard functions.}
\begin{align}
    \der \hat{\sigma}^{{}^{3} \mathrm{P}^{[8]}_{\langle J\rangle}} = \frac{1}{9} \left[ \der \hat{\sigma}^{{}^{3} \mathrm{P}^{[8]}_0} + 3 \der \hat{\sigma}^{{}^{3} \mathrm{P}^{[8]}_1} + 5 \der \hat{\sigma}^{{}^{3} \mathrm{P}^{[8]}_2}    \right] \,.
    \label{eq:SDC-average-3PJ}
\end{align}
We use $m_c =  M_{J/\psi}/2$ for the charm quark mass where $M_{J/\psi} = 3.1\ \mathrm{GeV}$. We compare the results in the CGC obtained in Sec.\,\ref{sec:projectionQQbar} with those obtained in the TMD and the improved TMD approximations in Sec.\,\ref{sec:TMD_factorization}. Before presenting our numerical results, we briefly describe the model used to compute the small-$x$ CGC distributions.

\subsection{Modeling the nonperturbative small-$x$ distributions}

The nonperturbative small-$x$ gluon distributions $\mathcal{G}^{\kappa}_{Y}(\pt,\rt,\rt')$ defined in Eqs.\,\eqref{eq:singlet_correlator} and \,\eqref{eq:octet_correlator}, and the WW gluon TMD $G^{\alpha\alpha'}_Y(\pt)$ defined in Eq.\,\eqref{eq:WW_lowx} respectively, can be computed from the correlator of light-like Wilson lines and their derivatives. Their rapidity (or energy) dependence is obtained by solving the JIMWLK \cite{Balitsky:1995ub,JalilianMarian:1996xn,JalilianMarian:1997dw,Kovner:2000pt,Iancu:2000hn,Iancu:2001ad,Ferreiro:2001qy} renormalization group equations.  We employ the McLerran-Venugopalan model \cite{McLerran:1993ka,McLerran:1993ni} for the initial conditions. In this preliminary study, we will not include the rapidity evolution and employ the so-called Gaussian approximation \cite{Blaizot:2004wv,Iancu:2011nj,Dominguez:2011br}, allowing us to express these correlators in terms of the two-point correlator (see Appendix\,\ref{app:Gaussian}). Furthermore, we assume translational invariance, thus the correlators depend only on the difference of transverse coordinates. We use the MV model of the dipole correlator,
\begin{align}
 S_Y^{(2)}(\Bt) = \exp\left[- \frac{1}{4} Q_{s}^2 B_\perp^2 \ln\left(\frac{1}{m B_\perp} + e \right)\right] \,,
\end{align}
where $m = 0.241\ \rm{GeV}$ and the saturation scale is $Q^2_{s,p} = 0.2 \ \rm{GeV}^2$ for a proton and $Q^2_{s,A} = 1.0 \ \rm{GeV}^2$ for a large nucleus. The small-$x$ distributions are then obtained following Eqs.\,\eqref{eq:color_octet_Gaussian},\,\eqref{eq:color_singlet_Gaussian},\,\eqref{eq:unpol_WW_Gaussian} and \eqref{eq:linpol_WW_Gaussian}.

\subsection{Numerical results for the short distance coefficients and the differential cross section}

We present the $p_\perp$ and $Q$-dependence of the short distance coefficients using two different saturation scales $Q^2_{s,p} = 0.2 \ \rm{GeV}^2$ (proton) and $Q^2_{s,A} = 1.0 \ \rm{GeV}^2$ (large nucleus). We focus on the case in which the photon is transversely polarized. The results for the longitudinally polarized photon and the off-diagonal elements (interference between different polarizations) are shown in Appendix \ref{app:additional_numerical_results}. Since we compare the CGC, ITMD, and TMD results, for simplicity, we normalize our results by the transverse area of the target:
\begin{align}
    \der \hat{N}^{\kappa} = \der \hat{\sigma}^{\kappa} / S_\perp \,. 
\end{align}
The expressions for the short distance coefficients obtained in the CGC, TMD, and ITMD are given by Eqs.\,\eqref{eq:Jpsi-gammaA-DIS}\,,\eqref{eq:Jpsi-gammaA-TMD} and \eqref{eq:Jpsi-gammaA-ITMD} respectively.
\begin{figure*}[t]
    \centering
    \includegraphics[width=0.49\textwidth]{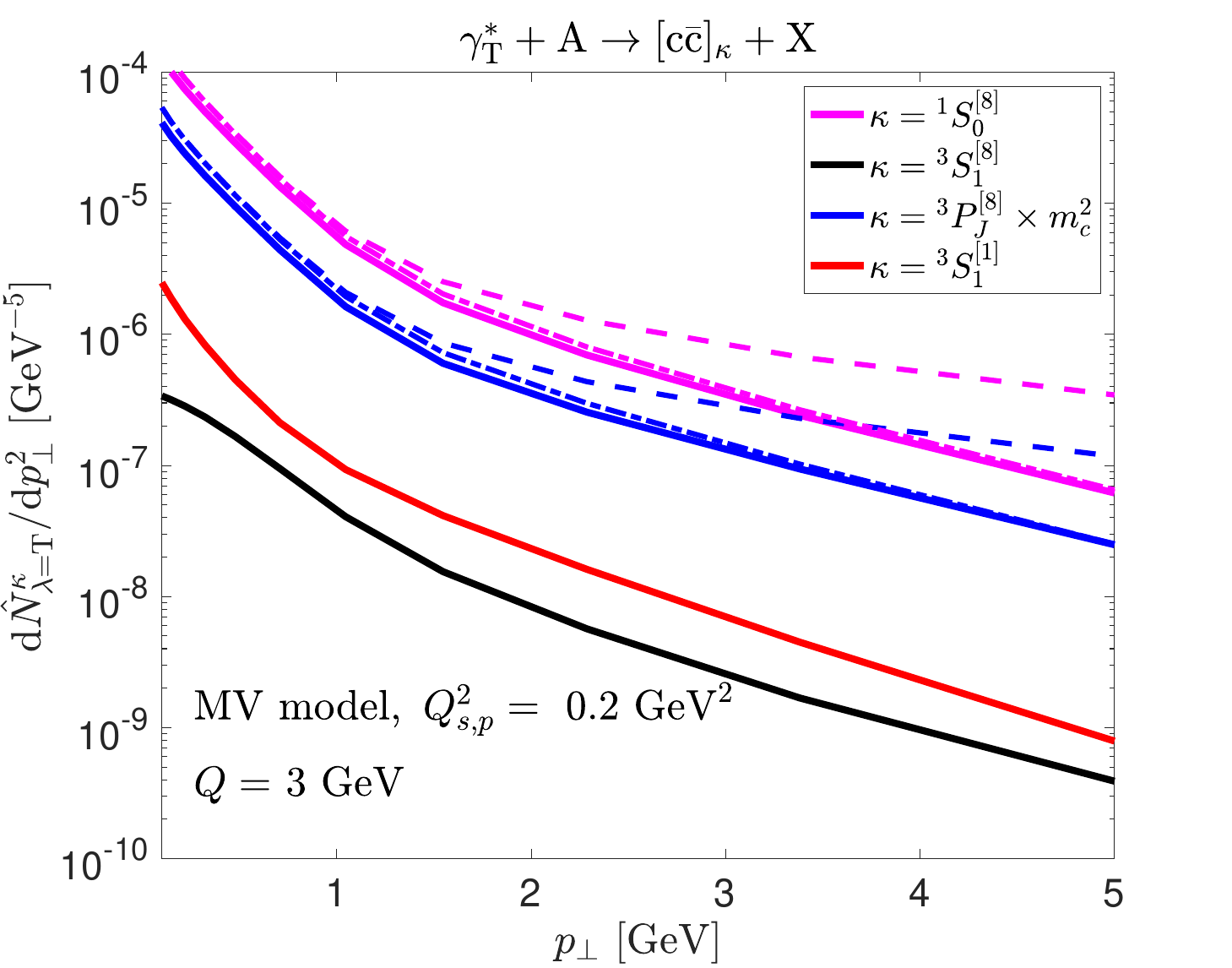}
    \includegraphics[width=0.49\textwidth]{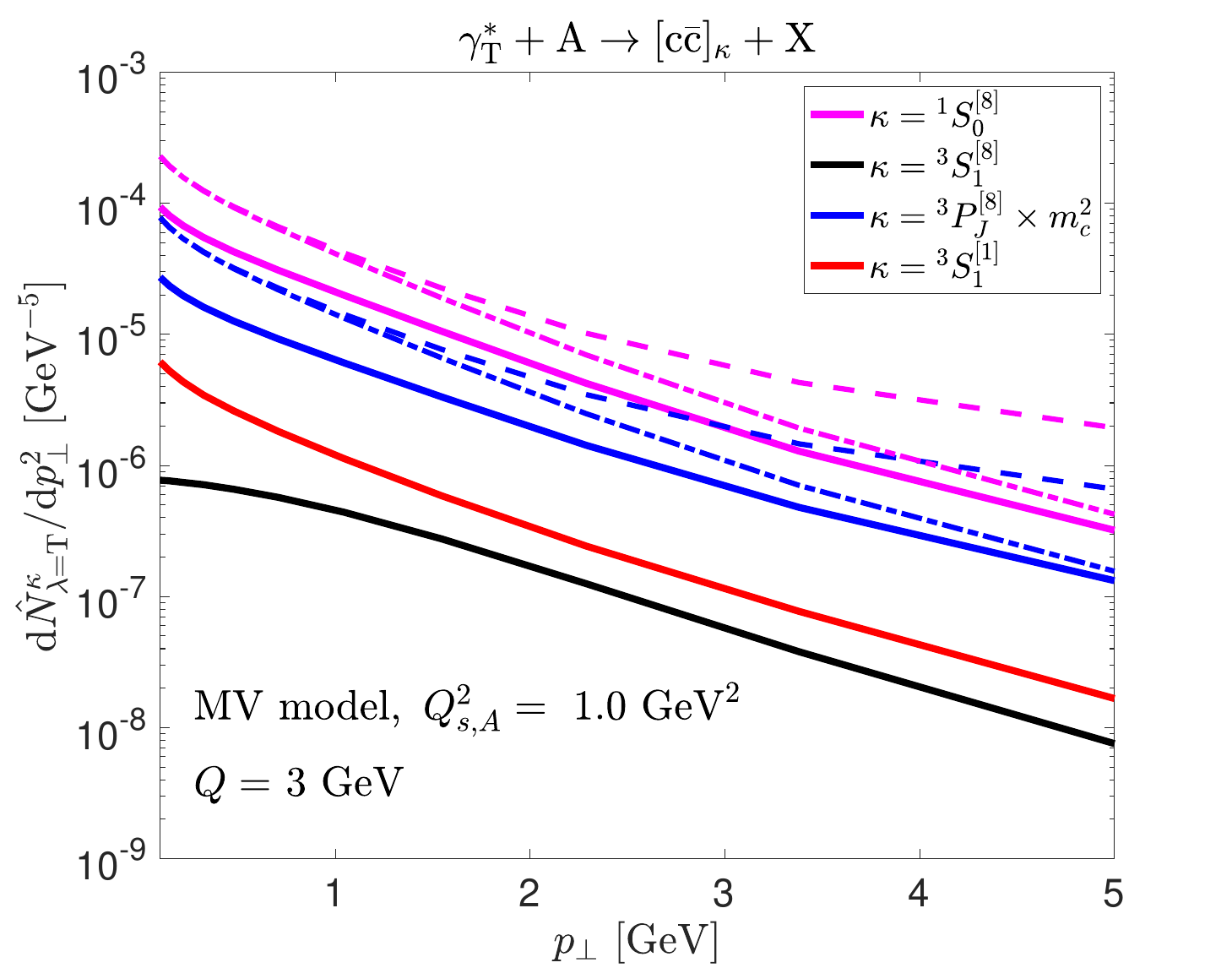}
    \includegraphics[width=0.49\textwidth]{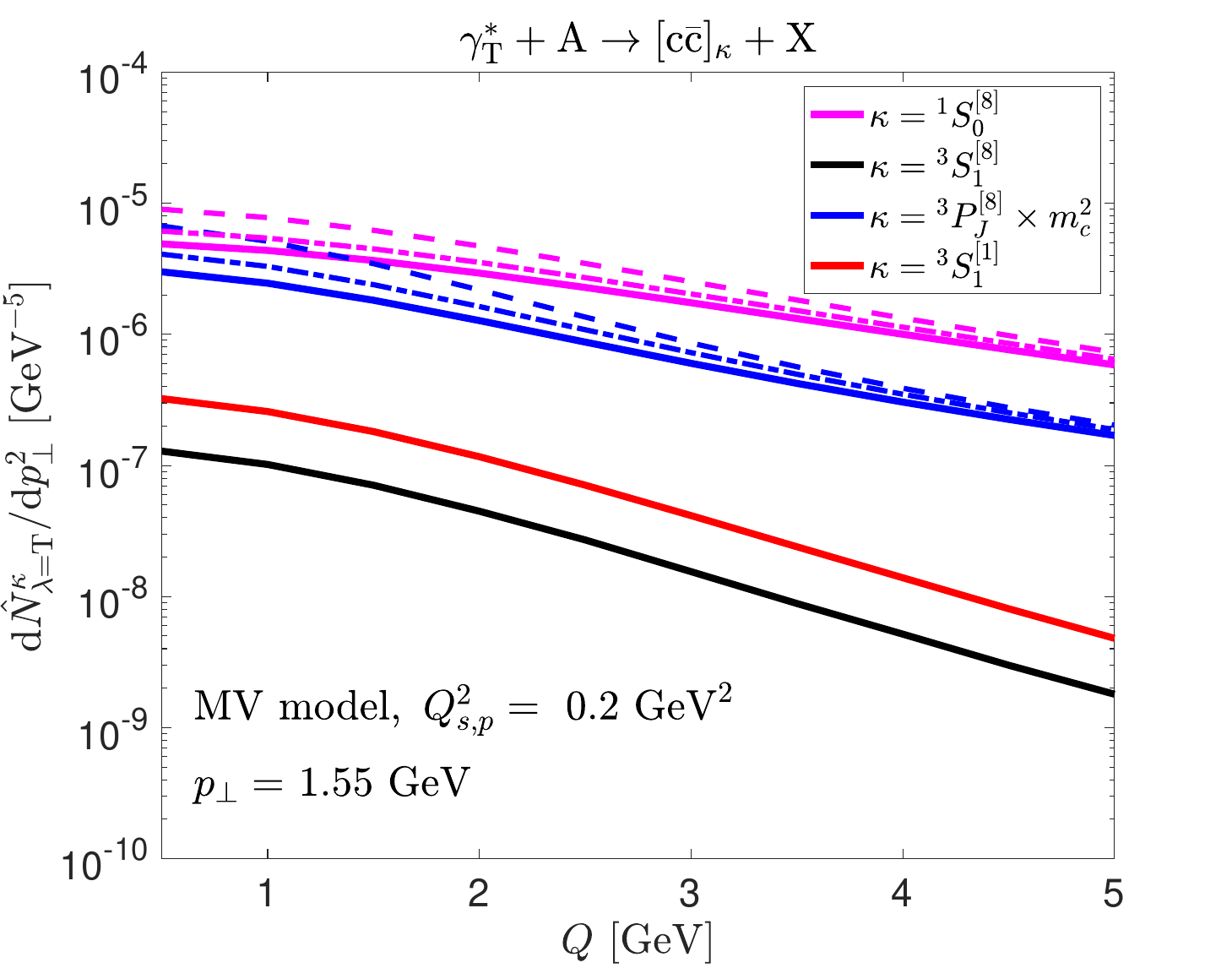}
    \includegraphics[width=0.49\textwidth]{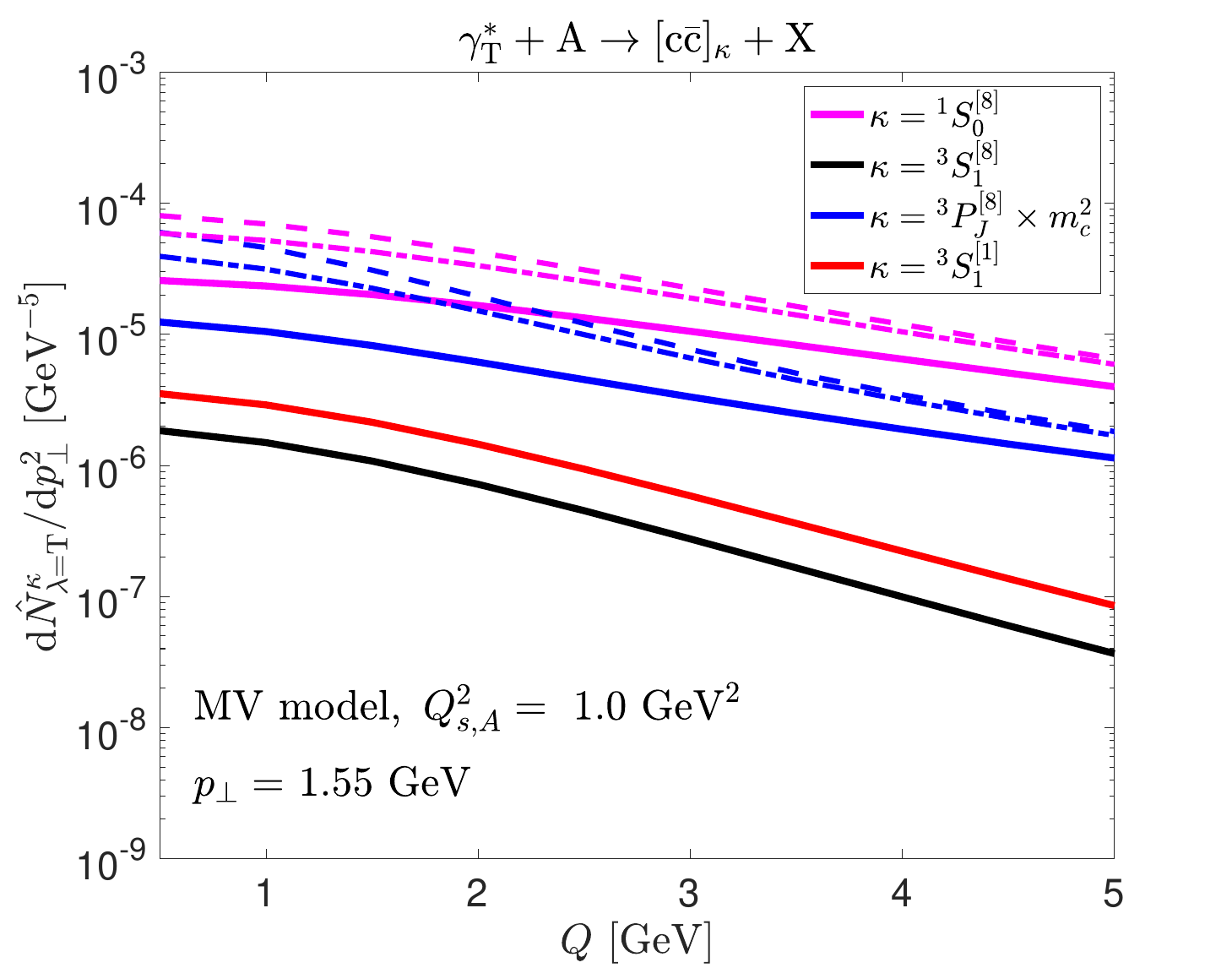}
    \caption{Upper panels: the $p_\perp$-dependence of the short distance coefficients at fixed virtuality $Q= 3.0\ \rm{GeV}$. Lower panels: The $Q$-dependence of the short distance coefficients at fixed transverse momentum $p_\perp = 1.55 \ \mathrm{GeV}$. 
    We show the results for the CGC (solid lines), the improved TMD (dashed-dotted), and the TMD (dashed). Panels on the right show the results at a $Q_s^2 = 0.2 \ \rm{GeV}^2$ (proton).  Panels on the left show the results at $Q_s^2 = 1.0 \ \rm{GeV}^2$ (large nucleus). The short distance coefficients corresponding to the P wave are multiplied by $m_c^2$. }
    \label{fig:SDC-Transverse-pTdep}
\end{figure*}
We note that, in the TMD framework, only $\kappa = {}^{1} \mathrm{S}_0^{[8]}$ and $\kappa = {}^{3} \mathrm{P}_J^{[8]}$ channels are non-vanishing, while in the full CGC result, all channels contribute\footnote{Unlike the TMD framework, where the channels ${}^{3} \mathrm{S}_{1}^{[1]}$ and ${}^{3} \mathrm{S}_{1}^{[8]}$ are forbidden by selection rules, the CGC can accommodate these channels via higher-twist corrections involving multiple gluon exchanges.}. In particular, the CGC has a non-vanishing contribution to the singlet channel ${}^{3} \mathrm{S}_{1}^{[1]}$ in which the virtual photon interacts with the nucleus via color singlet exchange, the so-called Pomeron. In the upper panels in Fig.\,\ref{fig:SDC-Transverse-pTdep} we show the $p_\perp$-dependence of the short distance coefficients. When the saturation scale is small, all three frameworks agree with each other in the small $p_\perp$ region ($\sim\ 1.5 \ \rm{GeV}$ ) as expected from the correlation expansion. The TMD result, where the $p_\perp$-dependence is completely determined by the WW gluon distribution, behaves as $1/p_\perp^2$ at large $p_\perp$. In contrast, the ITMD has an additional $p_\perp$-dependence on the hard function, which at large $p_\perp$ results in a $1/p_\perp^4$ behavior for the short distance coefficients, as expected from $k_\perp$-factorization. The CGC and ITMD results are in good agreement with each other throughout the entire $p_\perp$ range in the proton case (small saturation scale). However, large deviations are observed when the saturation scale is increased signaling the presence of genuine saturation corrections that are only captured in our full CGC calculation. These corrections significantly suppress production in the low-$p_\perp$ region. 

Next, we turn to the $Q$-dependence of the short distance coefficients shown in lower panels in Fig.\,\ref{fig:SDC-Transverse-pTdep}. In the high-$Q$ limit, we observe the convergence of all three frameworks for the $\kappa = {}^{1} \mathrm{S}_0^{[8]}$ and $\kappa = {}^{3} \mathrm{P}_J^{[8]}$ channels. At high $Q^2$ the kinematic and genuine saturation corrections are suppressed and the TMD approximation is adequate. In this regime, the $Q$-dependence is completely controlled by the hard function. The $\kappa = {}^{3} \mathrm{S}_1^{[1]}$ and $\kappa = {}^{3} \mathrm{S}_1^{[8]}$ channels
are power suppressed by an additional factor $1/Q^2$ as these require at least a two-gluon exchange (higher twist). When we either increase the saturation scale $Q_s$ or decrease the virtuality $Q$, the genuine higher-twist corrections in the CGC tend to suppress the short distance coefficients for $\kappa = {}^{1} \mathrm{S}_0^{[8]}$ and $\kappa = {}^{3} \mathrm{P}_J^{[8]}$. On the other hand, we note that the relative contribution of the color singlet becomes more important, which is consistent with the observation that diffractive events are more copious in the saturation regime \cite{Kowalski:2008sa}.

\begin{figure*}[t]
    \centering
    \includegraphics[width=0.49\textwidth]{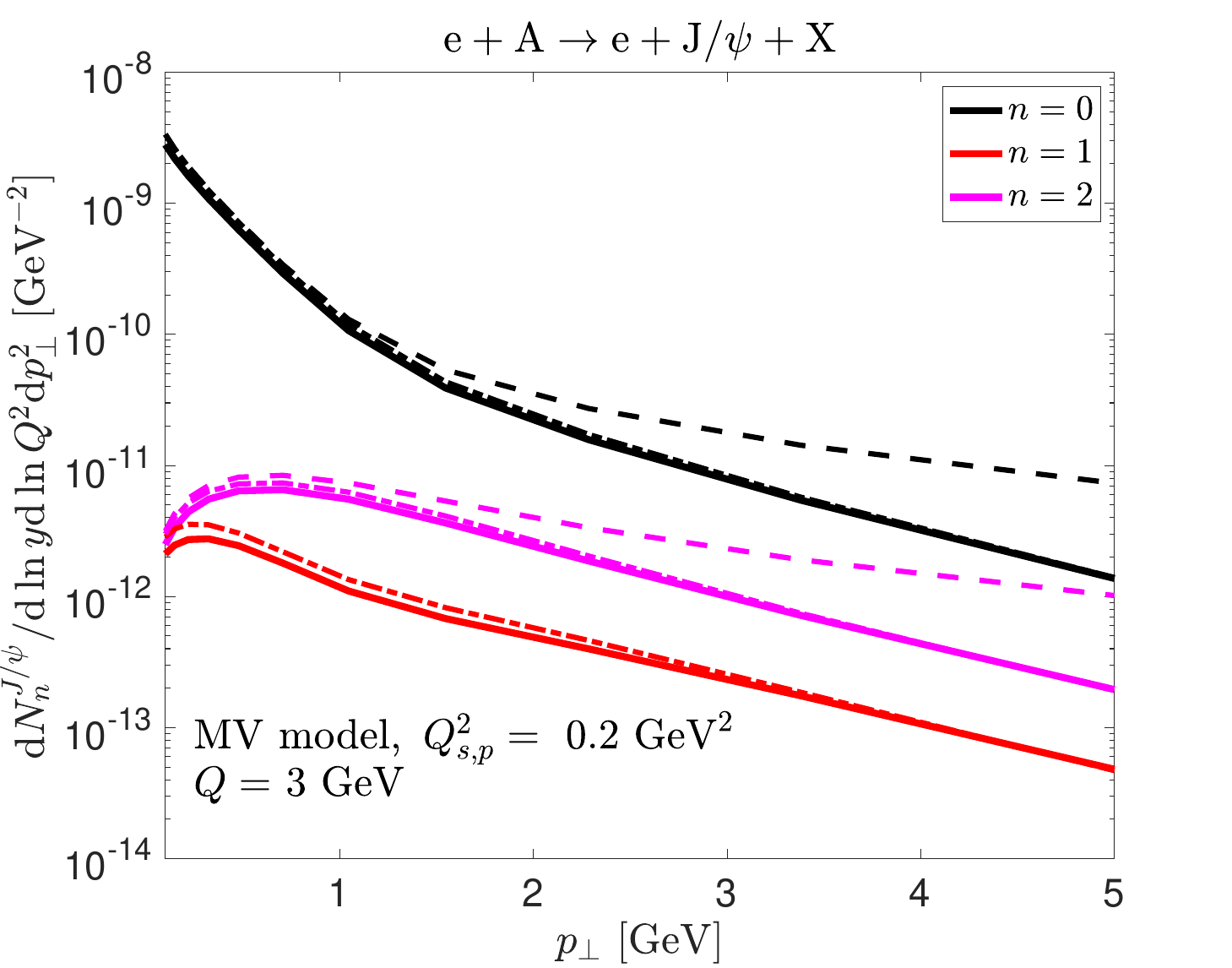}
    \includegraphics[width=0.49\textwidth]{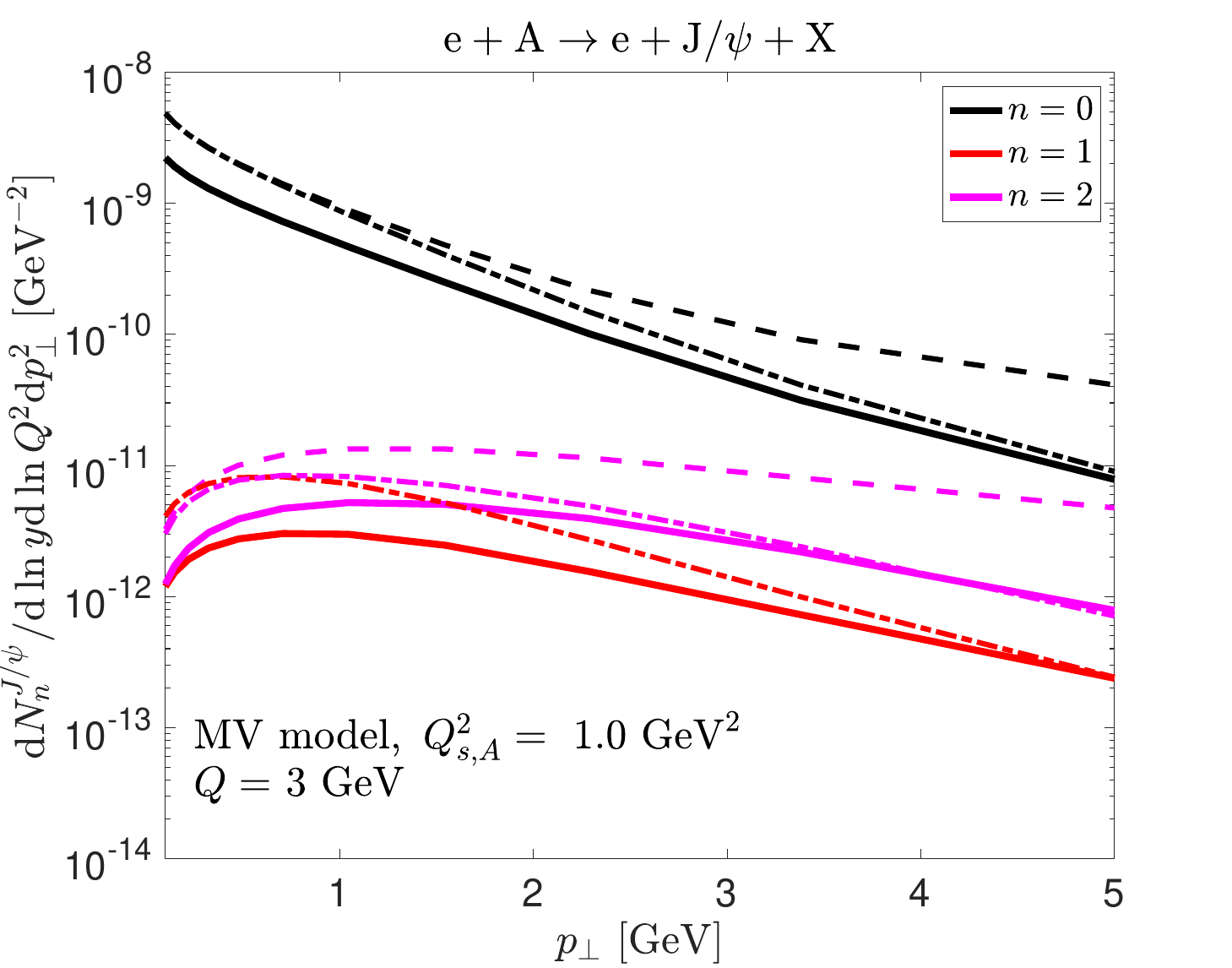}
    \includegraphics[width=0.49\textwidth]{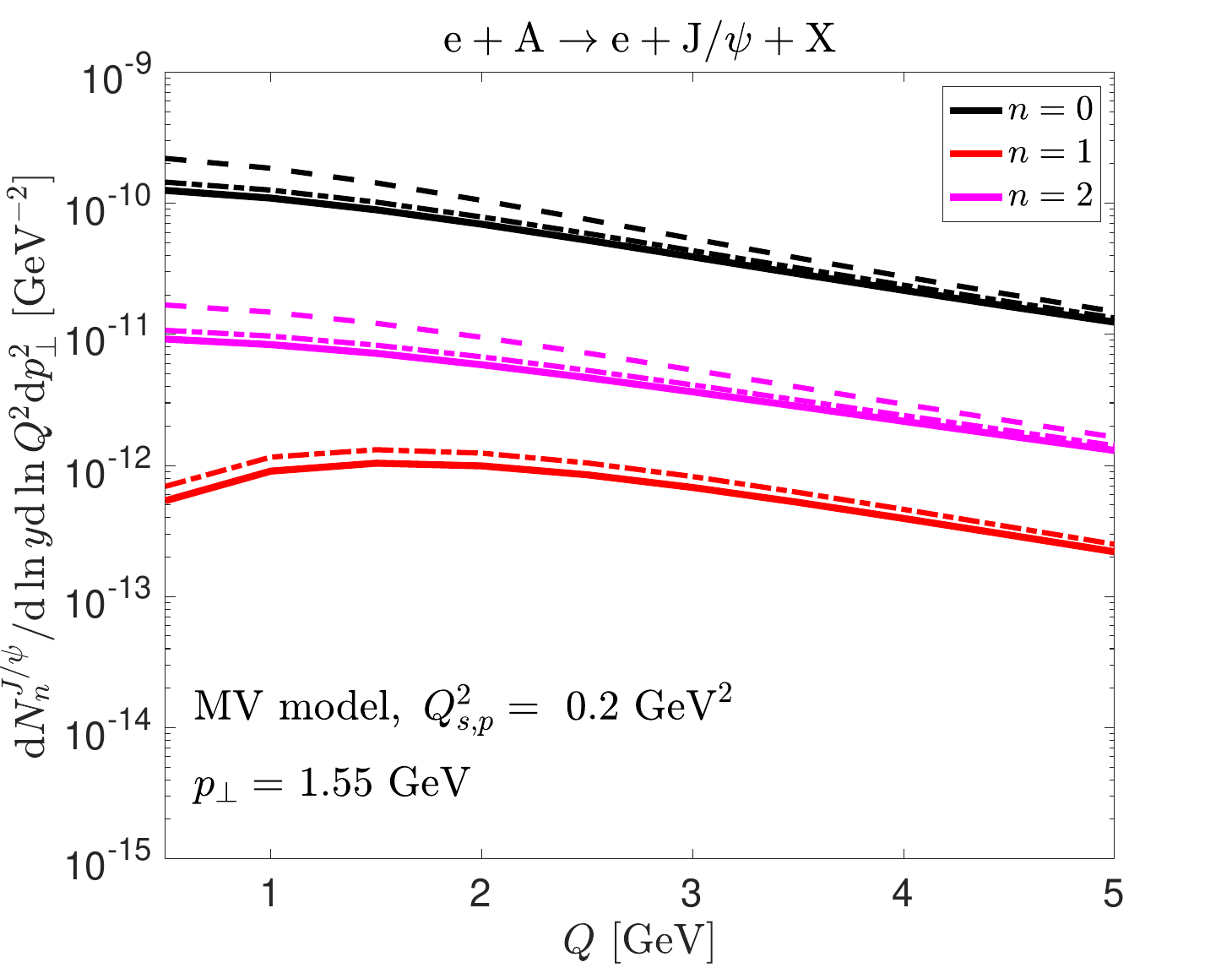}
    \includegraphics[width=0.49\textwidth]{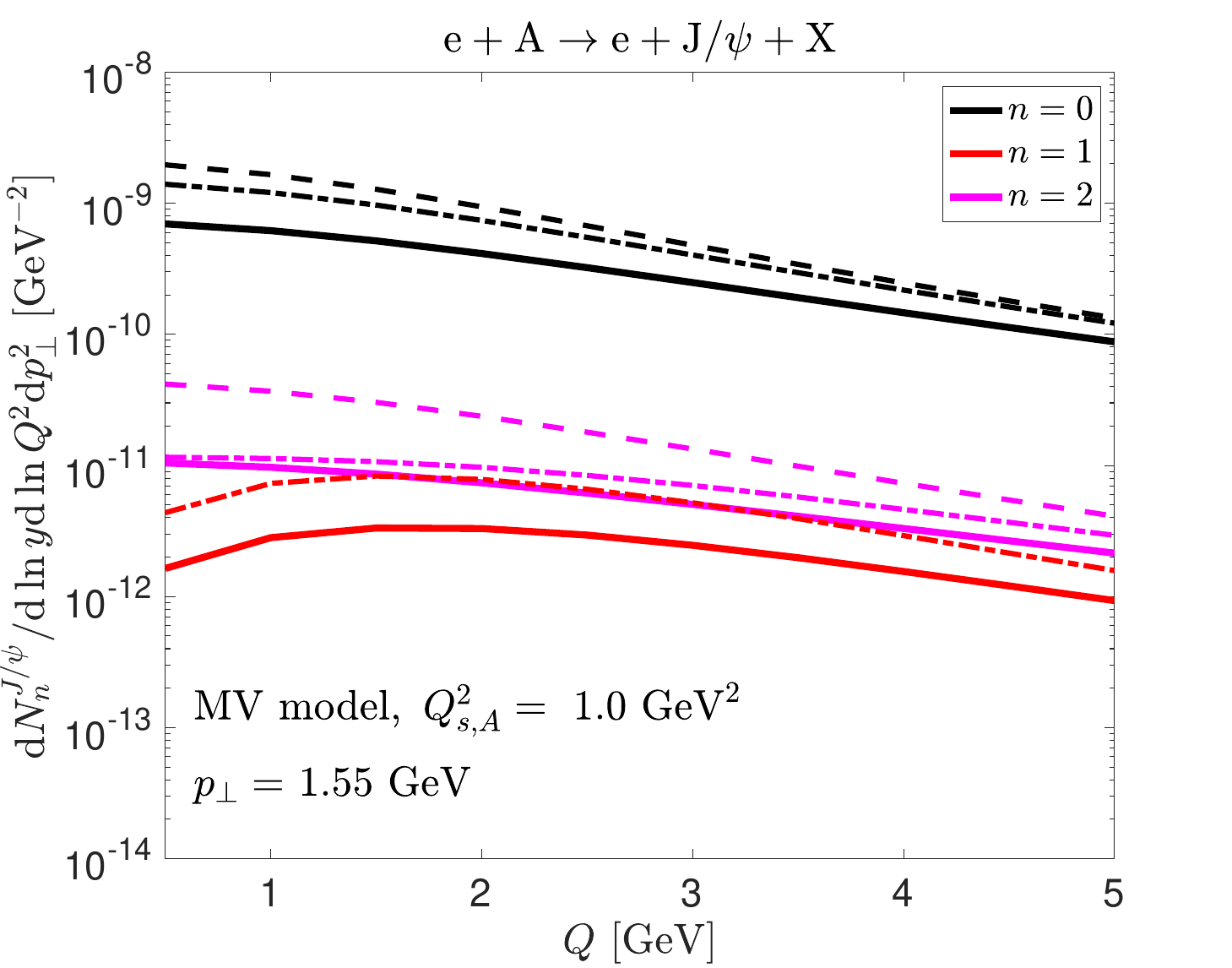}
    \caption{Upper panels: the $p_\perp$-dependence of the differential cross section at fixed virtuality $Q= 3.0\ \rm{GeV}$. Lower panels: The $Q$-dependence of the differential cross section at fixed transverse momentum $p_\perp = 1.55 \ \mathrm{GeV}$. 
    We show the results for the CGC (solid lines), the improved TMD (dashed-dotted), and the TMD (dashed). Panels on the right show the results at a $Q_s^2 = 0.2 \ \rm{GeV}^2$ (proton).  Panel on the left show the results at $Q_s^2 = 1.0 \ \rm{GeV}^2$ (large nucleus). }
    \label{fig:diff-Xsec-pTdep}
\end{figure*}

We now present the results for the differential cross section given by Eqs.\,\eqref{eq:CGC+NRQCD-DIS_quarkonium},\,\eqref{eq:diff-cross section-TMD} and \eqref{eq:diff-cross section-ITMD}. We normalize our results by the transverse area of the target and use the values for the long-distance matrix elements in \cite{Chao:2012iv}
\begin{align}
    \langle \Ocal_{ {}^{3} \mathrm{S}^{[1]}_{1} }^{J/\psi} \rangle &= 1.16/(2 N_c)\ \mathrm{GeV}^3 \,,\nonumber \\
    \langle \Ocal_{ {}^{1} \mathrm{S}^{[8]}_{0} }^{J/\psi} \rangle &= 0.089\ \mathrm{GeV}^3 \nonumber \,, \\
    \langle \Ocal_{ {}^{3} \mathrm{S}^{[8]}_{1} }^{J/\psi} \rangle &= 0.0030\ \mathrm{GeV}^3 \,,\nonumber \\
    \langle \Ocal_{ {}^{3} \mathrm{P}^{[8]}_{0} }^{J/\psi} \rangle/m_c^2 &= 0.0056\ \mathrm{GeV}^3 \,.
\end{align}
The fit to the LDMEs in \cite{Chao:2012iv} uses $m_c = 1.5 \pm 0.1\ \mathrm{GeV}$ compatible with our choice $m_c = M_{J/\psi}/2.$ While alternative fits for the LDMEs exist in the literature, our main objective in this preliminary numerical study is to compare results from the three different frameworks CGC, TMD, and ITMD, thus the specific values of the LDMEs are no critical.

We decompose our results in harmonics in the azimuthal angle between the electron and the produced $J/\psi$,
\begin{align}
    \frac{\der N_n^{eA\to e J/\psi +X}}{\der Q^2 \der y \der p_\perp^2 } 
    =& \int_{0}^{2\pi} \frac{\cos(n \phi_{e J/\psi}) 
    \der \phi_{e J/\psi}}{2\pi} \nonumber \\
    & \times \frac{\der N^{eA\to e J/\psi +X}}{\der Q^2 \der y \der p_\perp^2  \der \phi_{e J/\psi}} \,,
\end{align}
and we fix the inelasticity $y=0.8$.

The results for the $p_\perp$- and $Q^2$ dependence are shown in the upper and lower panels of Fig.\,\ref{fig:diff-Xsec-pTdep} respectively. The conclusions are similar to those observed in our comparison of the short distance coefficients. Namely, the TMD provides an excellent approximation when $p_\perp, Q_s \ll Q$ both for the azimuthally-averaged differential cross-section ($n=0$) as well as its elliptic modulation ($n=2$) \footnote{Note that the harmonic $n=1$ vanishes in the TMD framework, but it is non-zero both in the ITMD and the CGC.}. The ITMD framework extends the agreement to larger values of $p_\perp$ when the saturation scale is small, but it fails as the virtuality $Q^2$ is decreased or when the saturation $Q_s^2$ is increased. The genuine saturation corrections, which are not captured by the WW gluon TMD, contained in the full CGC calculation, suppress the production. The degree of suppression is larger at low values of $p_\perp$. This behavior was also observed for semi-inclusive dijet production in DIS in \cite{Mantysaari:2019hkq}.

\begin{figure}
    \centering
    \includegraphics[width=0.49\textwidth]{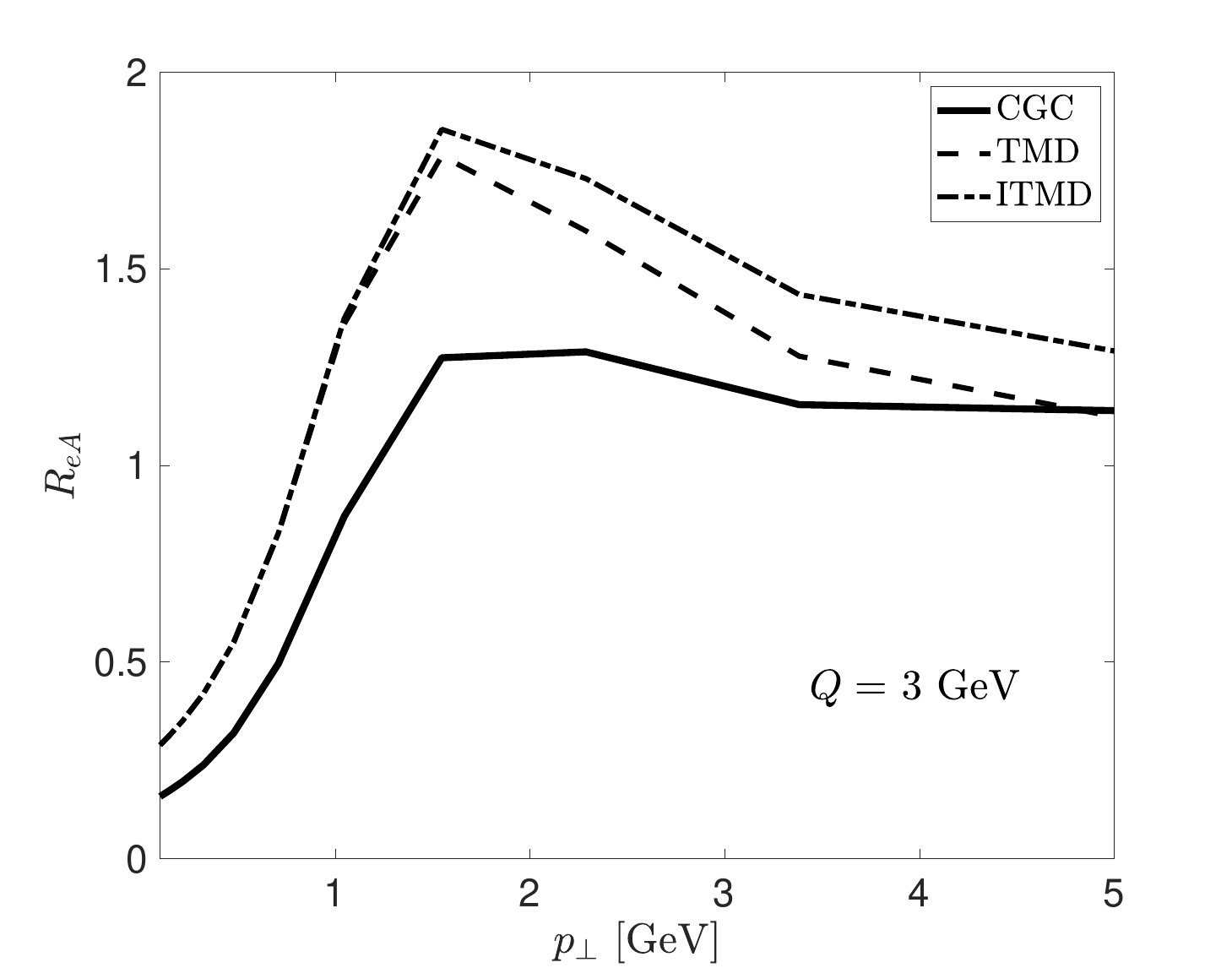}
    \caption{Nuclear modification factor (as defined in Eq.\,\eqref{eq:ReA}) for the azimuthally-averaged different cross section as a function of $p_\perp$ and at fixed virtuality $Q=3\ \rm{GeV}$.}
    \label{fig:RpA}
\end{figure}

Lastly, we compute a proxy for the nuclear modification factor, based on the azimuthally-averaged differential cross section:
\begin{align}
    R_{eA} = \frac{Q^2_{s,p}}{Q^2_{s,A}} \frac{\der N_n^{eA\to e J/\psi +X}}{\der Q^2 \der y \der p_\perp^2 }  \Bigg/ \frac{\der N_n^{ep\to e J/\psi +X}}{\der Q^2 \der y \der p_\perp^2 }  \,.
    \label{eq:ReA}
\end{align}
In the limit $Q_{s,p}, Q_{s,A} \ll p_\perp, Q$, where saturation effects are expected to be weak, $R_{eA} \to 1$. Our results for $R_{eA}$ are shown in Fig.\,\ref{fig:RpA}. The TMD result displays the characteristic broadening of the $p_\perp$ distribution, which reflects the migration of gluons to higher momentum modes in the WW gluon distribution. Effectively, this observable is proportional to the ratio of the nuclear to proton unpolarized WW gluon TMD. The results obtained in the improved TMD have a similar behavior; however, in this case, there is also a contribution from the linearly polarized WW gluon TMD. The broadening of the transverse momentum-dependent gluon distribution is a well-known property of the MV model \cite{Jalilian-Marian:2003rmm}; however, we expect that non-linear quantum evolution will result in a suppression of the nuclear modification ratio \cite{Kharzeev:2003wz,Albacete:2003iq,Caucal:2023fsf}. On the other hand, the inclusion of genuine saturation corrections in the CGC depletes the nuclear modification factor, even in the absence of evolution. We have verified that the degree of depletion increases as either the saturation scale is increased or the virtuality is reduced. A systematic treatment of the amount depletion must therefore take into account genuine saturation corrections as well as nonlinear small-$x$ evolution.

\section{Summary and Outlook}
\label{sec:summary_outlook}

In this manuscript, we obtain for the first time the differential cross section for direct quarkonium production in deep inelastic scattering at small $x$ using a joint framework of the Color Glass Condensate effective theory and nonrelativistic QCD. Our main result is given by Eq.\,\eqref{eq:CGC+NRQCD-DIS_quarkonium}. To derive this result, we first obtain the short distance coefficients for $\kappa =  {}^{1} \mathrm{S}^{[c]}_0\,,  {}^{3} \mathrm{S}^{[c]}_1\,, {}^{1} \mathrm{P}^{[c]}_1\,, {}^{3} \mathrm{P}^{[c]}_J$ for both color singlet $[c=1]$ and octet $[c=8]$ states. We consider separately the cases where the virtual photon is longitudinally and transversely polarized, as well as the interference between different polarizations.  Combined with the decomposition in Eq.\,\eqref{eq:decomposition_eA_gammaA}, these results allow us to compute the differential cross section at the level of electron-nucleus scattering. Our results for the short distance coefficients are shown in Eq.\,\eqref{eq:Jpsi-gammaA-DIS}. They are expressed in terms of polarizations and spin-dependent perturbative functions (the complete set of expressions is given in Appendix \ref{app:hard_CGC}).  The CGC distributions are built up of correlators of light-like Wilson lines given by Eqs.\,\eqref{eq:singlet_correlator} and \eqref{eq:octet_correlator} for color singlet and octet are respectively.

Furthermore, in the correlation expansion ($p_\perp\,, Q_s \ll Q\,, M_{J/\psi}$), our results are consistent with the TMD framework. In this limit, the differential cross section (see Eq.\,\eqref{eq:diff-cross section-TMD}) factorizes into a hard function and the small-$x$ Weizsäcker-Williams gluon TMD. Both the unpolarized and linearly polarized components of the WW gluon TMD contribute to the differential cross section and implicitly contain saturation effects as they are constructed from CGC correlators of light-like Wilson lines. We also consider the improved TMD expansion, which interpolates between the TMD and high-energy $k_\perp$ factorization frameworks and provides an adequate approximation to the CGC when $Q_s^2 \ll Q^2$ or $Q_s^2 \ll M_{J/\psi}^2$. 

Lastly, we perform a preliminary numerical study in the spirit of the work in \cite{Boussarie:2021ybe}, where we quantified the differences between the CGC, TMD, and ITMD frameworks. We focus on $J/\psi$ production and employ the MV model to compute the CGC distributions with two choices of saturation scale, $Q_{sp}^2 = 0.2\ \mathrm{GeV}^2$ 
and $Q_{sp}^2 = 1.0\ \mathrm{GeV}^2$, 
corresponding to protons and large nuclei respectively. We numerically evaluate the short distance coefficients relevant for $J/\psi$ production as well as the differential cross section and examine their dependencies on the transverse momentum of the $J/\psi$ and the DIS virtuality $Q^2$. We numerically confirm that the TMD and CGC results are consistent with each other in the regime $p_\perp\,, Q_s \ll Q\,, M_{J/\psi}$, and the ITMD provides a good approximation to the full CGC result when the saturation scale is smaller than the hard scale ($Q^2$ or $M_{J/\psi}^2$). However, when the saturation scale is comparable in magnitude to the hard scale, genuine saturation corrections only present in the full CGC result suppress the differential cross section, especially in the low $p_\perp$ region.

We conclude with an outlook on future studies that are suggested by our
work. A more realistic study of the initial conditions, the inclusion of small-$x$ evolution, an assessment of the uncertainties of the LDMEs, as well as a more precise calculation of the nuclear modification factor, would be desired before we confront our results with existing data from HERA \cite{ZEUS:2002src,H1:2002voc,H1:2010udv} and make predictions for the future EIC \cite{Accardi:2012qut,AbdulKhalek:2021gbh,Boer:2024ylx}. Furthermore, to obtain robust results, we plan to carry out the next-to-leading order corrections which include one-loop corrections in the CGC, as well as relativistic corrections to NRQCD. It has been found that in the exclusive production \cite{Lappi:2020ufv}, relativistic corrections can be as large as $30 \%$, especially at low values of $Q^2$. 

In the correlation limit, direct quarkonium production provides a window to access the small-$x$ WW gluon distribution, complementary to the back-to-back production of dijet and dihadron studies \cite{Zheng:2014vka,vanHameren:2021sqc,Caucal:2023fsf,Caucal:2024nsb}. In this limit, we also expect the emergence of large rapidity logarithms as well as large double and single Sudakov logarithms calling for a joint resummation \cite{Sun:2012vc} (for direct quarkonium production in proton-nucleus collisions, see \cite{Qiu:2013qka}). The first steps towards the one-loop corrections have been recently studied in \cite{Kang:2023doo} for the S-wave channel.

Furthermore, it would be interesting to perform a detailed analysis of our results in the photoproduction limit which can be accessed through ultra-peripheral collisions at RHIC and the LHC \footnote{A recent detailed analysis, in the collinear pQCD formalism, for the feasibility of studying inclusive quarkonium production in ultra-peripheral collisions at the LHC has been conducted in \cite{Lansberg:2024zap}.}. We expect that genuine saturation corrections will be large and thus necessitate the full CGC calculation. On the other hand, in this limit we expect soft gluon radiation to play a subdominant role, especially for charmonium production, providing an attractive avenue to study saturation through nuclear suppression effects.

It should be straightforward to study the polarization-dependence of the produced quarkonium which will manifest as correlations of the $J/\psi$ decay to dileptons. A similar study in the joint CGC+NRQCD framework has been conducted in proton-proton collisions \cite{Ma:2018qvc,Stebel:2021bbn}. Lastly, transverse spin asymmetries in polarized collisions provide an avenue to access the gluon Sivers function \cite{Sivers:1989cc,Boer:2015vso} with quarkonium production \cite{Yuan:2008vn,Rajesh:2018qks,Zheng:2018ssm,DAlesio:2019gnu,DAlesio:2020eqo}. It would be interesting to extend our framework and examine the consequences of gluon saturation to this observable at low-$x$ (see e.g. \cite{Goncalves:2017fkt} in proton-proton collisions). Such a program will require the implementation of sub-eikonal physics beyond the usual CGC framework \cite{Kovchegov:2012ga,Zhou:2013gsa,Kovchegov:2018znm,Yao:2018vcg,Li:2023tlw}.

\section*{Acknowledgements} We are grateful to Patricia Gimeno Estivill, Tuomas Lappi, Emilie Li, Jani Penttala, Maddox Spinelli, and Feng Yuan for valuable discussions. F.S. is supported by the Institute for Nuclear Theory’s U.S. DOE under Grant No. DE-FG02-00ER41132. Z.K. is supported by the National Science Foundation under Grant No.~PHY-1945471. The work of V.C. and R.V. was performed under the auspices of the U.S. Department of Energy by Lawrence Livermore National Laboratory under Contract DE-AC52-07NA27344, and through the Topical Collaboration in Nuclear Theory on ``Heavy-Flavor Theory (HEFTY) for QCD Matter" under award no. DE-SC0023547, and was supported by the LLNL-LDRD Program under Project No.~23-LW-036. This work is also supported by the U.S. Department of Energy, Office of Science, Office of Nuclear Physics, within the framework of the Saturated Glue (SURGE) Topical Theory Collaboration.

\clearpage
\onecolumngrid

\appendix

\begin{widetext}

\section{Spin projections of the S and P waves}
\label{app:projections}
In this appendix, we briefly outline the computation for the projections of $Q\overline{Q}$ into specific states of spin and orbital angular momentum.  
\subsection{NRQCD projector}
The covariant spin projectors defined in Eq.\,\eqref{eq:covariant-spin-projector} can be expressed as
\begin{align}
    \Pi^{00}(p,k) &=\frac{1}{\sqrt{8m_Q^3}} \left( \frac{\slashed{p}}{2} -\slashed{k} -m_Q \right) \gamma^5 \left( \frac{\slashed{p}}{2} + \slashed{k} +m_Q \right) \,,
    \label{eq:Pi00-projector}
\end{align}
and
\begin{align}
    \Pi^{1S_z}(p,k) &= \frac{\epsilon^*_{\rho}(S_z) }{\sqrt{8m_Q^3}} \left( \frac{\slashed{p}}{2} -\slashed{k} -m_Q \right)  \gamma^{\rho} \left( \frac{\slashed{p}}{2} +\slashed{k} +m_Q \right) \,,
    \label{eq:Pi1Sz-projector}
\end{align}
for the $S=0$ and $S=1$ states respectively. Since we are interested in unpolarized quarkonium production, the explicit expression for the polarization vectors $\epsilon^*_{\rho}(S_z)$ will not be needed, and it is sufficient to use the polarization summed relations \cite{Kang:2013hta}. We define 
\begin{align}
    \mathbb{P}_{\alpha\alpha'} = - g_{\alpha\alpha'} + \frac{p_{\alpha} p_{\alpha'}}{p^2} \,,
\end{align}
then
\begin{align}
    \sum_{L_z} \epsilon^*_{\rho}(L_z) \epsilon_{\rho'}(L_z) = \mathbb{P}_{\rho\rho'} \,, \label{eq:Projector3S1}
\end{align}
For the ${}^{3} \mathrm{P}_J$ states, it is convenient to define
\begin{align}
    \epsilon^{*}_{\rho\mu}(J,J_z) = \sum_{L_z,S_z}  \langle 1 L_z; S S_z | J J_z \rangle    \epsilon^{*}_{\mu}(L_z)  \epsilon^*_{\rho}(S_z) \,,
    \label{eq:tensor_p-wave}
\end{align}
then
\begin{align}
    \sum_{J_z} \epsilon^{*}_{\rho\mu}(0,J_z) \epsilon_{\rho'\mu'}(0,J_z) &= \frac{1}{3} \mathbb{P}_{\rho\mu}\mathbb{P}_{\rho'\mu'} \,,
    \label{eq:Projector3P0} \\
    \sum_{J_z} \epsilon^{*}_{\rho\mu}(1,J_z) \epsilon_{\rho'\mu'}(1,J_z) &= \frac{1}{2} \left( \mathbb{P}_{\rho \rho'}\mathbb{P}_{\mu \mu'} - \mathbb{P}_{\rho \mu'}\mathbb{P}_{\rho' \mu} \right) \,,
    \label{eq:Projector3P1} \\
    \sum_{J_z} \epsilon^{*}_{\rho\mu}(2,J_z) \epsilon_{\rho'\mu'}(2,J_z)&= \frac{1}{2} \left( \mathbb{P}_{\rho \rho'}\mathbb{P}_{\mu \mu'} + \mathbb{P}_{\rho \mu'}\mathbb{P}_{\rho' \mu} \right) - \frac{1}{3} \mathbb{P}_{\rho\mu}\mathbb{P}_{\rho'\mu'} \,.
    \label{eq:Projector3P2}
\end{align}
Lastly, since we will be using heavy-quark symmetry (see Eqs.\,\eqref{eq:heavy-quark-symmetry} and \eqref{eq:SDC-average-3PJ}) it is useful to employ the identity:
\begin{align}
    \sum_{J,J_z} \epsilon^{*}_{\rho\mu}(J,J_z) \epsilon_{\rho'\mu'}(J,J_z) = \mathbb{P}_{\rho\rho'}\mathbb{P}_{\mu \mu'} \,,
    \label{eq:Projector3PJ}
\end{align}
to compute the ``averaged" perturbative function in Eq.\,\ref{eq:pert-factor-3PJ-average}.  

\subsection{Computation of perturbative functions $\Fcal^{\lambda,\kappa,J_z}$}
In order to compute the perturbative part of the amplitude $\Fcal^{\lambda,\kappa,J_z}$ defined in Eq.\,\eqref{eq:F-pert-factor} we combine Eqs.\,\eqref{eq:NcalL} and \eqref{eq:NcalT} with the projectors in Eqs.\,\eqref{eq:Pi00-projector} and \eqref{eq:Pi1Sz-projector}, and use elementary identities of traces of gamma matrices.

\subsubsection{S wave}
For the S-wave ($L=0$), we have
\begin{align}
    \Fcal^{\lambda, \kappa,J_z}(p,\rt) 
    &=   \Tr\left[ \Pi^{J J_z}(p,0) \Ncal^{\lambda}(p, 0; \rt)  \right] \,, 
\end{align}
where we use $\langle 0 L_z; S S_z | J J_z \rangle = \delta_{0 L_z} \delta_{S J} \delta_{S_z J_z}$. 
The projectors in Eqs.\,\eqref{eq:Pi00-projector} and \eqref{eq:Pi1Sz-projector} at $k=0$ simplify to
\begin{align}
    \Pi^{00}(p,0)
    &=-\frac{1}{\sqrt{8m_Q}} \gamma^5 (\slashed{p} + 2m_Q) \,, \\
    \Pi^{1J_z}(p,0) &=-\frac{\epsilon^*_{\rho}(J_z) }{\sqrt{8m_Q}} \gamma^{\rho} (\slashed{p} + 2m_Q) \,.
\end{align}
The perturbative functions $\Ncal^{\lambda}$ in Eqs.\,\eqref{eq:NcalL} and \eqref{eq:NcalT} at $k=0$ (note $\xi=0$ since $k^+=0$) are
\begin{align}
    \Ncal^{\lambda=0}(p, 0; \rt) =& - \frac{1}{4} Q K_0(\overline{Q} |\rt|) \frac{\gamma^+}{q^+} \,,\\
    \Ncal^{\lambda=\pm 1}(p, 0; \rt) =&
    -\frac{i \rtL{\alpha}}{4|\rt|} \overline{Q} K_1(\overline{Q} |\rt|) \etL{\beta}^{\lambda}   [\gammatU{\alpha},\gammatU{\beta}]   \frac{\gamma^+}{q^+}  - \frac{1}{2} m_Q  K_0(\overline{Q} |\rt| ) \etL{\alpha}^{\lambda} \gammatU{\alpha} \frac{\gamma^+}{q^+} \,,
\end{align}
Then, for the  ${}^{1} \mathrm{S}_0$ state we have
\begin{align}
    \Fcal^{\lambda=0, {}^{1}\mathrm{S}_0,J_z=0}(p,\rt) & =0 \,, \\
    \Fcal^{\lambda=\pm 1, {}^{1}\mathrm{S}_0,J_z=0}(p,\rt)
    & = \frac{\etL{\beta}^{\lambda} }{\sqrt{2m_Q}}
    \frac{ \rtL{\alpha}}{|\rt|} \overline{Q} K_1(\overline{Q} |\rt|)   \epsilon_\perp^{\alpha\beta} \,,
\end{align}
where $\epsilon_\perp^{\alpha\beta}$ is the Levi-Civita tensor in 2 dimensions ($\epsilon_\perp^{12}=-\epsilon_\perp^{21}=1$).

Furthermore, for the ${}^{3} \mathrm{S}_1$ state we have
\begin{align}
    \Fcal^{\lambda, {}^{3}\mathrm{S}_1,J_z}(p,\rt) 
    & =  \widetilde{\Fcal}^{\lambda, {}^{3}\mathrm{S}_1,\rho}(p,\rt) \epsilon^*_{\rho}(J_z) \,, 
\end{align}
where
\begin{align}
    \widetilde{\Fcal}^{\lambda=0, {}^{3}\mathrm{S}_1,\rho}(p,\rt) 
    & =  \frac{1}{\sqrt{2} q^+} Q K_0(\overline{Q}|\rt|) \sqrt{m_Q} g^{\rho+}  \,, \\
    \widetilde{\Fcal}^{\lambda=\pm 1, {}^{3}\mathrm{S}_1,\rho}(p,\rt) 
    &  = -\frac{\etL{\alpha}^{\lambda}}{\sqrt{2q^+}}\sqrt{m_Q} K_0(\overline{Q} |\rt| )  ( g^{\rho+} \ptU{\alpha} - g^{\rho\alpha}_\perp p^+) \,.
\end{align}
\subsubsection{P wave}
For the P-wave ($L=1$) we have
\begin{align}
    \Fcal^{\lambda, \kappa,J_z}(p,\rt) =& \sum_{L_z,S_z}  \langle L L_z; S S_z | J J_z \rangle  \times \epsilon^{*}_{\mu}(L_z) \frac{\partial}{\partial k_\mu} \left\{ e^{-i \kt \cdot \rt} \Tr\left[ \Pi^{SS_z}(p,k)  \Ncal^{\lambda}(p, k; \rt)  \right]\right \} \Big |_{k=0} \,,
\end{align}
It is convenient to first carry out the derivative and set $k=0$ before taking trace of gamma matrices, then we have
\begin{align}
    & \frac{\partial}{\partial k_\mu} \left\{  \Pi^{SS_z}(p,k)  e^{-i \kt \cdot \rt}  \Ncal^{\lambda}(p, k; \rt) \right \} \Big |_{k=0} \nonumber \\ 
    &=\left[i \rtU{\mu} \Pi^{SS_z}(p,0)   + \frac{\partial}{\partial k_\mu} \left\{  \Pi^{SS_z}(p,k)  \right \} \Big |_{k=0}  \right]  \Ncal^{\lambda}(p, 0; \rt) +  \Pi^{SS_z}(p,0)  \frac{\partial}{\partial k_\mu}  \Ncal^{\lambda}(p, k; \rt) \Big |_{k=0} \,,
\end{align}
where we use
\begin{align}
    \frac{\partial}{\partial k_\mu} (\kt \cdot \rt) = \frac{\partial}{\partial k_\mu} \left(-g_{\perp}^{\alpha\beta} k_{\alpha} r_{\beta} \right) =  -g_{\perp}^{\alpha\beta} \delta^{\mu}_{\alpha} r_{\beta} =  -g_{\perp}^{\mu\beta}  r_{\beta} = - \rt^{\mu} \,.
\end{align}
We compute the derivatives of the projectors at $k=0$:
\begin{align}
    \frac{\partial}{\partial k_\mu} \Pi^{00}(p,k) \Big |_{k=0} &=  \frac{1}{\sqrt{32m_Q^3}} \gamma^5   [\gamma^{\mu},\slashed{p}] \,, \\
    \frac{\partial}{\partial k_\mu} \Pi^{1 S_z}(p,k) \Big |_{k=0} &=\frac{\epsilon^*_{\rho}(S_z) }{\sqrt{8m_Q^3}} \left[ \frac{1}{2} \gamma^\rho  [\gamma^{\mu},\slashed{p}] - g^{\mu\rho} (\slashed{p} + 2m_Q) \right] \,,
\end{align}
as well as the derivatives of the perturbative functions at $k=0$:
\begin{align}
    \frac{\partial}{\partial k_\mu}  \Ncal^{\lambda = 0}(p, k; \rt) \Big |_{k=0} &= 0 \,, \\
    \frac{\partial}{\partial k_\mu}  \Ncal^{\lambda=\pm 1}(p,k ; \rt) \Big|_{k=0} 
    & = - \frac{i}{p^+}  \frac{\rt \cdot \etU{\lambda}}{|\rt|}  \overline{Q} K_1(\overline{Q} |\rt|) g^{\mu+}  \frac{\gamma^+}{q^+} \,.  
\end{align}
It is worth noting that since we express the perturbative functions $\Ncal^{\lambda}(p, k; \rt)$ in Eqs.\,\eqref{eq:NcalL} and \eqref{eq:NcalT} in coordinate space and factor out the phase $e^{-i\kt\cdot\rt}$ from their definition, the only dependence on $k$ is on the momentum fraction $\xi$. 

Combining these results we find for the ${}^{1} \mathrm{P}_1$ state
\begin{align}
    \Fcal^{\lambda, {}^{1} \mathrm{P}_1,J_z}(p,\rt) 
    & =    \widetilde{\Fcal}^{\lambda, {}^{1} \mathrm{P}_1,\mu}(p,\rt) \epsilon^{*}_{\mu}(J_z)\,,
\end{align}
where we use $\langle 1 L_z; 0 S_z | J J_z \rangle = \delta_{0 S_z} \delta_{1 J} \delta_{L_z J_z}$, and
\begin{align}
    \widetilde{\Fcal}^{\lambda=0, {}^{1} \mathrm{P}_1,\mu}(p,\rt) 
    & = 0 \,, \\
    \widetilde{\Fcal}^{\lambda=\pm 1, {}^{1} \mathrm{P}_1,\mu}(p,\rt) 
    & =  i \frac{\etL{\beta}^{\lambda}}{\sqrt{2m_Q}} \left[  \frac{1}{|\rt|} \overline{Q} K_1(\overline{Q} |\rt|) \rtL{\alpha}  \rtU{\mu} \epsilon_\perp^{\alpha \beta} -  K_0(\overline{Q} |\rt| ) \delta_{\perp\alpha}^{\beta} \left(\epsilon_\perp^{\mu\alpha} - \frac{1}{q^+} p_{\nu} \epsilon_\perp^{\nu\alpha} g^{\mu+} \right)  \right]  \,.
\end{align}
Furthermore, for the ${}^{3} \mathrm{P}_{J=0,1,2}$ states we have
\begin{align}
    \Fcal^{\lambda, {}^{3} \mathrm{P}_J,J_z}(p,\rt) 
    & = \widetilde{\Fcal}^{\lambda, {}^{3} \mathrm{P}_J,\rho\mu}(p,\rt)  \epsilon^{*}_{\rho\mu}(J,J_z) \,, 
\end{align}
where $\epsilon^{*}_{\rho\mu}(J,J_z)$ was defined in Eq.\,\eqref{eq:tensor_p-wave}, and
\begin{align}
    \widetilde{\Fcal}^{\lambda=0, {}^{3} \mathrm{P}_J,\rho\mu}(p,\rt) 
    & =  \frac{1}{\sqrt{8 m_Q^3}} Q K_0(\overline{Q}|\rt|) \left( 2 i m_Q^2 \rtU{\mu} \frac{g^{\rho+}}{p^+}   +  p^{\rho} \frac{g^{\mu+}}{p^+} \right)     \,, \\
    \widetilde{\Fcal}^{\lambda=\pm 1, {}^{3}\mathrm{P}_J,\rho\mu}(p,\rt) & = \frac{\etL{\beta}^{\lambda}}{\sqrt{2m_Q^3}} \Bigg\{   m_Q^2 \rtU{\mu}  K_0(\overline{Q} |\rt| )  \left( - g^{\beta \rho}_{\perp}  + \frac{g^{\rho +}}{p^+} \ptU{\beta}  \right) -  \frac{\rtL{\alpha}}{|\rt|} \overline{Q} K_1(\overline{Q} |\rt|)  \left[ 4 m_Q^2   g_\perp^{\alpha\beta} \frac{g^{\mu+}}{p^+} \frac{g^{\rho+}}{p^+} \right.
     \nonumber \\
    &  \left.  + \ptU{\beta} \left( \frac{g^{\mu+}}{p^+} g_\perp^{\alpha \rho} - \frac{g^{\rho +}}{p^+} g_\perp^{\alpha \mu} \right) + \ptU{\alpha} \left(\frac{g^{\rho +}}{p^+} g_\perp^{\beta \mu} - \frac{g^{\mu+}}{p^+} g_\perp^{\beta \rho} \right) + \left( g_\perp^{\alpha\mu} g_\perp^{\beta \rho} - g_\perp^{\alpha \rho} g_\perp^{\beta \mu} \right) \right] \Bigg\}  \,.
\end{align}

\subsection{Computation of perturbative functions $\Gamma_{\lambda\lambda'}^{\kappa}$}
Following our results in the previous sections, the perturbative functions $\Gamma_{\lambda\lambda'}^{\kappa}(\pt,Q;\rt,\rt')$ defined in Eq.\,\eqref{eq:GammahardFactor} can be computed:
\begin{align}
    \Gamma_{\lambda\lambda'}^{{}^{1} \mathrm{S}_0}(\pt,Q;\rt,\rt') &= \alpha_{\mathrm{em}} e_Q^2 \Fcal^{\lambda, {}^{1} \mathrm{S}_0,J_z=0}(p,\rt) \Fcal^{\dagger \lambda', {}^{1} \mathrm{S}_0,J_z=0}(p,\rt)\,, \\
    \Gamma_{\lambda\lambda'}^{{}^{3} \mathrm{S}_1}(\pt,Q;\rt,\rt') &= \alpha_{\mathrm{em}} e_Q^2 \widetilde{\Fcal}^{\lambda, {}^{3} \mathrm{S}_1,\rho}(p,\rt) \widetilde{\Fcal}^{\dagger \lambda', {}^{3} \mathrm{S}_1,\rho'}(p,\rt) \frac{1}{3} \sum_{J_z} \epsilon^*_{\rho}(J_z) \epsilon_{\rho'}(J_z)\,,
     \\
    \Gamma_{\lambda\lambda'}^{{}^{1} \mathrm{P}_1}(\pt,Q;\rt,\rt') &= \alpha_{\mathrm{em}} e_Q^2 \widetilde{\Fcal}^{\lambda, {}^{1} \mathrm{P}_1,\mu}(p,\rt) \widetilde{\Fcal}^{\dagger \lambda', {}^{1} \mathrm{P}_1,\mu'}(p,\rt) \frac{1}{3}\sum_{J_z} \epsilon^*_{\mu}(J_z) \epsilon_{\mu'}(J_z) \,,\\
    \Gamma_{\lambda\lambda'}^{{}^{3} \mathrm{P}_{J}}(\pt,Q;\rt,\rt') &= \alpha_{\mathrm{em}} e_Q^2 \widetilde{\Fcal}^{\lambda, {}^{3} \mathrm{P}_J,\rho\mu}(p,\rt) \widetilde{\Fcal}^{\dagger \lambda', {}^{3} \mathrm{P}_J,\rho'\mu'}(p,\rt) \frac{1}{2J+1} \sum_{J_z} \epsilon^{*}_{\rho\mu}(J,J_z) \epsilon_{\rho'\mu'}(J,J_z) \,.
\end{align}
We also define the perturbative function,
\begin{align}
    \Gamma_{\lambda\lambda'}^{{}^{3} \mathrm{P}_{\langle J \rangle}} = \frac{1}{9}\left[ \Gamma_{\lambda\lambda'}^{{}^{3} \mathrm{P}_{0}} + 3\Gamma_{\lambda\lambda'}^{{}^{3} \mathrm{P}_{1}} + 5\Gamma_{\lambda\lambda'}^{{}^{3} \mathrm{P}_{2}}\right] \,, \label{eq:pert-factor-3PJ-average}  
\end{align}
corresponding to the short distance coefficient defined in Eq.\,\eqref{eq:SDC-average-3PJ} which was introduced to exploit the heavy quark symmetry of the LDMEs (see Eq.\,\eqref{eq:heavy-quark-symmetry}). Then we have
\begin{align}
    \Gamma_{\lambda\lambda'}^{{}^{3} \mathrm{P}_{\langle J \rangle}}(\pt,Q;\rt,\rt') &= \alpha_{\mathrm{em}} e_Q^2 \widetilde{\Fcal}^{\lambda, {}^{3} \mathrm{P}_J,\rho\mu}(p,\rt) \widetilde{\Fcal}^{\dagger \lambda', {}^{3} \mathrm{P}_J,\rho'\mu'}(p,\rt) \sum_{J} \sum_{J_z} \epsilon^{*}_{\rho\mu}(J,J_z) \epsilon_{\rho'\mu'}(J,J_z) \,.
\end{align}

The explicit expressions are given for these functions are provided in Appendix \ref{app:hard_CGC}.

\section{Complete expressions for the perturbative functions}
\label{app:hard_factors}
\subsection{Perturbative functions in the CGC}
\label{app:hard_CGC}

In the case of longitudinally polarized photons, the perturbative functions are:
\begin{align}
    \Gamma_{\mathrm{L}}^{{}^{3}\mathrm{S}_1}&=    \frac{\alpha_{\rm{em}} e_Q^2}{24 m_Q} Q K_0(\overline{Q}|\rt|) Q K_0(\overline{Q}|\rt'|) 
    \label{eq:Hard_gammaA_L0_3S1} \,, \\
    \Gamma_{\mathrm{L}}^{ {}^{3}\mathrm{P}_0}&=     \frac{\alpha_{\rm{em}} e_Q^2}{96 m_Q^3} (\pt \cdot \rt) (\pt \cdot \rt')  Q K_0(\overline{Q}|\rt|)  Q K_0(\overline{Q}|\rt'|) 
    \label{eq:Hard_gammaA_L0_3P0} \,, \\
    \Gamma_{\mathrm{L}}^{ {}^{3}\mathrm{P}_1}&=    \frac{\alpha_{\rm{em}} e_Q^2 }{48 m_Q} (\rt \cdot \rt')  Q K_0(\overline{Q}|\rt|)  Q K_0(\overline{Q}|\rt'|) 
    \label{eq:Hard_gammaA_L0_3P1} \,, \\
    \Gamma_{\mathrm{L}}^{ {}^{3}\mathrm{P}_2}&=     \frac{\alpha_{\rm{em}} e_Q^2}{80 m_Q^3} \left[ m_Q^2  (\rt \cdot \rt') + \frac{1}{3} (\pt \cdot \rt) (\pt \cdot \rt') \right]  Q K_0(\overline{Q}|\rt|)  Q K_0(\overline{Q}|\rt'|) 
    \label{eq:Hard_gammaA_L0_3P2} \,,
    \\
    \Gamma_{\mathrm{L}}^{ {}^{3}\mathrm{P}_{\langle J \rangle}}&=     \frac{\alpha_{\rm{em}} e_Q^2}{72 m_Q^3} \left[ m_Q^2  (\rt \cdot \rt') + \frac{1}{4} (\pt \cdot \rt) (\pt \cdot \rt') \right] Q K_0(\overline{Q}|\rt|)  Q K_0(\overline{Q}|\rt'|) \,.
    \label{eq:Hard_gammaA_L0_3PJ}
\end{align}  
The contributions of the ${}^{1}\mathrm{S}_0$ and ${}^{1}\mathrm{P}_1$ states vanish.

In the case of transversely polarized photons, the perturbative functions are:
\begin{align}
    \Gamma_{\mathrm{T}}^{{}^{1}\mathrm{S}_0} &=   \frac{\alpha_{\rm{em}} e_Q^2}{4 m_Q} \frac{\rt\cdot \rt'}{|\rt||\rtC|}    \overline{Q} K_1(\overline{Q} |\rt|)  \overline{Q} K_1(\overline{Q} |\rt'|)  \,, \label{eq:Hard_gammaA_T0_1S0}  \\
    \Gamma_{\mathrm{T}}^{ {}^{3}\mathrm{S}_1} &= \frac{\alpha_{\rm{em}} e_Q^2}{6 m_Q} m_Q K_0(\overline{Q}|\rt|) m_Q K_0(\overline{Q}|\rt'|) \,,
    \label{eq:Hard_gammaA_T0_3S1} \\
    \Gamma_{\mathrm{T}}^{ {}^{1}\mathrm{P}_1} &= \frac{\alpha_{\rm{em}} e_Q^2}{12m_Q^3} \Bigg\{ \frac{(\rt\cdot\rt')}{|\rt| |\rtC|} \overline{Q} K_1(\overline{Q} |\rt|) \overline{Q} K_1(\overline{Q} |\rtC|)  \left[ m_Q^2 (\rt\cdot\rt')  + \frac{1}{4}(\pt\cdot\rt)(\pt\cdot \rt') \right]  \nonumber \\
    & - \overline{Q}|\rt|  K_1(\overline{Q} |\rt|) m_Q^2 K_0(\overline{Q} |\rt'|) - m_Q^2 K_0(\overline{Q} |\rt|) \overline{Q}|\rt'|  K_1(\overline{Q} |\rt'|)  + 2 m_Q K_0(\overline{Q} |\rt|)  m_Q K_0(\overline{Q} |\rt'|) \Bigg\} \,, \\
    \Gamma_{\mathrm{T}}^{ {}^{3}\mathrm{P}_0} & =   \frac{\alpha_{\rm{em}} e_Q^2}{12 m_Q^3} \frac{(\rt\cdot \rt')}{|\rt||\rt'|} \left[\overline{Q} K_1(\overline{Q} |\rt|) + m_Q^2 |\rt| K_0(\overline{Q} |\rt|) \right] \left[\overline{Q} K_1(\overline{Q} |\rt'|) + m_Q^2 |\rt'| K_0(\overline{Q} |\rt'|) \right] \,,
    \label{eq:Hard_gammaA_T0_3P0} \\
    \Gamma_{\mathrm{T}}^{ {}^{3}\mathrm{P}_1} &=  \frac{\alpha_{\rm{em}} e_Q^2}{6 m_Q^3} \left\{  \frac{(\rt\cdot \rt')}{|\rt||\rt'|} \left[\overline{Q} K_1(\overline{Q} |\rt|) -\frac{1}{2} m_Q^2 |\rt| K_0(\overline{Q} |\rt|) \right] \left[\overline{Q} K_1(\overline{Q} |\rt'|) -\frac{1}{2}  m_Q^2 |\rt'| K_0(\overline{Q} |\rt'|) \right] \right. \nonumber \\
    & \left. + \frac{1}{8} (\pt \cdot \rt)(\pt \cdot \rt') m_Q K_0(\overline{Q} |\rt|) m_Q K_0(\overline{Q} |\rt'|) \right\} \,,
    \label{eq:Hard_gammaA_T0_3P1} \\
    \Gamma_{\mathrm{T}}^{ {}^{3}\mathrm{P}_2} &=  \frac{\alpha_{\rm{em}} e_Q^2}{30 m_Q^3} \left\{  \frac{(\rt\cdot \rt')}{|\rt||\rt'|} \left[\overline{Q} K_1(\overline{Q} |\rt|) -\frac{1}{2} m_Q^2 |\rt| K_0(\overline{Q} |\rt|) \right]  \left[\overline{Q} K_1(\overline{Q} |\rt'|) -\frac{1}{2}  m_Q^2 |\rt'| K_0(\overline{Q} |\rt'|) \right] \right. \nonumber \\
    & \left. + \frac{3}{2}\left[ m_Q^2 (\rt \cdot \rt') +\frac{1}{4} (\pt \cdot \rt)(\pt \cdot \rt') \right] m_Q K_0(\overline{Q} |\rt|) m_Q K_0(\overline{Q} |\rt'|) \right\} \,,
    \label{eq:Hard_gammaA_T0_3P2}  \\
    \Gamma_{\mathrm{T}}^{ {}^{3}\mathrm{P}_{\langle J \rangle}} &=  \frac{\alpha_{\rm{em}} e_Q^2}{12 m_Q^3}  \left\{ \frac{(\rt\cdot \rt')}{|\rt||\rt'|} \left[\overline{Q} K_1(\overline{Q} |\rt|) -\frac{1}{3} m_Q^2 |\rt| K_0(\overline{Q} |\rt|) \right] \left[\overline{Q} K_1(\overline{Q} |\rt'|) -\frac{1}{3}  m_Q^2 |\rt'| K_0(\overline{Q} |\rt'|) \right] \right. \nonumber \\
    & \left. + \frac{5}{9}\left[ m_Q^2 (\rt \cdot \rt') +\frac{3}{10} (\pt \cdot \rt)(\pt \cdot \rt') \right] m_Q K_0(\overline{Q} |\rt|) m_Q K_0(\overline{Q} |\rt'|) \right\} \,.
    \label{eq:Hard_gammaA_T0_3PJ} 
\end{align}
In the case of polarization changing photon ($TL$), the perturbative functions are:
\begin{align}
    \Gamma_{\mathrm{TL}}^{{}^{3}\mathrm{P}_0} &= -  \frac{\alpha_{\rm{em}} e_Q^2 }{24 \sqrt{2} m_Q^3} \frac{(\pt \cdot \rt)(\pt \cdot \rt')}{p_\perp |\rt|} \left[  m_Q^2 |\rt| K_0(\overline{Q} |\rt|) +  \overline{Q} K_1(\overline{Q} |\rt|) \right]  Q K_0(\overline{Q} |\rt'|) \,, \label{eq:Hard_gammaA_TL0_3P0} \\
    \Gamma_{\mathrm{TL}}^{{}^{3}\mathrm{P}_1} &=  \frac{\alpha_{\rm{em}} e_Q^2 }{48 \sqrt{2} m_Q^3} \frac{(\pt \cdot \rt)(\pt \cdot \rt')}{p_\perp |\rt|} m_Q^2 |\rt| K_0(\overline{Q} |\rt|) Q K_0(\overline{Q} |\rt'|) \,, \label{eq:Hard_gammaA_TL0_3P1} \\
    \Gamma_{\mathrm{TL}}^{ {}^{3}\mathrm{P}_2} &= -   \frac{\alpha_{\rm{em}} e_Q^2 }{60 \sqrt{2} m_Q^3} \frac{(\pt \cdot \rt)(\pt \cdot \rt')}{p_\perp |\rt|} \left[\overline{Q} K_1(\overline{Q} |\rt|) + \frac{1}{4} m_Q^2 |\rt| K_0(\overline{Q} |\rt|)  \right] Q K_0(\overline{Q} |\rt'|) \,, \label{eq:Hard_gammaA_TL0_3P2}  \\
    \Gamma_{\mathrm{TL}}^{{}^{3}\mathrm{P}_{\langle J \rangle}} &= -   \frac{\alpha_{\rm{em}} e_Q^2 }{72 \sqrt{2} m_Q^3} \frac{(\pt \cdot \rt)(\pt \cdot \rt')}{p_\perp |\rt|} \overline{Q} K_1(\overline{Q} |\rt|) ] Q K_0(\overline{Q} |\rt'|) \label{eq:Hard_gammaA_TL0_3PJ} \,.  
\end{align}
The contribution of ${}^{1}\mathrm{S}_0$, ${}^{3}\mathrm{S}_1$ and ${}^{1}\mathrm{P}_1$ states vanish.

In the case of polarization flipping photon ($T \lambda=+1,  T\lambda'=-1$), the perturbative functions are:
\begin{align}
    \Gamma_{\mathrm{T}\mathrm{flip}}^{{}^{1}\mathrm{S}_0} &= - \frac{\alpha_{\rm{em}} e_Q^2 }{4 m_Q} \Pi_{\perp}^{\beta\beta'}(\pt)\frac{\rtL{\beta} \rtCL{\beta'}}{|\rt||\rt'|}  \overline{Q} K_1(\overline{Q} |\rt|) \overline{Q} K_1(\overline{Q} |\rt'|) \label{eq:Hard_gammaA_Tflip0_1S0} \,, \\
    \Gamma_{\mathrm{T}\mathrm{flip}}^{{}^{1}\mathrm{P}_1}&= -\frac{\alpha_{\rm{em}} e_Q^2}{12 m_Q^3} \Pi_{\perp}^{\beta\beta'}(\pt) \left\{ \frac{\rtL{\beta} \rtCL{\beta'}}{|\rt||\rt'|}  \overline{Q} K_1(\overline{Q} |\rt|) \overline{Q} K_1(\overline{Q} |\rt'|) \left[ m_Q^2 (\rt\cdot\rt')  + \frac{1}{4}(\pt\cdot\rt)(\pt\cdot \rt') \right]   \right. \nonumber \\
    & \left. - \frac{\rtL{\beta}\rtL{\beta'}}{|\rt|} \overline{Q} K_1(\overline{Q}|\rt|)  m_Q^2 K_0(\overline{Q}|\rtC|)  + \frac{\rtCL{\beta}\rtCL{\beta'}}{|\rtC|} m_Q^2 K_0(\overline{Q}|\rt|)  \overline{Q} K_1(\overline{Q}|\rtC|) 
    \right\}\,, \\
    \Gamma_{\mathrm{T}\mathrm{flip}}^{{}^{3}\mathrm{P}_0} &=  \frac{\alpha_{\rm{em}} e_Q^2 }{12 m_Q^3} \Pi_{\perp}^{\beta\beta'}(\pt)\frac{\rtL{\beta} \rtCL{\beta'}}{|\rt||\rt'|} \left[\overline{Q} K_1(\overline{Q} |\rt|)  + m_Q^2 |\rt| K_0(\overline{Q} |\rt|) \right] \left[\overline{Q} K_1(\overline{Q} |\rt'|)  + m_Q^2 |\rt'| K_0(\overline{Q} |\rt'|) \right] \label{eq:Hard_gammaA_Tflip0_3P0} \,, \\
    \Gamma_{\mathrm{T}\mathrm{flip}}^{{}^{3}\mathrm{P}_1} &=  -\frac{\alpha_{\rm{em}} e_Q^2 }{6 m_Q^3} \Pi_{\perp}^{\beta\beta'}(\pt)\frac{\rtL{\beta} \rtCL{\beta'}}{|\rt||\rt'|}   \left[\overline{Q} K_1(\overline{Q} |\rt|)  -\frac{1}{2} m_Q^2 |\rt| K_0(\overline{Q} |\rt|) \right] \left[\overline{Q} K_1(\overline{Q} |\rt'|)  -\frac{1}{2}  m_Q^2 |\rt'| K_0(\overline{Q} |\rt'|) \right] \label{eq:Hard_gammaA_Tflip0_3P1} \,, \\
    \Gamma_{\mathrm{T}\mathrm{flip}}^{{}^{3}\mathrm{P}_2} &= \frac{\alpha_{\rm{em}} e_Q^2 }{30 m_Q^3} \Pi_{\perp}^{\beta\beta'}(\pt)\frac{\rtL{\beta} \rtCL{\beta'}}{|\rt||\rt'|} \left[\overline{Q} K_1(\overline{Q} |\rt|)  -\frac{1}{2} m_Q^2 |\rt| K_0(\overline{Q} |\rt|) \right] \left[\overline{Q} K_1(\overline{Q} |\rt'|)  -\frac{1}{2}  m_Q^2 |\rt'| K_0(\overline{Q} |\rt'|) \right] \label{eq:Hard_gammaA_Tflip0_3P2} \,,  \\
    \Gamma_{\mathrm{T}\mathrm{flip}}^{{}^{3}\mathrm{P}_{\langle J \rangle}} &= \frac{\alpha_{\rm{em}} e_Q^2 }{36 m_Q^3} \Pi_{\perp}^{\beta\beta'}(\pt)\frac{\rtL{\beta} \rtCL{\beta'}}{|\rt||\rt'|}   
    \Big\{ m_Q^2 |\rt| K_0(\overline{Q} |\rt|) m_Q^2 |\rt'| K_0(\overline{Q} |\rt'|)  \nonumber \\
    &  -\left[\overline{Q} K_1(\overline{Q} |\rt|)  - m_Q^2 |\rt| K_0(\overline{Q} |\rt|) \right] \left[\overline{Q} K_1(\overline{Q} |\rt'| - m_Q^2 |\rt'| K_0(\overline{Q} |\rt'|) )  \right]  \Big \} \label{eq:Hard_gammaA_Tflip0_3PJ} \,.
\end{align}
The contribution of the ${}^{3}\mathrm{S}_{1}$ state vanishes.

\subsection{Hard functions in TMD}
\label{sec:hard_factor_TMD}
Let us define the prefactor:
\begin{align}
    \Acal= \frac{\alpha_{s} \alpha_{\rm{em}}  e_Q^2(2\pi)^4}{(N_c^2-1)} \,,
\end{align}
and the mass of the quarkonium $M_{\mathcal{Q}} =  2 m_Q$. We will express the hard functions in terms of the tensors $\delta_{\perp,\alpha\alpha'}$ and $\Pi_{\perp \alpha \alpha'}(\pt)$, the latter was introduced in Eq.\,\eqref{eq:Pi-projector}.

Following the heavy-quark symmetry, we define the hard function:
\begin{align}
    H^{{}^{3} \mathrm{P}_{\langle J \rangle}}_{\lambda\lambda',\alpha\alpha'} = \frac{1}{9} \left[ H^{{}^{3} \mathrm{P}_{0}}_{\lambda\lambda',\alpha\alpha'} + 3 H^{{}^{3} \mathrm{P}_{1}}_{\lambda\lambda',\alpha\alpha'} + 5 H^{{}^{3} \mathrm{P}_{2}}_{\lambda\lambda',\alpha\alpha'} \right] \,. 
\end{align}

In the case of longitudinally polarized photons, the hard functions are:
\begin{align}
    H_{\mathrm{L},\alpha \alpha'}^{{}^{3}\mathrm{P}_1}(Q) &=   \frac{64 \Acal Q^2}{3 M_{\mathcal{Q}}(Q^2+M_{\mathcal{Q}}^2)^4} \delta_{\perp \alpha \alpha'} \,,
    \label{eq:Hard_corre_gammaA_L0_3P1} \\
    H_{\mathrm{L},\alpha \alpha'}^{ {}^{3}\mathrm{P}_2}(Q) &=  \frac{64 \Acal Q^2}{5 M_{\mathcal{Q}}(Q^2+M_{\mathcal{Q}}^2)^4} \delta_{\perp \alpha \alpha'} \,,
    \label{eq:Hard_corre_gammaA_L0_3P2}
    \\
    H_{\mathrm{L},\alpha \alpha'}^{ {}^{3}\mathrm{P}_{\langle J \rangle}}(Q) & = \frac{128 \Acal Q^2}{9 M_{\mathcal{Q}}(Q^2+M_{\mathcal{Q}}^2)^4} \delta_{\perp \alpha \alpha'} \,.
    \label{eq:Hard_corre_gammaA_L0_3PJ}
\end{align} 
The contributions of ${}^{1} \mathrm{S}_{0}$, ${}^{3} \mathrm{S}_{1}$, ${}^{1} \mathrm{P}_{1}$ and ${}^{3} \mathrm{P}_{0}$ states vanish. 

In the case of transversely polarized photons, the hard functions are:
\begin{align}
    H_{\mathrm{T},\alpha \alpha'}^{ {}^{1}\mathrm{S}_0}(Q) & =  \frac{4\Acal}{M_{\mathcal{Q}}(Q^2 + M_{\mathcal{Q}}^2)^2} \delta_{\perp \alpha \alpha'} \,, \label{eq:Hard_corre_gammaA_T0_1S0}  \\
    H_{\mathrm{T},\alpha \alpha'}^{ {}^{3}\mathrm{P}_0}(Q) & = \frac{16 \Acal \left(Q^2 + 3 M_{\mathcal{Q}}^2 \right)^2}{3 M_{\mathcal{Q}}^3 (Q^2+M_{\mathcal{Q}}^2)^4} \delta_{\perp \alpha \alpha'} \,,
    \label{eq:Hard_corre_gammaA_T0_3P0} \\
    H_{\mathrm{T},\alpha \alpha'}^{ {}^{3}\mathrm{P}_1}(Q) &= \frac{32 \Acal Q^4}{3 M_{\mathcal{Q}}^3 (Q^2+M_{\mathcal{Q}}^2)^4} \delta_{\perp \alpha \alpha'} \,,
    \label{eq:Hard_corre_gammaA_T0_3P1} \\
    H_{\mathrm{T},\alpha \alpha'}^{ {}^{3}\mathrm{P}_2}(Q) &=  \frac{32 \Acal \left(Q^4 + 6 M_{\mathcal{Q}}^4 \right)}{15 M_{\mathcal{Q}}^3 (Q^2+M_{\mathcal{Q}}^2)^4} \delta_{\perp \alpha \alpha'} \,,
    \label{eq:Hard_corre_gammaA_T0_3P2}  \\
    H_{\mathrm{T},\alpha \alpha'}^{ {}^{3}\mathrm{P}_{\langle J \rangle}}(Q) & = \frac{16 \Acal \left[ 3 Q^4 + 2 M_{\mathcal{Q}}^2 Q^2 + 7 M_{\mathcal{Q}}^4 \right]}{9 M_{\mathcal{Q}}^3 (Q^2+M_{\mathcal{Q}}^2)^4} \delta_{\perp \alpha \alpha'} \,.
    \label{eq:Hard_corre_gammaA_T0_3PJ} 
\end{align}  
The contributions of ${}^{3} \mathrm{S}_{1}$,  and ${}^{1} \mathrm{P}_{1}$ states vanish.\\

There are no contributions to the polarization changing photon ($TL$) case.\\

In the case of polarization flipping photon ($T \lambda=+1,  T\lambda'=-1$), the hard functions are:
\begin{align}
    H_{\mathrm{Tflip},\alpha \alpha'}^{ {}^{1}\mathrm{S}_0}(Q) &= -\frac{4 \Acal}{M_{\mathcal{Q}}(Q^2 + M_{\mathcal{Q}}^2)^2}\Pi_{\perp \alpha \alpha'}(\pt)  \,, \label{eq:Hard_corre_gammaA_Tflip0_1S0}  \\
    H_{\mathrm{Tflip},\alpha \alpha'}^{ {}^{3}\mathrm{P}_0}(Q) & = \frac{16 \Acal \left(Q^2 + 3 M_{\mathcal{Q}}^2 \right)^2}{3 M_{\mathcal{Q}}^3 (Q^2+M_{\mathcal{Q}}^2)^4}
    \Pi_{\perp \alpha \alpha'} (\pt) \,, \label{eq:Hard_corre_gammaA_Tflip0_3P0} \\
    H_{\mathrm{Tflip},\alpha \alpha'}^{ {}^{3}\mathrm{P}_1}(Q) & = - \frac{32 \Acal Q^4 }{3 M_{\mathcal{Q}}^3 (Q^2+M_{\mathcal{Q}}^2)^4}\Pi_{\perp \alpha \alpha'}(\pt) \,,
    \label{eq:Hard_corre_gammaA_Tflip0_3P1} \\
    H_{\mathrm{Tflip},\alpha \alpha'}^{ {}^{3}\mathrm{P}_2}(Q) & = \frac{32 \Acal Q^4 }{15 M_{\mathcal{Q}}^3 (Q^2+M_{\mathcal{Q}}^2)^4}\Pi_{\perp \alpha \alpha'}(\pt) \,,
    \label{eq:Hard_corre_gammaA_Tflip0_3P2}  \\
    H_{\mathrm{Tflip},\alpha \alpha'}^{ {}^{3}\mathrm{P}_{\langle J \rangle}}(Q) & = \frac{16 \Acal \left(3 M_{\mathcal{Q}}^4 + 2 M_{\mathcal{Q}}^2 Q^2 - Q^4 \right)}{9 M_{\mathcal{Q}}^3 (Q^2+M_{\mathcal{Q}}^2)^4}  \Pi_{\perp \alpha \alpha'}(\pt) \,.
    \label{eq:Hard_corre_gammaA_Tflip0_3PJ} 
\end{align} 
The contributions of ${}^{3} \mathrm{S}_{1}$,  and ${}^{1} \mathrm{P}_{1}$ states vanish. 

\subsection{Hard functions in ITMD}
\label{app:hard_factors_ITMD}
In the equations below we define the variable
\begin{align}
    \chi^2 = \frac{p_\perp^2}{Q^2 + M_{\mathcal{Q}}^2}\,,
\end{align}
and the auxiliary functions:
\begin{align}
    I_{1a}(\chi) & = \frac{1}{2\left(1+\chi^2\right)^2} +\frac{1}{2\chi^2} \arctan^{2}\left( \chi \right) \,, \label{eq:I22a} \\
    I_{1b}(\chi) & = \frac{1}{2\left(1+\chi^2\right)^2} -\frac{1}{2\chi^2} \arctan^{2}\left( \chi \right) \,, 
    \label{eq:I22b}
\end{align}
\begin{align}
    I_{2a}(\chi)  & = \frac{1}{2(1+\chi^2)^4}+\frac{1}{8(1+\chi^2)^2} + \frac{1}{8\chi^2} \arctan^{2}\left( \chi\right) + \frac{1}{ 4\chi (1+\chi^2)} \arctan\left( \chi \right) \,, \label{eq:I33a} \\
    I_{2b}(\chi)  & = \frac{1}{2(1+\chi^2)^4}-\frac{1}{8(1+\chi^2)^2} - \frac{1}{8\chi^2} \arctan^{2}\left( \chi\right) - \frac{1}{ 4\chi (1+\chi^2)} \arctan\left( \chi \right) 
    \label{eq:I33b} \,,
\end{align}
\begin{align}
    I_{3a}(\chi) & = \frac{1}{2\left(1 + \chi^2\right)^3} + \frac{1}{4\chi(1+\chi^2)} \arctan\left( \chi \right) + \frac{1}{4\chi^2} \arctan^{2}\left( \chi \right) \,, \label{eq:I23a}  \\
    I_{3b}(\chi) & = \frac{1}{2\left(1 + \chi^2\right)^3} - \frac{1}{4\chi(1+\chi^2)} \arctan\left( \chi \right) - \frac{1}{4\chi^2} \arctan^{2}\left( \chi \right) \,.
    \label{eq:I23b}
\end{align}
Following the heavy-quark symmetry, we define the hard function:
\begin{align}
    \Hcal^{{}^{3} \mathrm{P}_{\langle J \rangle}}_{\lambda\lambda',\alpha\alpha'} = \frac{1}{9} \left[ \Hcal^{{}^{3} \mathrm{P}_{0}}_{\lambda\lambda',\alpha\alpha'} + 3 \Hcal^{{}^{3} \mathrm{P}_{1}}_{\lambda\lambda',\alpha\alpha'} + 5 \Hcal^{{}^{3} \mathrm{P}_{2}}_{\lambda\lambda',\alpha\alpha'} \right] \,. 
\end{align}

In the case of longitudinally polarized photons, the hard functions are:
\begin{align}
    \Hcal^{{}^{3} \mathrm{P}_0}_{\mathrm{L},\alpha\alpha'}(Q,\pt) & = \frac{\Acal Q^2}{24 m_Q^3 \overline{Q}^6} \frac{\chi^2}{(1 +\chi^2)^4}  \Big\{ \delta_{\perp\alpha\alpha'} +  \Pi_{\perp\alpha\alpha'}(\pt) \Big\}\,, \\
    \Hcal^{{}^{3} \mathrm{P}_1}_{\mathrm{L}, \alpha\alpha'}(Q,\pt) & = \frac{\Acal Q^2}{24 m_Q \overline{Q}^8}  \Big\{ I_{2a}(\chi) \delta_{\perp\alpha\alpha'} +  I_{2b}(\chi)  \Pi_{\perp\alpha\alpha'}(\pt) \Big\}\,, \\
    \Hcal^{{}^{3} \mathrm{P}_2}_{\mathrm{L},\alpha\alpha'}(Q,\pt) & = \frac{\Acal Q^2}{40 m_Q \overline{Q}^8}  \left\{ \left[ I_{2a}(\chi)  + \frac{2\chi^2}{3(1 +\chi^2)^4}\frac{\overline{Q}^2}{m_Q^2} \right]\delta_{\perp\alpha\alpha'} +  \left[ I_{2b}(\chi)  + \frac{2\chi^2}{3(1 +\chi^2)^4}\frac{\overline{Q}^2}{m_Q^2} \right]\Pi_{\perp\alpha\alpha'}(\pt) \right\}\,, \\
    \Hcal^{{}^{3} \mathrm{P}_{\langle J \rangle}}_{\mathrm{L},\alpha\alpha'}(Q,\pt) & = \frac{\Acal Q^2}{36 m_Q \overline{Q}^8}  \left\{ \left[ I_{2a}(\chi)  + \frac{\chi^2}{2(1 +\chi^2)^4}\frac{\overline{Q}^2}{m_Q^2} \right]\delta_{\perp\alpha\alpha'} +  \left[ I_{2b}(\chi)  + \frac{\chi^2}{2(1 +\chi^2)^4}\frac{\overline{Q}^2}{m_Q^2} \right]\Pi_{\perp\alpha\alpha'}(\pt) \right\} \,.
\end{align}
The contribution of ${}^{1} \mathrm{S}_0$, ${}^{3} \mathrm{S}_1$ and ${}^{1} \mathrm{P}_1$ states vanish.

In the case of transversely polarized photons, the hard functions are:
\begin{align}
    \Hcal^{ {}^{1}\mathrm{S}_0}_{\mathrm{T},\alpha\alpha'}(Q,\pt) = \frac{\Acal}{8m_Q \overline{Q}^4} & \Big\{ I_{1a}(\chi) \delta_{\perp\alpha\alpha'} + I_{1b}(\chi) \Pi_{\perp\alpha\alpha'}(\pt) \Big\}\,, \\
    \Hcal^{ {}^{3}\mathrm{P}_0}_{\mathrm{T},\alpha\alpha'}(Q,\pt) 
    = \frac{\Acal}{24 m_Q^3\overline{Q}^4} & \Bigg\{ \left[ I_{1a}(\chi)  +\frac{4m_Q^2}{\overline{Q}^2} I_{3a}(\chi)  + \frac{4m_Q^4}{\overline{Q}^4} I_{2a}(\chi) \right]\delta_{\perp\alpha\alpha'}\nonumber \\
    & + \left[ I_{1b}(\chi)  +\frac{4m_Q^2}{\overline{Q}^2} I_{3b}(\chi)  + \frac{4m_Q^4}{\overline{Q}^4} I_{2b}(\chi) \right] \Pi_{\perp\alpha\alpha'}(\pt) \Bigg\} \,, \\
    \Hcal^{ {}^{3}\mathrm{P}_1}_{\mathrm{T},\alpha\alpha'}(Q,\pt) 
    = \frac{\Acal}{12 m_Q^3\overline{Q}^4} & \Bigg\{ \left[ I_{1a}(\chi)  -\frac{2m_Q^2}{\overline{Q}^2} I_{3a}(\chi)  + \frac{m_Q^4}{\overline{Q}^4} I_{2a}(\chi)    +  \frac{m_Q^2}{ \overline{Q}^2}  \frac{\chi^2}{\left(1+\chi^2\right)^4} \right]\delta_{\perp\alpha\alpha'}\nonumber \\
    & + \left[ I_{1b}(\chi)   -\frac{2m_Q^2}{\overline{Q}^2} I_{3b}(\chi)  + \frac{m_Q^4}{\overline{Q}^4} I_{2b}(\chi)  +  \frac{m_Q^2}{ \overline{Q}^2}  \frac{\chi^2}{\left(1+\chi^2\right)^4} \right] \Pi_{\perp\alpha\alpha'}(\pt) \Bigg\} \,,\\
    \Hcal^{ {}^{3}\mathrm{P}_2}_{\mathrm{T},\alpha\alpha'}(Q,\pt) 
    = \frac{\Acal}{60 m_Q^3\overline{Q}^4} & \Bigg\{ \left[ I_{1a}(\chi)  -\frac{2m_Q^2}{\overline{Q}^2} I_{3a}(\chi)  + \frac{7m_Q^4}{\overline{Q}^4} I_{2a}(\chi)    +  \frac{3m_Q^2}{ \overline{Q}^2}  \frac{\chi^2}{\left(1+\chi^2\right)^4} \right]\delta_{\perp\alpha\alpha'}\nonumber \\
    & + \left[ I_{1b}(\chi)   -\frac{2m_Q^2}{\overline{Q}^2} I_{3b}(\chi)  + \frac{7m_Q^4}{\overline{Q}^4} I_{2b}(\chi)  +  \frac{3m_Q^2}{ \overline{Q}^2}  \frac{\chi^2}{\left(1+\chi^2\right)^4} \right] \Pi_{\perp\alpha\alpha'}(\pt) \Bigg\} \,,\\
    \Hcal^{ {}^{3}\mathrm{P}_{\langle J \rangle}}_{\mathrm{T},\alpha\alpha'}(Q,\pt) 
    = \frac{\Acal}{24 m_Q^3\overline{Q}^4} & \Bigg\{ \left[ I_{1a}(\chi)  -\frac{4m_Q^2}{3\overline{Q}^2} I_{3a}(\chi)  + \frac{8 m_Q^4}{3 \overline{Q}^4} I_{2a}(\chi)    +  \frac{4m_Q^2}{3 \overline{Q}^2}  \frac{\chi^2}{\left(1+\chi^2\right)^4} \right]\delta_{\perp\alpha\alpha'}\nonumber \\
    & + \left[ I_{1b}(\chi)   -\frac{4m_Q^2}{3\overline{Q}^2} I_{3b}(\chi)  + \frac{8 m_Q^4}{3 \overline{Q}^4} I_{2b}(\chi)  +  \frac{4m_Q^2}{3 \overline{Q}^2}  \frac{\chi^2}{\left(1+\chi^2\right)^4} \right] \Pi_{\perp\alpha\alpha'}(\pt) \Bigg\} \,.
\end{align}
The contribution of ${}^{3} \mathrm{S}_1$ and ${}^{1} \mathrm{P}_1$ states vanish.

In the case of polarization changing photon ($TL$), the hard functions are:
\begin{align}
    \Hcal^{ {}^{3}\mathrm{P}_0}_{\mathrm{TL},\alpha\alpha'}(Q,\pt) &= - \frac{\Acal Q}{24 \sqrt{2} m_Q^3}  \frac{1}{\overline{Q}^5} \frac{\chi}{\left(1+\chi^2\right)^3} \left[\frac{2 m_Q^2}{\overline{Q}^2} \frac{1}{1+\chi^2} + 1 \right]\Big\{ \delta_{\perp\alpha\alpha'} + \Pi_{\perp\alpha\alpha'}(\pt) \Big\} \,,  \\
    \Hcal^{ {}^{3}\mathrm{P}_1}_{\mathrm{TL},\alpha\alpha'}(Q,\pt) &=  \frac{\Acal Q}{24 \sqrt{2} m_Q}  \frac{1}{\overline{Q}^7} \frac{\chi}{\left(1+\chi^2\right)^4}  \Big\{ \delta_{\perp\alpha\alpha'} + \Pi_{\perp\alpha\alpha'}(\pt) \Big\} \,,  \\
    \Hcal^{ {}^{3}\mathrm{P}_2}_{\mathrm{TL},\alpha\alpha'}(Q,\pt) &= - \frac{\Acal Q}{60 \sqrt{2} m_Q^3}  \frac{1}{\overline{Q}^5}\frac{\chi}{\left(1+\chi^2\right)^3}\left[\frac{m_Q^2}{2 \overline{Q}^2} \frac{1}{1+\chi^2} + 1 \right] \Big\{ \delta_{\perp\alpha\alpha'} + \Pi_{\perp\alpha\alpha'}(\pt) \Big\} \,, \\
    \Hcal^{ {}^{3}\mathrm{P}_{\langle J \rangle}}_{\mathrm{TL},\alpha\alpha'}(Q,\pt) &=  - \frac{\Acal Q}{72 \sqrt{2} m_Q^3}  \frac{1}{\overline{Q}^5}\frac{\chi}{\left(1+\chi^2\right)^3} \Big\{ \delta_{\perp\alpha\alpha'} + \Pi_{\perp\alpha\alpha'}(\pt) \Big\} \,. 
\end{align}
The contributions of ${}^{1} \mathrm{S}_0$, ${}^{3} \mathrm{S}_1$ and ${}^{1} \mathrm{P}_1$ states vanish.

In the case of polarization flipping photon ($T \lambda=+1,  T\lambda'=-1$), the hard functions are:
\begin{align}
    \Hcal^{ {}^{1}\mathrm{S}_0}_{\mathrm{Tflip},\alpha\alpha'}(Q,\pt) = - \frac{\Acal}{8 m_Q \overline{Q}^4} &\Big\{ I_{1b}(\chi) \delta_{\perp\alpha\alpha'} +  I_{1a}(\chi) \Pi_{\perp\alpha\alpha'}(\pt) \Big\} \,, \\
    \Hcal^{ {}^{3}\mathrm{P}_0}_{\mathrm{Tflip},\alpha\alpha'}(Q,\pt)= \frac{\Acal}{24 m_Q^3 \overline{Q}^4}  
    & \left\{ \left[I_{1b}(\chi)+\frac{4 m_Q^2}{\overline{Q}^2} I_{3b}(\chi)  +\frac{4 m_Q^4}{\overline{Q}^4} I_{2b}(\chi)  \right] \delta_{\perp\alpha\alpha'} 
    \right. \nonumber \\
    & + \left. \left[I_{1a}(\chi)+\frac{4 m_Q^2}{\overline{Q}^2} I_{3a}(\chi) +\frac{4m_Q^4}{\overline{Q}^4} I_{2a}(\chi)  \right]\Pi_{\perp\alpha\alpha'}(\pt)   \right\} \,, \\
    \Hcal^{ {}^{3}\mathrm{P}_1}_{\mathrm{Tflip},\alpha\alpha'}(Q,\pt)= -\frac{\Acal}{12 m_Q^3 \overline{Q}^4}  
    & \left\{ \left[I_{1b}(\chi)-\frac{2 m_Q^2}{\overline{Q}^2} I_{3b}(\chi)  +\frac{m_Q^4}{\overline{Q}^4} I_{2b}(\chi)  \right] \delta_{\perp\alpha\alpha'} 
    \right. \nonumber \\
    & + \left. \left[I_{1a}(\chi)-\frac{2 m_Q^2}{\overline{Q}^2} I_{3a}(\chi) +\frac{m_Q^4}{\overline{Q}^4} I_{2a}(\chi)  \right]\Pi_{\perp\alpha\alpha'}(\pt)   \right\} \,, \\
    \Hcal^{ {}^{3}\mathrm{P}_2}_{\mathrm{Tflip},\alpha\alpha'}(Q,\pt)= \frac{\Acal}{60 m_Q^3 \overline{Q}^4}  
    & \left\{ \left[I_{1b}(\chi)-\frac{2 m_Q^2}{\overline{Q}^2} I_{3b}(\chi)  +\frac{m_Q^4}{\overline{Q}^4} I_{2b}(\chi)  \right] \delta_{\perp\alpha\alpha'} 
    \right. \nonumber \\
    & + \left. \left[I_{1a}(\chi)-\frac{2 m_Q^2}{\overline{Q}^2} I_{3a}(\chi) +\frac{m_Q^4}{\overline{Q}^4} I_{2a}(\chi)  \right]\Pi_{\perp\alpha\alpha'}(\pt)   \right\} \,, \\
    \Hcal^{ {}^{3}\mathrm{P}_{\langle J \rangle}}_{\mathrm{Tflip},\alpha\alpha'}(Q,\pt)= \frac{\Acal}{72 m_Q^3 \overline{Q}^4}  
    &\left\{ \left[\frac{4 m_Q^2}{\overline{Q}^2} I_{3b}(\chi)  -I_{1b}(\chi) \right] \delta_{\perp\alpha\alpha'} + \left[\frac{4 m_Q^2}{\overline{Q}^2} I_{3a}(\chi)  -I_{1a}(\chi) \right] \Pi_{\perp\alpha\alpha'}(\pt)   \right\} \,.
\end{align}
The contribution of ${}^{3} \mathrm{S}_1$ and ${}^{1} \mathrm{P}_1$ states vanish.

\section{Gaussian approximation for Wilson line correlators}
\label{app:Gaussian}
The Gaussian approximation allows us to express any multi-point correlator of light-like Wilson line (and their derivatives) in terms of the dipole correlator and its derivatives \cite{Blaizot:2004wv,Iancu:2011nj,Dominguez:2011br}. 

In this approximation, the quadrupole correlator is
\begin{align}
    & S^{(4)}_Y(\xt,\yt;\yt',\xt') =\  S^{(2)}_Y(\xt,\yt) S^{(2)}_Y(\yt',\xt') \times \mathrm{exp}\left(-\frac{N_c}{4}F_Y(\xt,\yt';\yt,\xt') + \frac{1}{2N_c}F_Y(\xt,\yt;\yt',\xt')\right) \nonumber \\
    &\times \left[ \left( \frac{\sqrt{\Delta_Y} + F_Y(\xt,\yt';\yt,\xt')}{2\sqrt{\Delta_Y}} -\frac{ F_Y(\xt,\yt;\yt',\xt')}{\sqrt{\Delta_Y}} \right)\mathrm{exp}\left(\frac{N_c}{4}\sqrt{\Delta_Y}\right) \right.  \nonumber \\
    &\left. + \left( \frac{\sqrt{\Delta_Y} - F_Y(\xt,\yt';\yt,\xt')}{2\sqrt{\Delta_Y}} +\frac{ F_Y(\xt,\yt;\yt',\xt')}{\sqrt{\Delta_Y}} \right)\mathrm{exp}\left(-\frac{N_c}{4}\sqrt{\Delta_Y}\right) \right] \,,
    \label{eq:quadrupole_gaussian}
\end{align}
and the double-dipole correlator is
\begin{align}
    & S^{(2,2)}_Y(\xt,\yt;\yt',\xt') =\  S^{(2)}_Y(\xt,\yt) S^{(2)}_Y(\yt',\xt') \times \mathrm{exp}\left(-\frac{N_c}{4}F_Y(\xt,\yt';\yt,\xt') + \frac{1}{2N_c}F_Y(\xt,\yt;\yt',\xt')\right) \nonumber \\
    &\times \left[ \left( \frac{\sqrt{\Delta_Y} + F_Y(\xt,\yt';\yt,\xt')}{2\sqrt{\Delta_Y}} -\frac{ F_Y(\xt,\yt;\yt',\xt')}{N_c^2 \sqrt{\Delta_Y}} \right)\mathrm{exp}\left(\frac{N_c}{4}\sqrt{\Delta_Y}\right) \right.  \nonumber \\
    &\left. + \left( \frac{\sqrt{\Delta_Y} - F_Y(\xt,\yt';\yt,\xt')}{2\sqrt{\Delta_Y}} +\frac{ F_Y(\xt,\yt;\yt',\xt')}{N_c^2 \sqrt{\Delta_Y}} \right)\mathrm{exp}\left(-\frac{N_c}{4}\sqrt{\Delta_Y}\right) \right] \,, 
    \label{eq:dipole-dipole_gaussian}
\end{align}
where
\begin{align}
    \Delta_Y &= F_Y^2(\xt,\yt';\yt,\xt')+ \frac{4}{N_c^2} F_Y(\xt,\yt;\yt',\xt') F_Y(\xt,\xt';\yt',\yt) \,,
\end{align}
and
\begin{align}
    F_Y(\xt,\yt;\yt',\xt')& = \frac{1}{C_F} \ln\left[\frac{S^{(2)}_Y(\xt-\yt')   S^{(2)}_Y(\yt-\xt')}{S^{(2)}_Y(\xt-\xt') S^{(2)}_Y(\yt-\yt')}\right]\,.
\end{align}

\subsection{Color Octet}
In the Gaussian approximation, the correlator corresponding to the color octet contribution in Eq.\,\eqref{eq:octet_correlator} is obtained using Eqs.\,\eqref{eq:quadrupole_gaussian} and \eqref{eq:dipole-dipole_gaussian}:
\begin{align}
    &\overline{\Xi}_{Y}^{[8]}(\xt,\yt,\yt'\xt') = \frac{N_c}{N_c^2-1} \left[ S^{(4)}_Y(\xt,\yt;\yt',\xt') - S^{(2,2)}_Y(\xt,\yt;\yt',\xt') \right]
    \nonumber \\
    &= \frac{2 C_F}{N_c^2-1} S^{(2)}_Y(\xt,\yt) S^{(2)}_Y(\yt',\xt') \ \mathrm{exp}\left(-\frac{N_c}{4}F_Y(\xt,\yt';\yt,\xt') + \frac{1}{2N_c}F_Y(\xt,\yt;\yt',\xt')\right) \nonumber \\
    &\times \frac{ F_Y(\xt,\yt;\yt',\xt')}{\sqrt{\Delta_Y}} \left[\mathrm{exp}\left(-\frac{N_c}{4}\sqrt{\Delta_Y}\right)- \mathrm{exp}\left(\frac{N_c}{4}\sqrt{\Delta_Y}\right)  \right] \,.
    \label{eq:color_octet_Gaussian}
\end{align}

\subsection{Color Singlet}
Before we compute the color structure corresponding to singlet contribution in Eq.\eqref{eq:singlet_correlator}, it is convenient to split it into elastic and inelastic pieces
\begin{align}
    \overline{\Xi}_{Y}^{[1]}(\xt,\yt;\yt'\xt') &= \overline{\Xi}_{Y}^{[1],\mathrm{el}}(\xt,\yt;\yt'\xt')  + \overline{\Xi}_{Y}^{[1],\mathrm{inel}}(\xt,\yt;\yt'\xt') \,, 
\end{align}
where
\begin{align}
    \overline{\Xi}_{Y}^{[1],\mathrm{el}}(\xt,\yt;\yt'\xt') = N_c \left[ S_Y^{(2)}(\xt,\yt)-1\right]\left[ S_Y^{(2)}(\yt',\xt')-1\right] \,, \label{eq:singlet_correlator_elastic}
\end{align}
\begin{align}
    \overline{\Xi}_{Y}^{[1],\mathrm{inel}}(\xt,\yt;\yt'\xt') = N_c \left[ S_Y^{(2,2)}(\xt,\yt;\yt',\xt') - S_Y^{(2)}(\xt,\yt) S_Y^{(2)}(\yt',\xt') \right] \,.
    \label{eq:singlet_correlator_inelastic}
\end{align}
Since the elastic piece in Eq.\,\eqref{eq:singlet_correlator_elastic} is independent of $\bt-\bt'$ its contribution to the cross section is proportional to $\delta^{(2)}(\pt)$, so we can drop it. Thus it is sufficient to consider only the inelastic piece
\begin{align}
    & \overline{\Xi}_{Y}^{[1],\mathrm{inel}}(\xt,\yt;\yt'\xt') = N_c \left[ S^{(2,2)}_Y(\xt,\yt;\yt',\xt') - S^{(2)}_Y(\xt,\yt) S^{(2)}_Y(\yt',\xt') \right]
    \nonumber \\
    &= N_c S^{(2)}_Y(\xt,\yt) S^{(2)}_Y(\yt',\xt') \Bigg\{ \mathrm{exp}\left(-\frac{N_c}{4}F_Y(\xt,\yt';\yt,\xt') + \frac{1}{2N_c}F_Y(\xt,\yt;\yt',\xt')\right) \nonumber \\
    & \times \left[ \left( \frac{\sqrt{\Delta_Y} + F_Y(\xt,\yt';\yt,\xt')}{2\sqrt{\Delta_Y}} -\frac{ F_Y(\xt,\yt;\yt',\xt')}{N_c^2 \sqrt{\Delta_Y}} \right)\mathrm{exp}\left(\frac{N_c}{4}\sqrt{\Delta_Y}\right) \right.  \nonumber \\
    &\left. + \left( \frac{\sqrt{\Delta_Y} - F_Y(\xt,\yt';\yt,\xt')}{2\sqrt{\Delta_Y}} +\frac{ F_Y(\xt,\yt;\yt',\xt')}{N_c^2 \sqrt{\Delta_Y}} \right)\mathrm{exp}\left(-\frac{N_c}{4}\sqrt{\Delta_Y}\right) \right] -1 \Bigg\} \,.
    \label{eq:color_singlet_Gaussian}
\end{align}

\subsection{Weizsäcker-Williams gluon distribution}
In the Gaussian approximation, one can show (see appendix A in \cite{Lappi:2017skr})
\begin{align}
    \alpha_s G^{\alpha \alpha'}_Y(\pt) = 2 C_F \int \frac{\der^2 \bt \der^2 \bt'}{(2\pi)^4} e^{-i\pt\cdot \Bt} \frac{\partial^2 \Gamma_Y(\bt,\bt')}{\partial \btL{\alpha} \partial \btCL{\alpha'} } \frac{  \left[1-\exp\left( -\frac{C_A}{C_F} \Gamma_Y(\bt,\bt')\right) \right] }{\Gamma_Y(\bt,\bt')} \,,
\end{align}
where
\begin{align}
    \Gamma_Y(\bt,\bt') = -\ln\left(S^{(2)}_Y(\bt,\bt') \right)\,.
\end{align}
Furthermore, assuming translational invariance:
\begin{align}
    G^{0}_Y(p_\perp) = \frac{2 C_F S_\perp}{\alpha_s (2\pi)^3}  \int B_\perp \der B_\perp J_0(p_\perp B_\perp) & \frac{1}{\Gamma_Y(\Bt)} \left[\frac{\partial^2 \Gamma_Y(\Bt)}{\partial B_\perp^2} + \frac{1}{B_\perp} \frac{\partial \Gamma_Y(\Bt)}{\partial B_\perp} \right]  \left[1-\exp\left( -\frac{C_A}{C_F} \Gamma_Y(\Bt)\right) \right]\,, \label{eq:unpol_WW_Gaussian} \\
    h^{0}_Y(p_\perp) = -\frac{2 C_F S_\perp}{\alpha_s (2\pi)^3}  \int B_\perp \der B_\perp J_2(p_\perp B_\perp) & \frac{1}{\Gamma_Y(\Bt)} \left[\frac{\partial^2 \Gamma_Y(\Bt)}{\partial B_\perp^2} - \frac{1}{B_\perp} \frac{\partial \Gamma_Y(\Bt)}{\partial B_\perp} \right]  \left[1-\exp\left( -\frac{C_A}{C_F} \Gamma_Y(\Bt)\right) \right]\,.
    \label{eq:linpol_WW_Gaussian}
\end{align}
where $\Bt = \bt-\bt'$ and $S_\perp$ is the transverse area of the target, $\int \der^2\bt \der^2 \bt' =  S_\perp \int \der^2 \Bt$. Here have abused the notation, and written the translational invariant correlator $\Gamma_Y(\Bt=\bt-\bt') = \Gamma_Y(\bt,\bt')$.

\section{Additional numerical results for the short distance coefficients}
\label{app:additional_numerical_results} 
In this appendix, we provide the results for the short distance coefficients when the photon is longitudinally polarized and for the off-diagonal matrix elements LT and Tflip corresponding to different polarizations of the photon in the amplitude and complex conjugate amplitude. The $p_\perp$ and $Q$ dependencies are shown in Figs.\,\eqref{fig:SDC-Longitudinal-Offdiagonal-pTdep} and \eqref{fig:SDC-Longitudinal-Offdiagonal-Qdep} respectively.

\begin{figure}[H]
    \centering
    \includegraphics[width=0.49\textwidth]{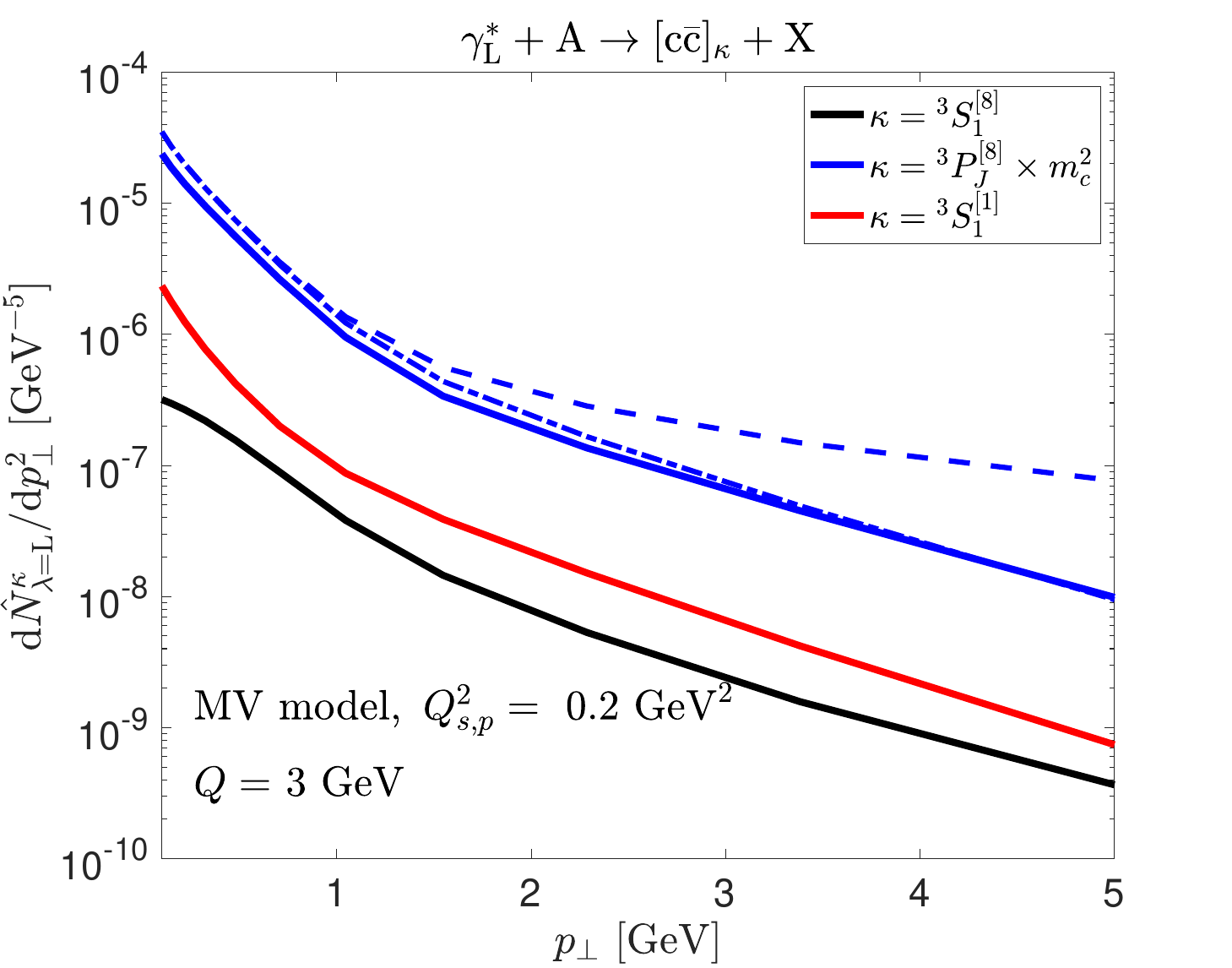}
    \includegraphics[width=0.49\textwidth]{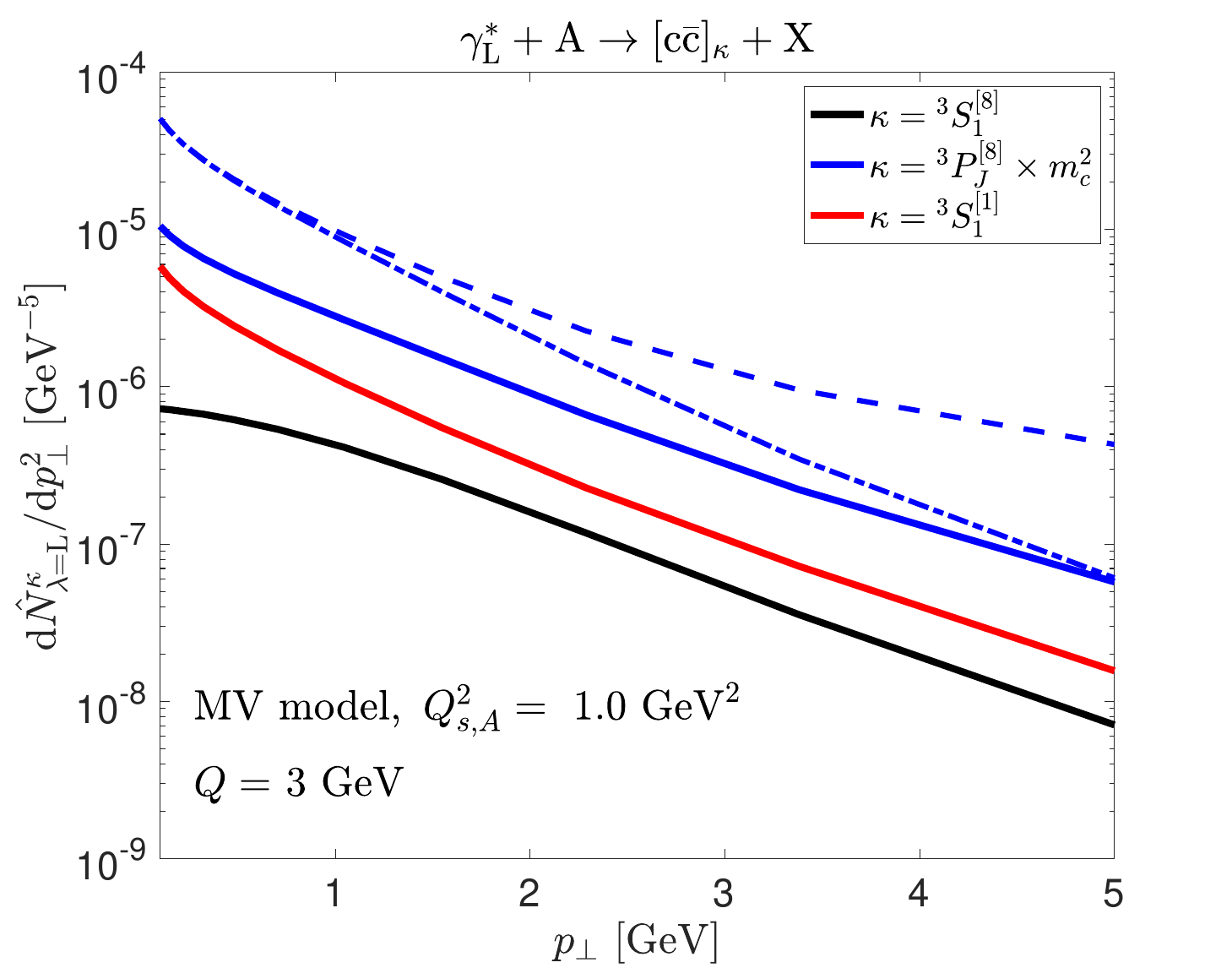}
    \includegraphics[width=0.49\textwidth]{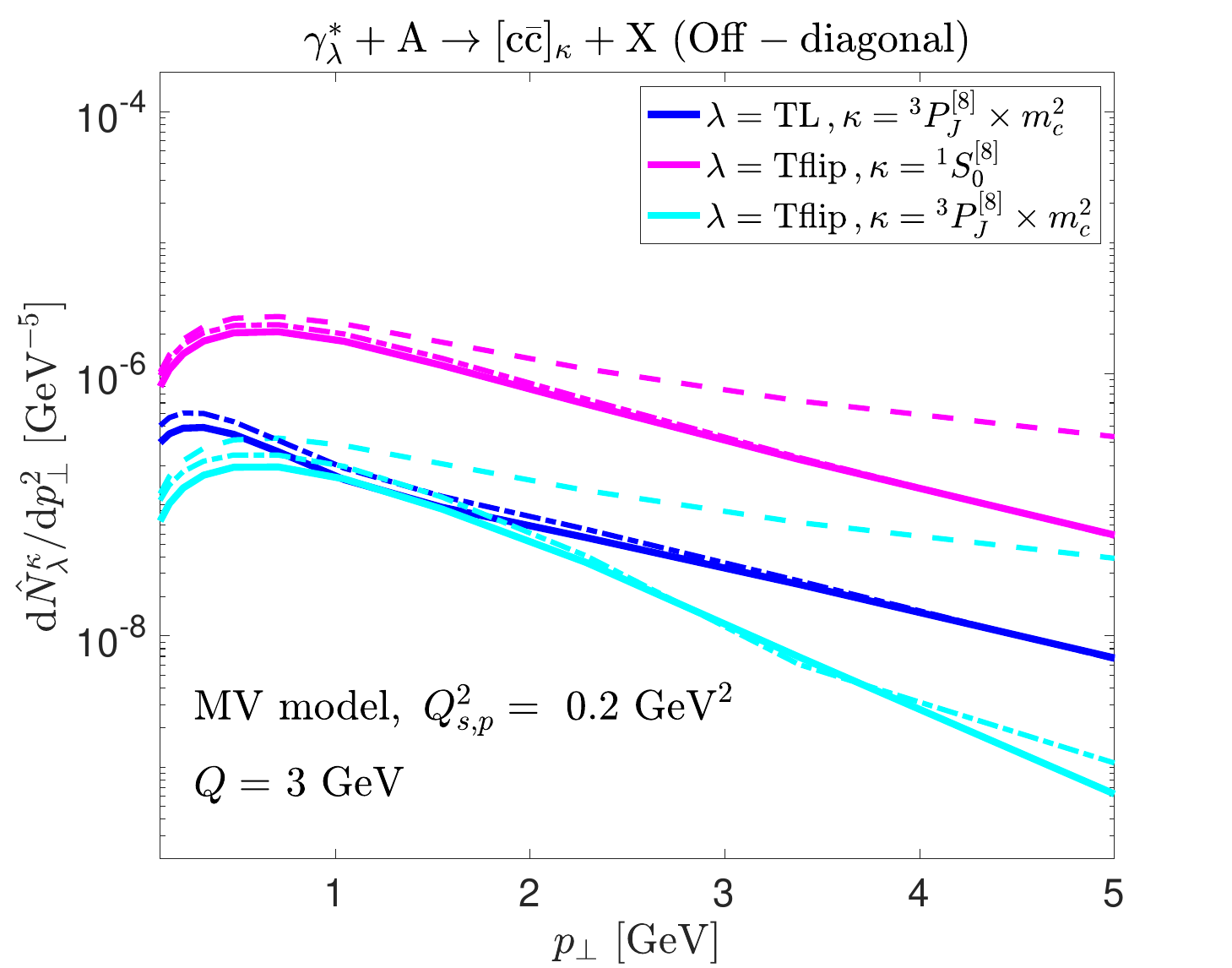}
    \includegraphics[width=0.49\textwidth]{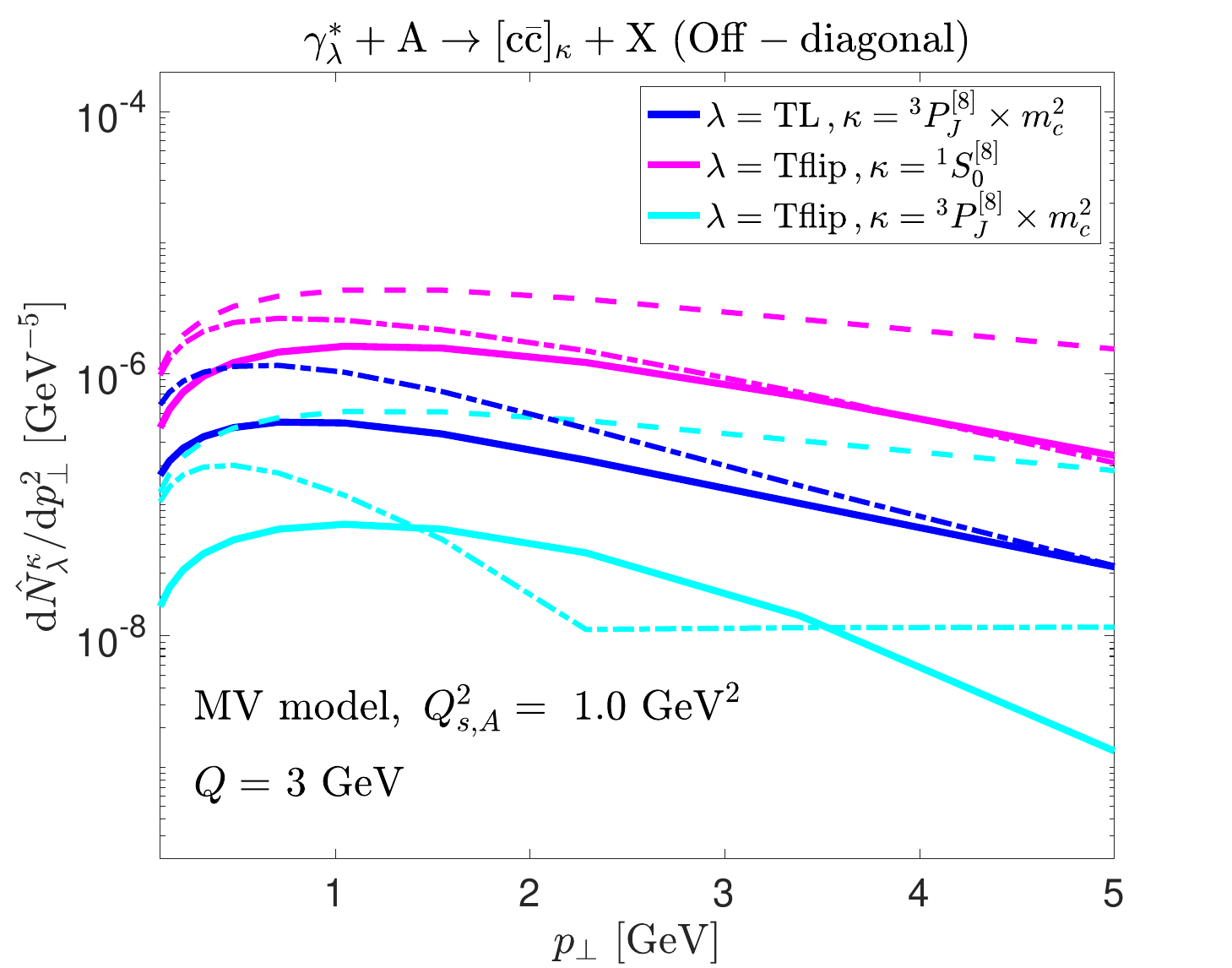}
    \caption{The $p_\perp$-dependence of the short distance coefficients computed in the CGC (solid lines), the improved TMD (dashed-dotted), and the TMD (dashed). All results are shown at fixed virtuality $Q= 3.0\ \rm{GeV}$. The upper panels show the results when the photon is longitudinally polarized, and the lower panels show the result for the absolute value of the off-diagonal matrix elements (interference of polarization). Panels on the left show the results at $Q_s^2 = 0.2 \ \rm{GeV}^2$ (proton).  Panels on the right show the results at $Q_s^2 = 1.0 \ \rm{GeV}^2$ (large nucleus).  The short distance coefficients corresponding to the P wave are multiplied by $m_c^2$. }
    \label{fig:SDC-Longitudinal-Offdiagonal-pTdep}
\end{figure}

\begin{figure}[H]
    \centering
    \includegraphics[width=0.49\textwidth]{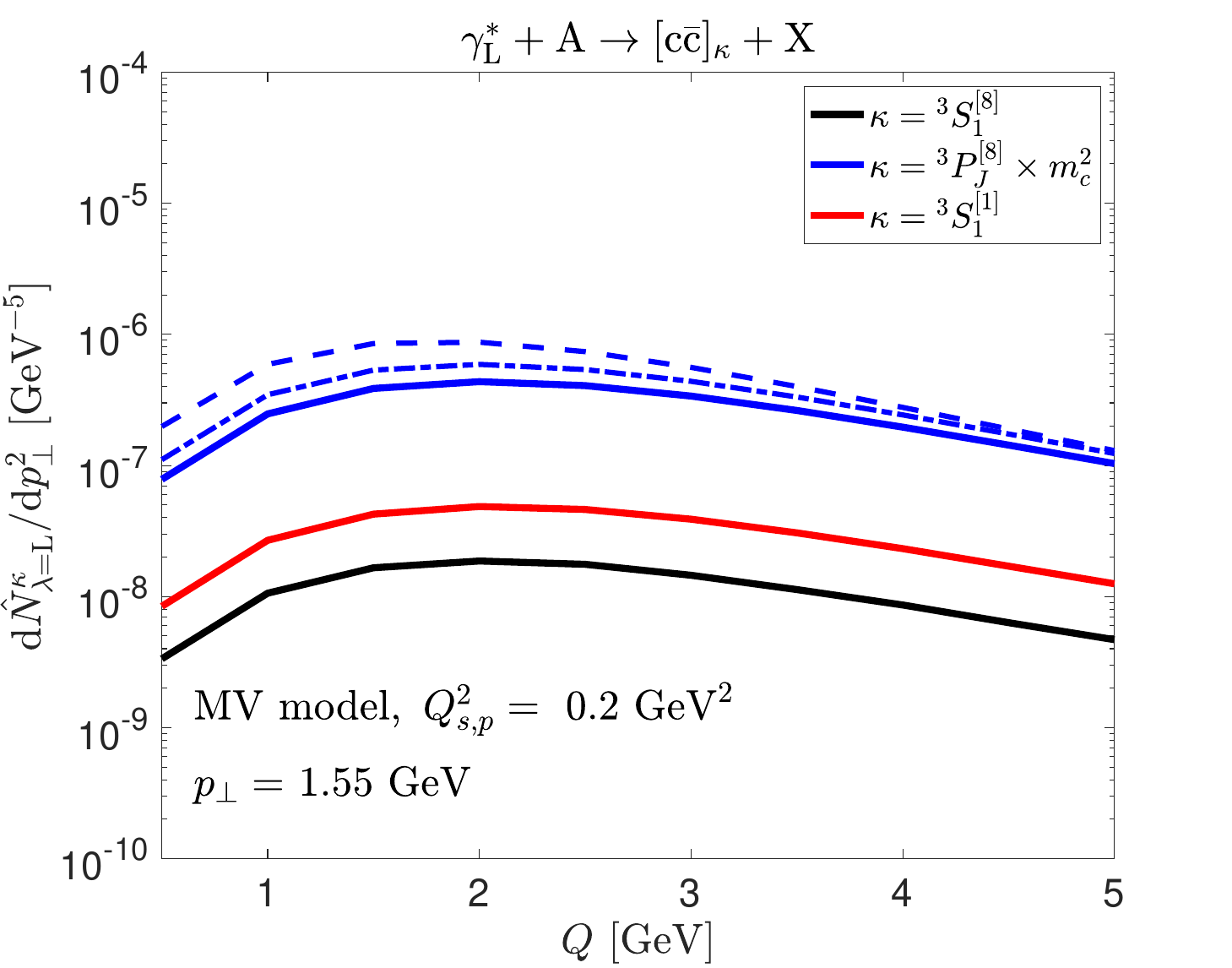}
    \includegraphics[width=0.49\textwidth]{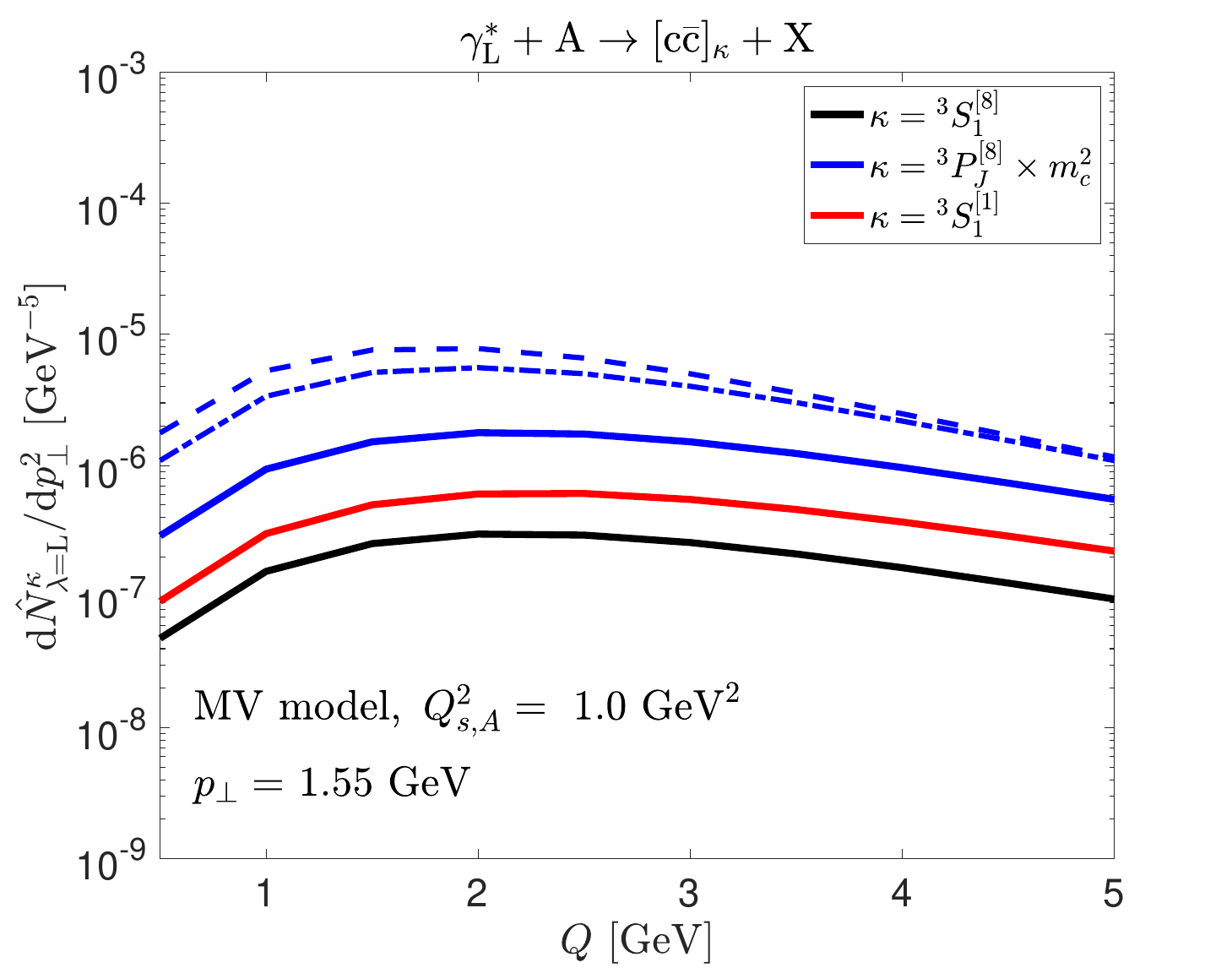}
    \includegraphics[width=0.49\textwidth]{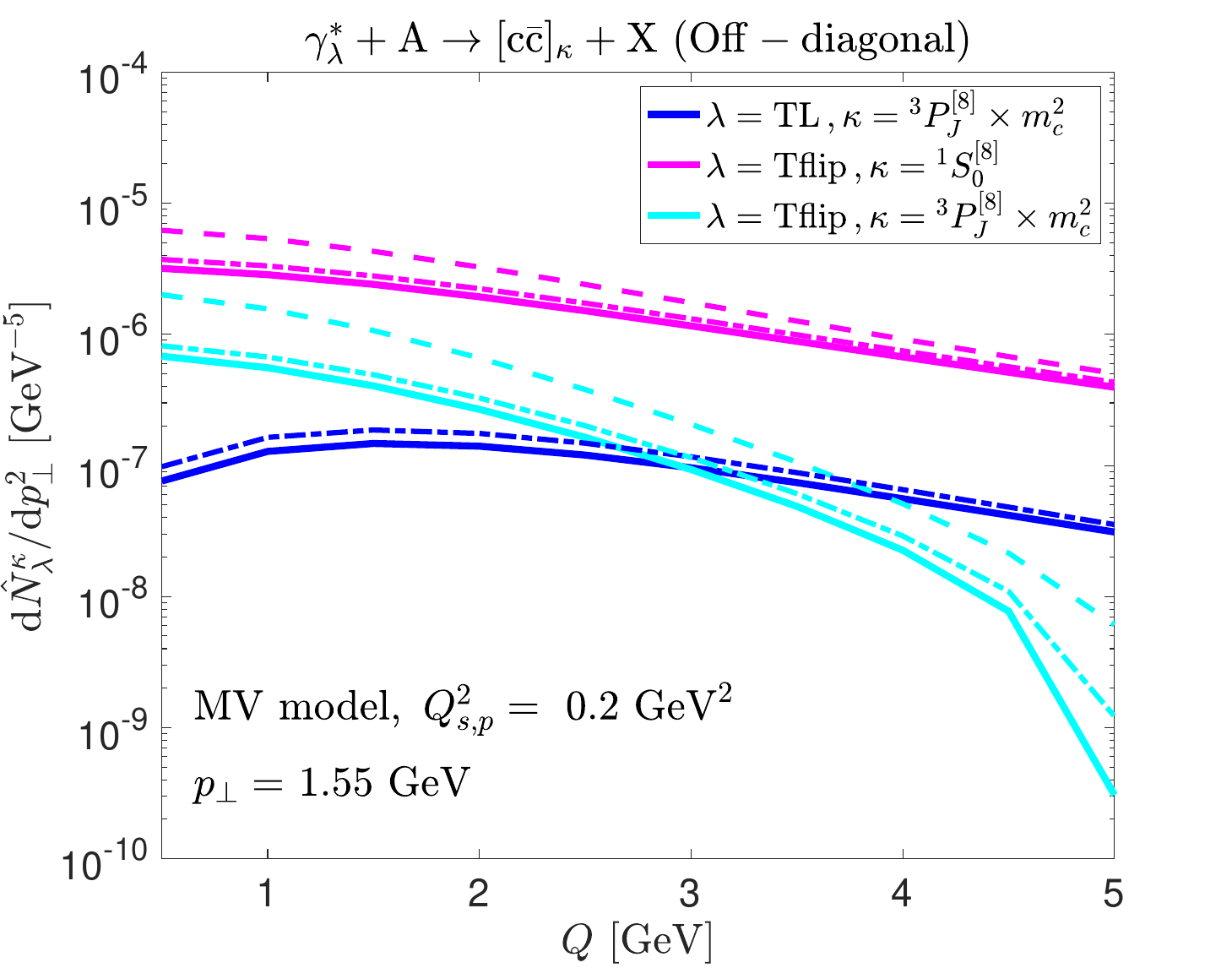}
    \includegraphics[width=0.49\textwidth]{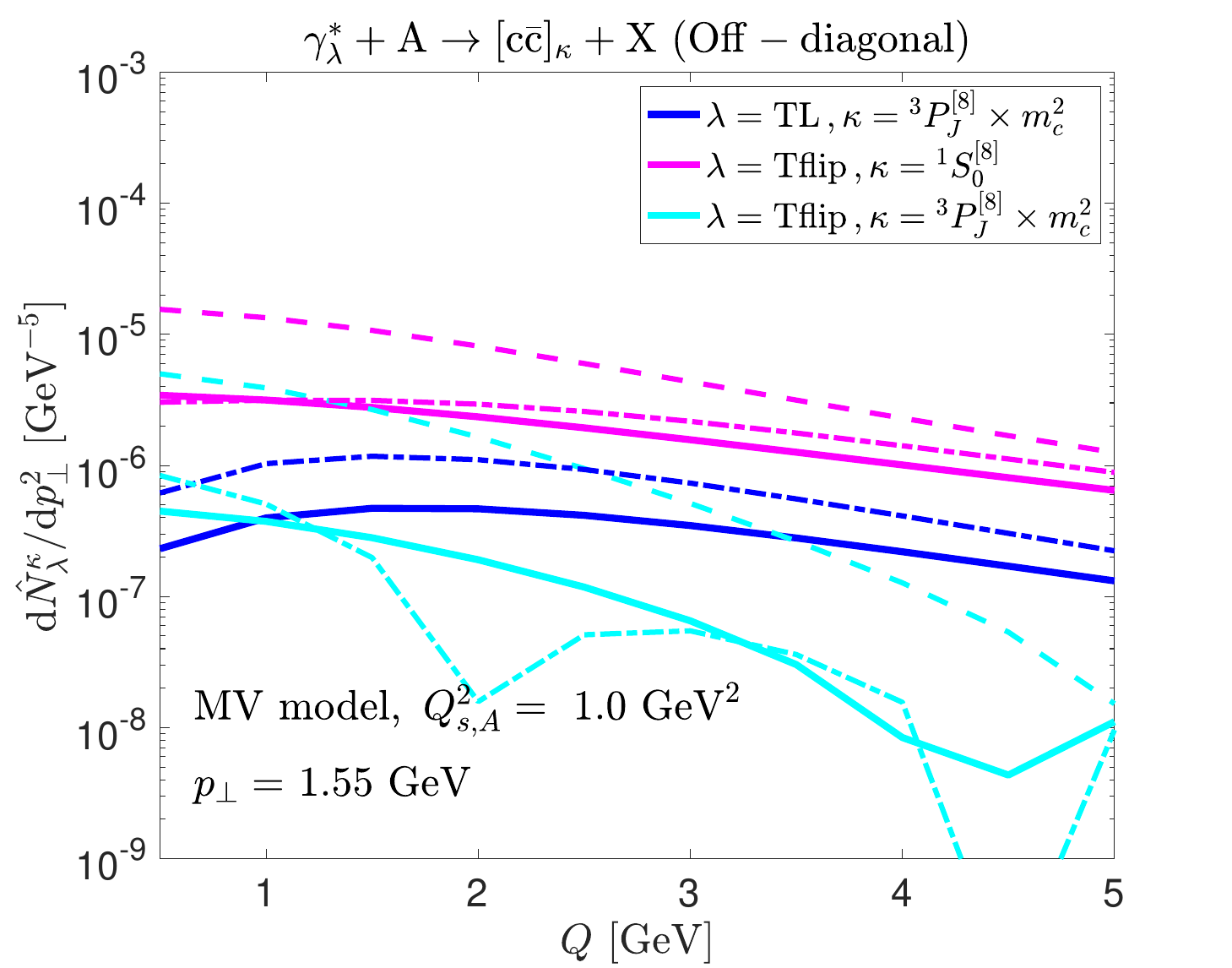}
    \caption{The $Q$-dependence of the short distance coefficients computed in the CGC (solid lines), the improved TMD (dashed-dotted), and the TMD (dashed). All results are shown at fixed virtuality $Q= 3.0\ \rm{GeV}$. The upper panels show the results when the photon is longitudinally polarized, and the lower panels show the result for the absolute value of the off-diagonal matrix elements (interference of polarization). Panels on the left show the results at $Q_s^2 = 0.2 \ \rm{GeV}^2$ (proton).  Panels on the right show the results at $Q_s^2 = 1.0 \ \rm{GeV}^2$ (large nucleus). The short distance coefficients corresponding to the P wave are multiplied by $m_c^2$. }
    \label{fig:SDC-Longitudinal-Offdiagonal-Qdep}
\end{figure}

\end{widetext}
\bibliography{ref}

\end{document}